\documentclass[fleqn,usenatbib]{mnras}

\usepackage{graphicx}	
\usepackage{amsmath}	
\usepackage{amssymb}	
\usepackage{multicol}        
\usepackage{pdflscape}	
\usepackage{lipsum}

\usepackage{newtxtext,newtxmath}

\DeclareRobustCommand{\VAN}[3]{#2}
\let\VANthebibliography\thebibliography
\def\thebibliography{\DeclareRobustCommand{\VAN}[3]{##3}\VANthebibliography}

\usepackage{xspace}	
\usepackage{color}
\usepackage{lineno}
\usepackage{fleqn,natbib}
\usepackage{flushend}
\usepackage{cuted}
\usepackage{rotating}
\usepackage{xcolor}
\usepackage{dcolumn}
\usepackage{multicol}
\usepackage{adjustbox}
\usepackage{hyperref}

\newcommand{\thisgrb}{GRB~210619B\xspace}
\newcommand{\fermi}{{\em Fermi}\xspace}
\newcommand{\kw}{{\em Konus}-Wind\xspace}

\newcommand{\fermiT}{{T$_{\rm 0}$}\xspace}
\newcommand{\keV}{{\rm keV}\xspace}

\newcommand{\swift}{{\em Swift}\xspace}
\newcommand{\tninty}{{$T_{\rm 90}$}\xspace}
\newcommand{\mvts}{{$t_{\rm mvts}$}\xspace}

\newcommand{\Ep}{$E_{\rm p}$\xspace}
\newcommand{\sw}[1]{\texttt{#1}}
\graphicspath{{./}{figures/}}

\usepackage{color}
\usepackage{amsmath}
\usepackage{multirow,longtable,supertabular}
\usepackage{float}
\usepackage{tablefootnote}
\usepackage{threeparttable}
\title[Peculiar spectral evolution of \thisgrb]{Multi-wavelength study of the luminous \thisgrb observed with {\it Fermi} and ASIM}

\author[Caballero-Garc\'{i}a et~al.]{M.~D. Caballero-Garc\'{i}a$^{1}$\thanks{E-mail:
mcaballero@iaa.es}, Rahul Gupta$^{2,3}$\thanks{E-mail: rahulbhu.c157@gmail.com}, S. B. Pandey$^{2}$, S. R. Oates$^{4}$, M. Marisaldi$^{5,6}$, A. Ramsli$^{5}$, \newauthor Y.-D. Hu$^{1}$, 
A. J. Castro-Tirado$^{1,7}$, R. S\'anchez-Ram\'{\i}rez$^{1}$, P. H. Connell$^{8}$, F. Christiansen$^{9}$, A. Kumar Ror$^{2}$, \newauthor  A. Aryan$^{2,3}$,  
 J.-M. Bai$^{10}$, M. A. Castro-Tirado$^{1,7}$, Y.-F. Fan$^{10}$,
E. Fern{\'a}ndez-Garc{\'\i}a$^{1}$, 
A. Kumar$^{2,11}$, \newauthor
A. Lindanger$^{5}$, A. Mezentsev$^{5}$, 
J. Navarro-Gonz\'alez$^{8}$, T. Neubert$^{9}$, N. {\O}stgaard$^{5}$, I. P\'erez-Garc\'{i}a$^{1}$, \newauthor
V. Reglero$^{8}$, D. Sarria$^{5}$, T. R. Sun$^{1}$, D.-R. Xiong$^{10}$, J. Yang$^{12}$, Y.-H. Yang$^{12}$, and B.-B. Zhang$^{12,13}$ 
\\
$^{1}$ Instituto de Astrof\'isica de Andaluc\'ia (IAA-CSIC), Glorieta de la Astronom\'ia s/n, E-18008, Granada \\
$^{2}$ Aryabhatta Research Institute of Observational Sciences (ARIES), Manora Peak, Nainital-263002, India \\
$^{3}$ Department of Physics, Deen Dayal Upadhyaya Gorakhpur University, Gorakhpur-273009, India \\
$^{4}$ School of Physics and Astronomy \& Institute for Gravitational Wave Astronomy, University of Birmingham, B15 2TT, UK \\
$^{5}$ Birkeland Centre for Space Science, Department of Physics and Technology, University of Bergen, Norway \\
$^{6}$ National Institute for Astrophysics, Osservatorio di Astrofisica e Scienzia dello Spazio, Bologna, Italy \\
$^{7}$ Unidad Asociada al CSIC, Departamento de Ingenier\'ia de Sistemas y Autom\'atica, Escuela de Ingenier\'ias, Universidad de M\'alaga, M\'alaga, Spain \\
$^{8}$ Image Processing Laboratory, University of Valencia, Paterna, Valencia, Spain \\
$^{9}$ National Space Institute, Technical University of Denmark, Kgs. Lyngby, Denmark \\
$^{10}$ Yunnan Observatories, Chinese Academy of Sciences, Kunming, 650216, China \\
$^{11}$ School of Studies in Physics and Astrophysics, Pandit Ravishankar Shukla University, Raipur, Chattisgarh-492010, India \\
$^{12}$ School of Astronomy and Space Science, Nanjing University, Nanjing 210093, China \\
$^{13}$ Key Laboratory of Modern Astronomy and Astrophysics (Nanjing University), Ministry of Education, Nanjing 210093, China }

\date{Accepted XXX. Received YYY; in original form ZZZ}

\pubyear{2022}

\begin{document}
\label{firstpage}
\pagerange{\pageref{firstpage}--\pageref{lastpage}}
\maketitle

\begin{abstract}
We report on detailed multi-wavelength observations and analysis of the very bright and long \thisgrb, detected by the Atmosphere-Space Interactions Monitor (ASIM) installed on the International Space Station ({\it ISS}) and the Gamma-ray Burst Monitor (GBM) on-board the \fermi mission. Our main goal is to understand the radiation mechanisms and jet composition of \thisgrb. With a measured redshift of $z$ = 1.937, we find that \thisgrb falls within the 10 most luminous bursts observed by \fermi so far. The energy-resolved prompt emission light curve of \thisgrb exhibits an extremely bright hard emission pulse followed by softer/longer emission pulses. The low-energy photon indices ($\alpha_{\rm pt}$) values obtained using the time-resolved spectral analysis of the burst suggest a transition between the thermal (during harder pulse) to non-thermal (during softer pulse) outflow. We examine the correlation between spectral parameters and find that both peak energy and $\alpha_{\rm pt}$ exhibit the flux tracking pattern. The late time broadband photometric dataset can be explained within the framework of the external forward shock model with $\nu_m$ $< \nu_c$ $< \nu_{x}$ (where $\nu_m$, $\nu_c$, and $\nu_{x}$ are the synchrotron peak, cooling-break, and X-ray frequencies, respectively) spectral regime supporting a rarely observed hard electron energy index ($p<$ 2). We find moderate values of host extinction of E(B-V) = 0.14 $\pm$ 0.01 mag for the Small Magellanic Cloud (SMC) extinction law. In addition, we also report late-time optical observations with the 10.4\,m GTC placing deep upper limits for the host galaxy ($z$=1.937), favouring a faint, dwarf host for the burst.

\end{abstract}

\begin{keywords}
{gamma-ray burst: general, gamma-ray burst: individual: \thisgrb, methods: data analysis}
\end{keywords}



\section{Introduction}

Gamma-ray bursts (GRBs) are the brightest and most explosive electromagnetic transients in the Universe. GRBs originate at cosmological distances with energy releases of $10^{51}-10^{53}$\,erg\,s$^{-1}$ and are not well-understood in terms of physical models more than five decades after their serendipitous discovery \citep{1973ApJ...182L..85K, 2015PhR...561....1K}. The multi-wavelength emission (from very high energy through to the radio) that follows the gamma-ray emission (the ``afterglow") partly satisfies the predictions of the ``standard" relativistic fireball model \citep{1993ApJ...405..278M}. 

The progenitors that produce the long-duration ($>2$\,s) GRBs are considered to be the collapse of massive stars \citep{1993ApJ...405..273W, 2003Natur.423..847H, 2006Natur.442.1008C}. They are believed to be powered by a central engine (stellar-mass black holes or millisecond magnetars; \citealt{2020MNRAS.496.2910P}). Furthermore, optical/near-IR studies of GRB afterglows and their hosts have provided insight into key aspects related to their underlying physical mechanisms and possible progenitors.

GRB prompt emission is observed when the relativistic jets dissipate the energy and accelerate the particles either via internal shocks or magnetic reconnection \citep{2015AdAst2015E..22P}. These processes radiate and provide a non-thermal spectrum, probably synchrotron radiation \citep{2020NatAs...4..174B, 2020NatAs...4..210Z}. However, synchrotron emission is not consistent with the observed spectra for a large number of GRBs \citep{2015AdAst2015E..22P, 2014IJMPD..2330002Z}. The key feature of the prompt emission spectra of GRBs is the sub-MeV peak, which can be explained well using the empirical \sw{Band} function \citep{Band:1993}. In addition to non-thermal synchrotron emission (\sw{Band}) originated at the optically thin region, a quasi-thermal blackbody component (produced by the photosphere) is also observed in some GRBs \citep{2005ApJ...625L..95R, 2011ApJ...727L..33G, 2014IJMPD..2330002Z}. According to the classical fireball model, initially, the jetted outflow is optically thick, and due to expansion, the flow becomes transparent at a certain point (also known as photospheric radius, r$_{\rm ph}$), and thermal radiation emitted towards the observer, results as a subdominant/dominant part of the entire observed prompt gamma-ray spectra \citep{2017SSRv..207...87B}. Moreover, some spectra show a few additional features, such as a sub-GeV spectral cut-off \citep{2018ApJ...864..163V, 2020ApJ...903....9C}, a low energy break in the spectrum \citep{2019A&A...625A..60R, 2022MNRAS.511.1694G}, and multiple components \citep{2019ApJ...876...76T}. Thus, it is challenging to explain the radiation physics of GRBs using a single model (different spectral components are associated with different emissions originated at different regions).

Time-resolved spectral analysis is a unique method for understanding the temporal evolution of modelled spectral parameters and also providing insights into the origin of the prompt emission \citep{2015AdAst2015E..22P}. The prompt light curve of GRBs generally consists of multiple pulses produced by irregular central engine activity or multiple internal shocks, and the temporal evolution of spectral parameters (mainly peak energy, \Ep ) among the pulses. There are four possible patterns of \Ep evolution reported in the literature: (i) `soft-to-hard (STH)' pattern, where \Ep becomes harder with time \citep{1994ApJ...422..260K}; (ii) `hard-to-soft (HTS)' pattern, where \Ep becomes softer with time \citep{1986ApJ...301..213N}; (iii) `flux-tracking (FT)' pattern, where \Ep becomes harder when the flux increases or vice versa \citep{1999ApJ...512..693R};  (iv) random pattern, where \Ep evolves randomly \citep{1994ApJ...422..260K}. The evolution pattern of \Ep has been used to explain the observed spectral lag in the bursts light curves \citep{2018ApJ...869..100U}. In addition to \Ep evolution, the low energy power-law index ($\alpha_{\rm pt}$) also varies over time; however, it does not have any particular pattern \citep{1997ApJ...479L..39C}. Recently, \cite{2019ApJ...886...20Y} studied 37 single pulses of 38 GRBs observed by \fermi GBM till 2018 and found that most of the individual pulses ($\sim$ 60\%) have a harder $\alpha_{\rm pt}$ than the synchrotron line of death ($\alpha_{\rm pt}$ = -2/3, i.e. LOD; \citealt{1998ApJ...506L..23P}). More recently, \cite{2021ApJS..254...35L} performed the time-resolved spectral analysis of multi-pulsed GRBs observed by \fermi GBM and found that the typical pattern of the pulses becomes softer with time, mostly with $\alpha_{\rm pt}$ becoming smaller, suggesting a transition from the photosphere to synchrotron emission. For some bright bursts, some authors  \citep{2019ApJ...884..109L, 2021MNRAS.505.4086G, 2022MNRAS.511.1694G} examined the evolution pattern of spectral parameters and noticed that \Ep and $\alpha_{\rm pt}$ both show a flux tracking (``double-tracking'') behavior.

In this paper, we studied in detail the prompt emission and afterglow through the analysis of one of the most luminous bursts (\thisgrb). It was promptly detected by the Modular X- and Gamma-ray Sensor \citep{2019SSRv..215...23O} on the Atmosphere-Space Interactions Monitor ({\it ASIM}; \citealt{2019arXiv190612178N}) installed on the International Space Station ({\it ISS}) and the Gamma-ray Burst Monitor (GBM; \citealt{2009ApJ...702..791M}) onboard the \fermi mission. The GBM discovered this bright and long GRB which was also observed by ASIM.

We also report our optical follow-up observations using many observatories, including the host galaxy search from the observations with the {\it Gran Telescopio de Canarias} (10.4\,m) GTC (Canary Islands, Spain). This GRB detection by ASIM (plus \swift-BAT and \fermi-GBM) having peculiar properties along with the late-time optical follow-up observations using the 10.4\,m GTC motivated us for a detailed analysis of this luminous \thisgrb.

The paper is organized as follows. In \S~\ref{prompt}, we give the details of high energy observations, data analysis, results, and discussion of the prompt emission of \thisgrb. In \S~\ref{afterglow}, we give the details of broadband observations, data analysis, results, and discussion of the afterglow of \thisgrb. Finally, a brief summary and conclusion is given in \S~\ref{summaryandconclusions}. We used a standard cosmology model (H$_{0}$= 70 \,km $\rm s^{-1}$ $\rm Mpc^{-1}$, ${\Omega}_{\rm M}$= 0.27, and ${\Omega}_{\lambda}$= 0.73) to calculate the restframe parameters.

\section{High energy emission of \thisgrb}
\label{prompt}

\subsection{Observations and data analysis}

\thisgrb was initially detected by the Gamma-Ray Burst Monitor (GBM) and the Burst Alert Telescope (BAT; \citealt{2005SSRv..120..143B}), being on-board the \fermi and \swift satellites, respectively \citep{2021GCN.30279....1P, 2021GCN.30261....1D}. This section presents the prompt observational and data analysis details carried out by \fermi, ASIM-{\it ISS}, and \swift ~missions.

\subsubsection{\fermi GBM}
\label{GBM}

\begin{figure}
\centering
\includegraphics[scale=0.35]{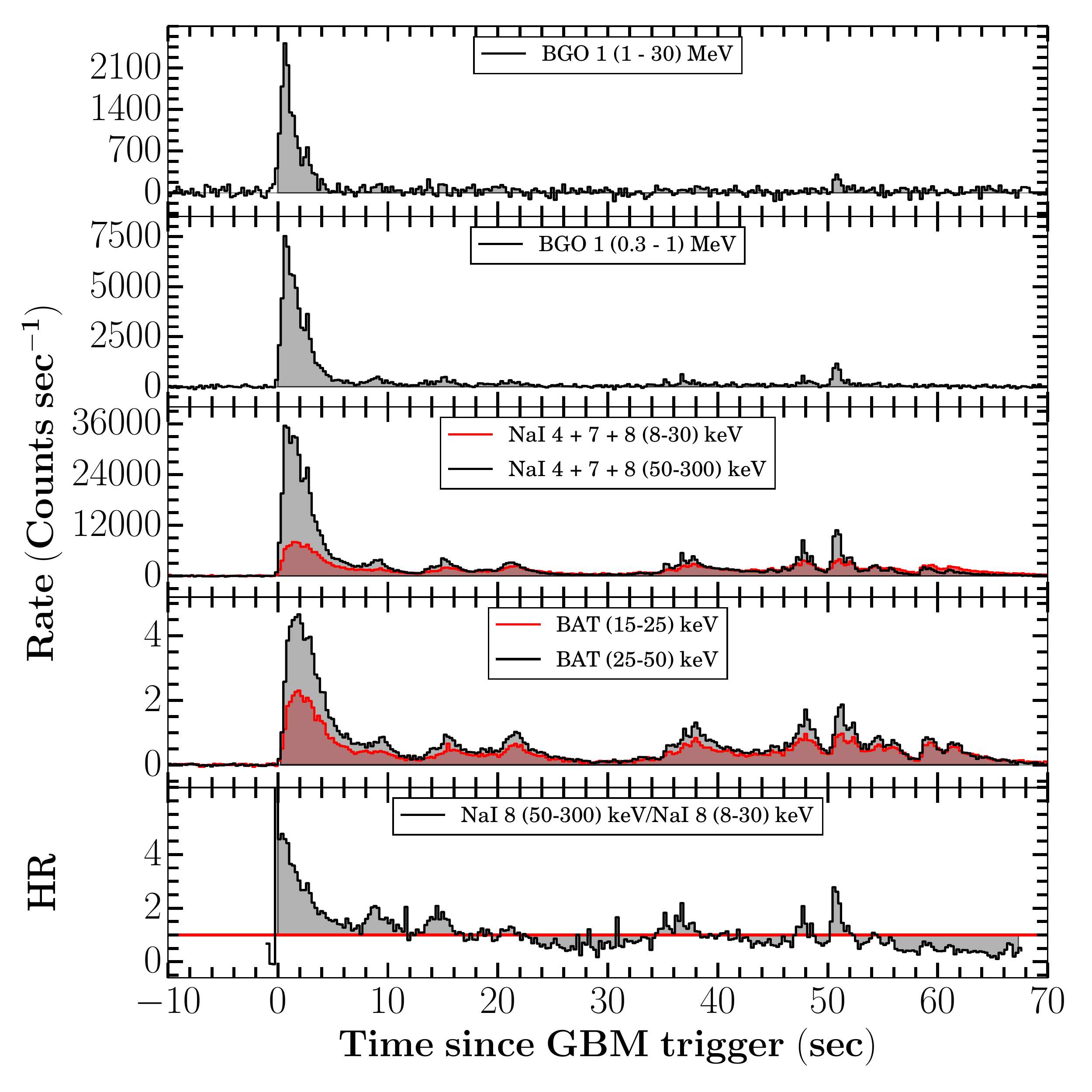}
\caption{{\bf Prompt light curve of \thisgrb:} The top four panels show the background-subtracted multi-channel prompt emission light curve of \thisgrb obtained using \fermi GBM and \swift BAT observations (256\,ms bin size). The grey and red shading regions show the time-interval (\fermiT to \fermiT+ 67.38\,s) of the light curve used for the time-averaged spectral analysis. The bottom-most panel shows the hardness ratio, obtained using the ratio of the count-rate light curve in the hard and soft energy range. The solid red line indicates the hardness ratio equal to one.}
\label{promptlc}
\end{figure}

The \fermi GBM detected \thisgrb (see the light curve in Figure~\ref{promptlc}) at 23:59:25.60 UT on 19 June 2021 \citep{2021GCN.30279....1P}. We used the time-tagged events (TTE) mode \fermi GBM data\footnote{\url{https://heasarc.gsfc.nasa.gov/W3Browse/fermi/fermigbrst.html}} for the temporal and spectral analysis of \thisgrb. The TTE mode observations have a good temporal and spectral resolution. For the GBM temporal and spectral analysis, we considered the three brightest sodium iodide (NaI) and one of the brightest bismuth germanate (BGO) detectors. We have used \sw{RMFIT} version 4.3.2 software\footnote{\url{https://fermi.gsfc.nasa.gov/ssc/data/analysis/rmfit/}} to create the prompt emission light curve of \thisgrb in different energy channels. The background-subtracted \fermi GBM light curve of \thisgrb in different energy channels along with the hardness-ratio (HR) evolution is shown in Figure~\ref{promptlc}. The shaded region in Figure~\ref{promptlc} shows the time-interval (\fermiT to \fermiT+67.38\,s) for the time-averaged spectral analysis.

In addition to this, for the spectral analysis, we have used the same NaI and BGO detectors and reduced the spectrum using the \sw{gtburst}\footnote{\url{https://fermi.gsfc.nasa.gov/ssc/data/analysis/scitools/gtburst.html}} software. We selected the total emission interval (\fermiT to \fermiT+67.38\,s) for the time-averaged spectral analysis using a Bayesian block algorithm. For the spectral modelling of both time-averaged as well as time-resolved spectra, we have used the Multi-Mission Maximum Likelihood framework \citep[\sw{3ML}\footnote{\url{https://threeml.readthedocs.io/en/latest/}}]{2015arXiv150708343V} software since it is optimized for Bayesian analysis\footnote{\url{https://threeml.readthedocs.io/en/stable/xspec_users.html}}. We selected the GBM spectrum over 8-900 \keV and 250-40000 \keV for NaI detectors and BGO detector, respectively. In addition, we have ignored the NaI K-edge (30-40\,keV) energy range for both time-averaged and the time-resolved spectral analysis. For the time-averaged spectral modelling of \thisgrb, we have used the empirical \sw{Band} function and \sw{Cutoff power-law} models. We adopted the deviation information criterion (DIC; \citealt{doi:10.1080/01621459.1995.10476572, spiegelhalter2002bayesian}) to know the best fit function among the \sw{Band}, and \sw{Cutoff power-law} models and selected the best model with the least DIC value ($\Delta$ DIC $<$ -10; \citealt{2019ApJ...886...20Y, 2021ApJS..254...35L}). This suggests that the \sw{Band} function is better fitting with respect to the \sw{Cutoff power-law} model. To know about the best fit model, we utilized the following condition:

 $\Delta$DIC = $\Delta$DIC$_{\rm Band}$ - $\Delta$DIC$_{\rm CPL}$

The negative values of $\Delta DIC$ indicate that there is an improvement in the spectral fit, and if the difference of the deviation information criterion is less than - 10, in such scenario, the existence of \sw{Band} function in the spectrum is confirmed. The time-averaged spectral parameters for both models are tabulated in Table \ref{tab:TAS}.

\begin{table*}
\scriptsize
\caption{The time-averaged (\fermiT to \fermiT+ 67.38\,s) spectral analysis results of \thisgrb. The best fit model (\sw{Band}) is highlighted with boldface.}
\begin{center}
\label{tab:TAS}
\begin{tabular}{cccccccc}
\hline
\textbf{Model} & \multicolumn{4}{c|}{\textbf{Spectral parameters}} & \textbf{-Log(posterior)} & \textbf{DIC} & \textbf{$\rm \bf \Delta~ {DIC}$} \\ \hline 
\sw{\bf Band} &\bf $\bf \alpha_{\rm \bf pt}$=  -0.90$\bf ^{+0.01}_{-0.01}$ & \multicolumn{2}{c|}{\bf $\bf \beta_{\rm \bf pt}$= -2.05$\bf ^{+0.02}_{-0.02}$} &\bf \Ep = 216.08$^{+4.20}_{-4.20}$ &\bf -4347.76 &\bf 8693.99 & --\\ \hline 
\sw{CPL} & $\it \alpha_{\rm pt}$=  -1.03$^{+0.01}_{-0.01}$ & \multicolumn{3}{c|}{$E_{0}$ =320.30$^{+5.74}_{-5.63}$} & -4633.92 & 9251.24 & -557.25\\ \hline 
\end{tabular}
\end{center}
\end{table*}

{\bf Time-resolved spectral analysis:} We performed the time-resolved spectroscopy of \thisgrb using the same NaI and BGO detectors used for temporal/time-integrated analysis and the \sw{3ML} software. To select the temporal bins for the time-resolved spectral analysis, we used Bayesian blocks binning algorithm \citep{2013arXiv1304.2818S}. \cite{2014MNRAS.445.2589B} suggested that the Bayesian blocks algorithm is crucial to find the finest temporal bins and intrinsic spectral evolution. However, it could result in low signal-to-noise ratio due to high variability in the light curve. Therefore, we further selected only those bins which have statistical significance (S) $\geq$ 30. After implementing the Bayesian blocks and signal-to-noise ratio algorithm, we find a total of 59 spectra for the time-resolved spectroscopy of \thisgrb (see Table B1 of the appendix). We modelled each of the time-resolved spectra using individual empirical \sw{Band}, and \sw{Cutoff power-law} models.

\subsubsection{ASIM}
\label{ASIM_spectra}

The Atmosphere-Space Interactions Monitor (ASIM; \citealt{2019arXiv190612178N}) triggered on \thisgrb on June 19 2021 \citep{2021GCN.30315....1M}. The ASIM reference time used in the rest of the analysis is T$_{\rm 0, ASIM}$ = 23:59:24.915550 UT. At trigger time, it was local day time, so only High-Energy Detector (HED) data are available for this event. The burst direction was $135^{\circ}$ off-axis with respect to the Modular X- and Gamma-ray Sensor (MXGS; \citealt{2019SSRv..215...23O}) pointing direction. This is outside the instrument's nominal field of view, but the instrument is sensitive to the full solid angle because of its characteristics and high energy range. Photons interaction in the surrounding material and in the Columbus module are accounted for in the Monte Carlo simulation model of the instrument performed using the GEANT4 simulation toolkit \citep{Agostinelli2003}. This model was used to generate the Detector Response Matrix (DRM) of the instrument used for the spectral analysis.

Instrumental effects are carefully accounted for in HED data analysis by the implementation of the ``safety time" criterion, described in details in \cite{Lindanger2021}. This procedure removes any count which is closer in time to the previous count in the same detector than a specific time interval, ranging from 0 and $30~\mu s$ depending on the energy of the previous count. This procedure is introduced to guarantee the best possible energy estimate, which is to some extent affected by the energy of the previous counts. This effect is particularly relevant for Terrestrial Gamma-ray Flashes (TGFs), which exhibit very high fluxes on time scales of few hundreds of microseconds. In the case of \thisgrb, this effect is almost negligible and the ``safety time" criterion removes only about 0.7\% of the counts.

\begin{figure}
\centering
\includegraphics[angle=0,scale=0.35]{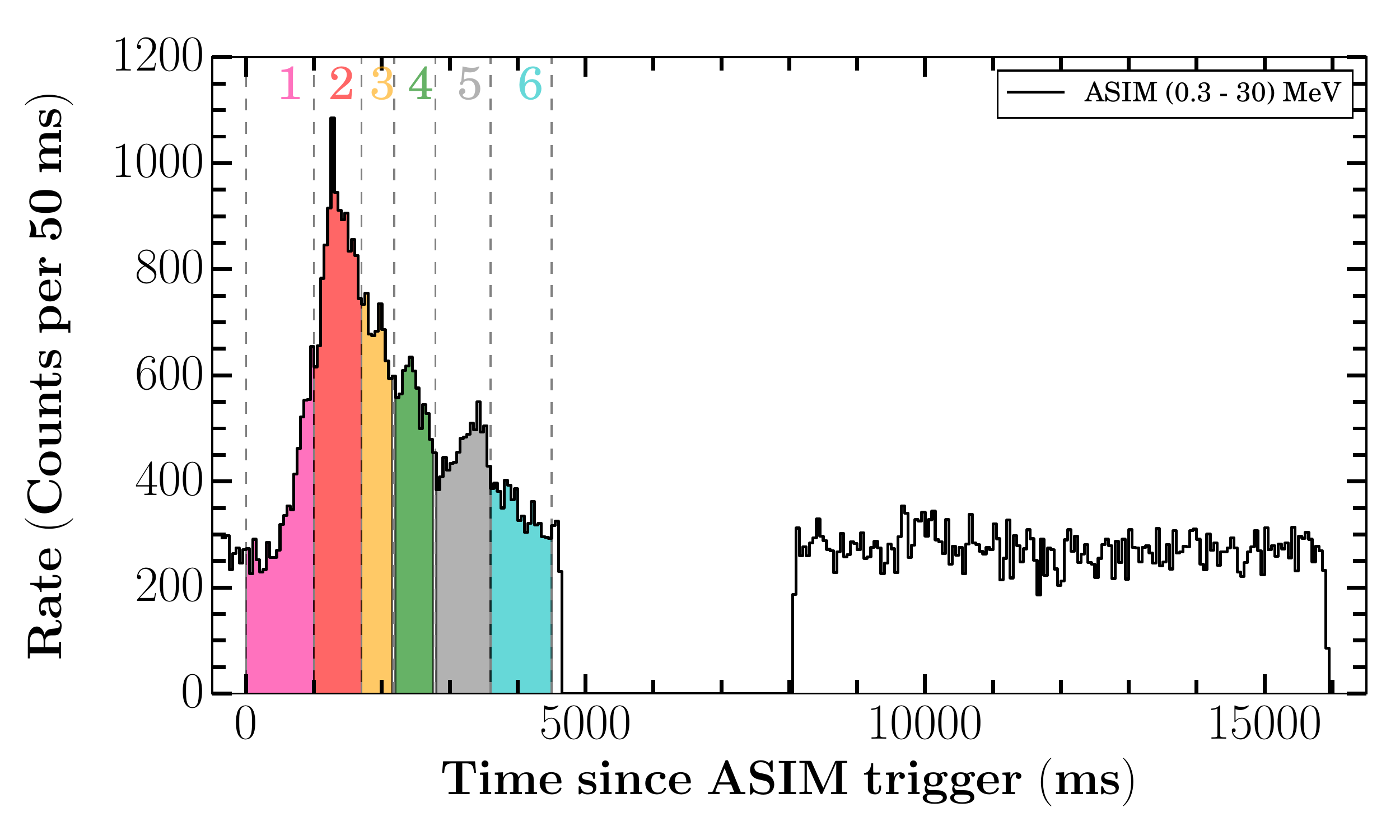}
\caption{The prompt emission (0.3 - 30)\,MeV light curve of the burst as seen by ASIM/HED. Vertical colored shaded regions (lines) separate the time intervals used for spectral analysis, labeled by numbers on top of the panel.}
\label{ASIM_LC_SED}
\end{figure}

The top panel of Figure~\ref{ASIM_LC_SED} shows the light curve of ASIM HED data in the (0.3-30\,MeV) energy range and with a time bining of 50\,ms. Because of the characteristics of the ASIM trigger logic, only part of the burst has been collected. A data gap is evident between 4.5 and 8\,s after ASIM reference time (T$_{\rm 0, ASIM}$). Only limited data are available prior to T$_{\rm 0, ASIM}$, therefore the background estimation for spectral analysis is based on the data in the time interval ($-420$ to 0\,ms) with respect to T$_{\rm 0, ASIM}$. To assess the spectral evolution of the burst, ASIM HED data were divided into seven-time intervals, as shown in the top panel of Figure~\ref{ASIM_LC_SED}. The choice of the intervals has been made in order to include in separate intervals all the peaks evident in the light curve (intervals 2, 3, 4, and 5). The spectral fitting has been carried out using the X-ray Spectral Fitting Package \citep[\sw{XSPEC};][]{1996ASPC..101...17A} software (version 12.12). To estimate the flux, the convolution model {\tt cflux} has been used. All time intervals from 1 to 6 can be well fit with a simple power law in the energy range 0.5 to 10\,MeV. The best-fit parameters and the fluxes in the energy range 0.5 to 10\,MeV are shown in Table~\ref{tab:asim}.

\subsubsection{\swift BAT and other gamma-ray detectors}

The BAT detected \thisgrb (see the mask-weighted light curve in Figure~\ref{promptlc}) at 23:59:25 UT on 19 June 2021 at the following location in the sky: RA = 319.713, DEC = +33.860 deg. (J2000) with an error radius of 3\,arcmin \citep{2021GCN.30261....1D}. The \swift BAT and \fermi GBM positions were consistent for \thisgrb.  

We downloaded the \swift BAT observations data of \thisgrb from the \swift database portal\footnote{\url{https://www.swift.ac.uk/swift_portal/}}. We processed the BAT data using the HEASOFT software (version 6.25). We reduced the \swift BAT data following the methodology presented in \cite{2021MNRAS.505.4086G} to obtain the energy-resolved mask-weighted light curve. The BAT light curves in the 25-50 \keV and 50-100 \keV energy ranges are shown in Figure~\ref{promptlc}.

The prompt emission of \thisgrb was also detected by the Gravitational wave high energy Electromagnetic Counterpart All sky Monitor ({\it GECAM}; \citealt{2021GCN.30264....1Z}), the \kw Experiment \citep{2021GCN.30276....1S}, the {\it CALET} Gamma-ray Burst Monitor (CGBM; \citealt{2021GCN.30284....1K}), the Mikhail Pavlinsky ART-XC telescope on board the {\it Spektr}-{\it RG} observatory \citep{2021GCN.30283....1L}, and the SPI-ACS/{\it INTEGRAL} \citep{2021GCN.30304....1M} satellites.

\subsubsection{Joint BAT-GBM-ASIM spectral analysis}

We performed joint spectral analysis of \swift BAT, \fermi GBM, and ASIM observations during the main emission pulse of \thisgrb, using the {\tt XSPEC} software. We used the Chi-squared statistics since the data-sets are well sampled. For this purpose, we created the time-sliced (six bins) spectra using \swift BAT and \fermi GBM data for the same temporal segments of ASIM (time-resolved analysis) discussed in section \ref{ASIM_spectra} for the joint spectral analysis. We obtained the continuous spectra in
the overlapping energy regions of \swift BAT (15-150 \,keV), \fermi GBM and BGO (8 \,keV-40\,MeV), and ASIM (500 \,keV to 10\,MeV) instruments. For the spectral fitting, we ignored the NaI K-edge (30-40 \keV). Therefore, we used the data in the spectral ranges of 8-30,40-900\, KeV; 250\,keV-40\,MeV; 15-150\,KeV and 500\,-10 MeV for the \fermi/GBM, \fermi/BGO, \swift/BAT and ASIM, respectively. The spectra were rebined using the {\tt grppha} task to have at least 20 counts for each background-subtracted spectral bin. 

We initially fitted each spectrum with two different models: 1) Band, 2) cutoff power-law models. In all the time intervals except 1 and 6 (where we got ${\chi}^{2}/{\nu}=1.45,1.57$ for ${\nu}=353,363$, respectively), we obtain bad fits ${\chi}^{2}/{\nu}{\gtrsim}2.0$
with the CPL ({\tt cutoffpl} in XSPEC) model (see Table ~\ref{Joint_spectralanalysis_Table}). On the other hand all the joint BAT-GBM-ASIM spectra are reasonably well fitted using a \sw{Band} 
function (${\chi}^{2}/{\nu}=1.0-1.5$, see Table ~\ref{Joint_spectralanalysis_Table}). \par

However there are still noticeable wavy residuals towards the high energy data points. We tried to improve them by adding a {\tt power-law} component in the previous model and did not get any significant improvement (see Table ~\ref{Joint_spectralanalysis_Table}). We also added a {\tt power-law} component
to the {\tt cutoffpl} model and whilst results improved the fits were statistically unacceptable (with the exception of Obs.~1 and 6, see comments above). We notice that
Obs.~1 and 6 are the ones showing the lowest (and similar) fluxes so it is difficult to disentangle between models in statistical terms (both {\tt Band} and {\tt cutoffpl} give similar
fits for these two observations). Therefore, we conclude that the {\tt Band} is the best fit for Obs.~1-6 presented in this paper. 

\begin{figure}
\centering
\includegraphics[scale=0.35]{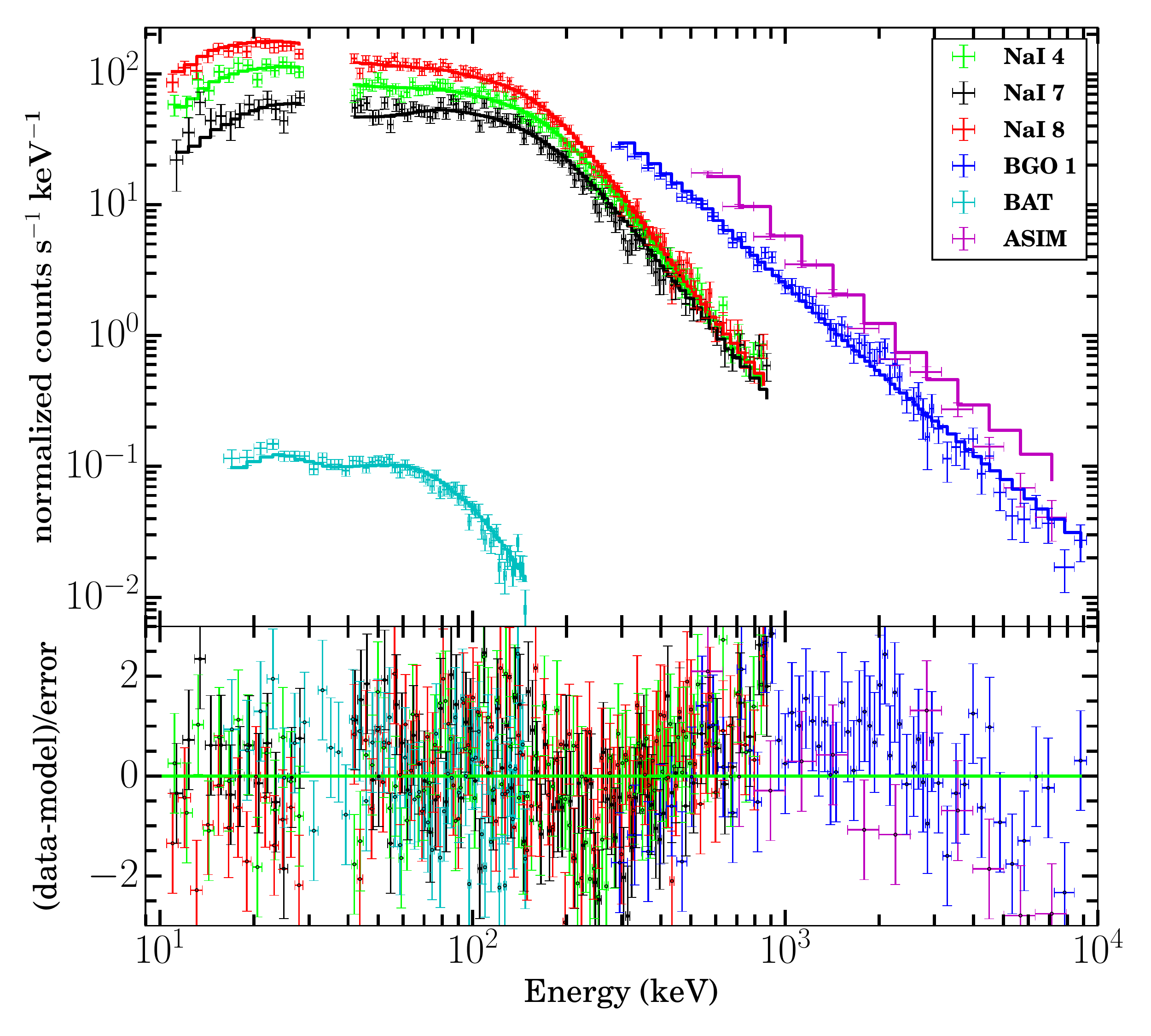}
\caption{{Top panel:} The best fit (\sw{Band}) joint \swift BAT (15-150 \keV), \fermi GBM (8 \keV-40 MeV), and ASIM (500 \keV to 10 MeV) spectrum for the temporal bin 2 (T$_{\rm 0, ASIM}$+1.00 to T$_{\rm 0, ASIM}$+ 1.70\,s) of \thisgrb. {Bottom panel:} The corresponding residuals of the model fitting.}
\label{TRS_joint}
\end{figure}

The joint \swift BAT, \fermi GBM, and ASIM count rate spectrum along with the best
fit model and residuals for one of the time bins (T$_{\rm 0, ASIM}$+1.00 to
T$_{\rm 0, ASIM}$+ 1.70\,s, i.e., the spectrum with the highest flux, Obs.~2) is shown in Figure \ref{TRS_joint}. The spectral parameters and the fluxes obtained for all the six bins are given in
Table~\ref{Joint_spectralanalysis_Table}.
The fluxes are calculated in the interval 8\,keV-10\,MeV using the multiplicative model
{\tt cflux} in {\tt XSPEC}. The cross-calibration constants between the instruments were left as free parameters in the fitting process. The resulting values for GBM NaI, BGO and ASIM HED are consistent with unity within few 10\%.

\begin{table*}
\caption{Joint \swift BAT, \fermi GBM, and ASIM spectral analysis results of the bright emission pulse of \thisgrb using \sw{Band}, \sw{CPL} and \sw{Band+pow}, \sw{CPL+pow} functions. Flux values (in erg $\rm cm^{-2}$ $\rm s^{-1}$) are calculated in 8 \keV-10\,MeV energy range. T$_{\rm start}$ and T$_{\rm stop}$ have been referred with respect to the ASIM reference time (T$_{\rm 0, ASIM}$). The errors given are $1{\sigma}$.}
\label{Joint_spectralanalysis_Table}
\begin{scriptsize}

\begin{center}
\begin{tabular}{|c|c|ccccc|ccc|} \hline
T$_{\rm start}$ (s) & T$_{\rm stop}$ (s)  & \boldmath $\it \alpha_{\rm pt}$ & \boldmath $\it \beta_{\rm pt}$ & \boldmath \Ep (\keV) &  \bf (Flux $\times 10^{-06}$)  & \bf $\chi^2_{\nu}$  (d.o.f.) & \bf {\tt Band+pow} & \bf {\tt cutoffpl}  & \bf {\tt cutoffpl+pow}  \\ \hline
0.00& 1.00 & $-0.41{\pm}0.08$      & $-1.79_{-0.06}^{+0.04}$ & $590_{-70}^{+100}$ & $24.1_{-1.6}^{+1.7}$    &  1.09\,(352)  &  1.10\,(350) & 1.45\,(353)  & 1.56\,(351)    \\ \hline
1.00& 1.70 & $-0.36{\pm}0.02$      & $-2.12{\pm}0.02$      & $422{\pm}14$         & $121.0_{-2.6}^{+2.7}$   &  1.30\, (417) &  1.31 (415)  & 3.25 (418)   & 2.17 (416)             \\ \hline
1.70& 2.18 & $-0.52{\pm}0.02$      & $-2.37_{-0.04}^{+0.03}$ & $380_{-14}^{+15}$  & $87.6{\pm}2.3$          &  1.20\, (381) &  1.21 (379)  & 1.86 (382)   & 1.62 (380)              \\ \hline
2.18& 2.79 & $-0.51{\pm}0.02$      & $-2.40{\pm}0.04$      & $317{\pm}10$         & $69.0_{-1.8}^{+1.9}$    &  1.30\, (382) &  1.31 (380)  & 1.93 (383)   &  1.73 (381)           \\ \hline
2.79& 3.60 & $-0.49{\pm}0.02$      & $-2.28{\pm}0.03$      & $250_{-7}^{+8}$      & $51.2{\pm}1.5$    &    1.48 (389) &   1.49 (387) & 2.29 (390)   &  2.02 (388)           \\ \hline
3.60& 4.50 & $-0.61{\pm}0.03$      & $-2.37_{-0.06}^{+0.05}$ & $202{\pm}7$        & $23.9_{-0.9}^{+1.0}$    &    1.24 (362) &   1.25 (360) & 1.57 (363)   &  1.53 (361)           \\ \hline
\end{tabular}
\end{center}
\end{scriptsize}
\end{table*}

\subsection{Results}

The prompt emission results of \thisgrb obtained using \fermi and ASIM data are presented in this section.

\subsubsection{Light curve and time-integrated spectrum}

\begin{figure}
\centering
\includegraphics[scale=0.45]{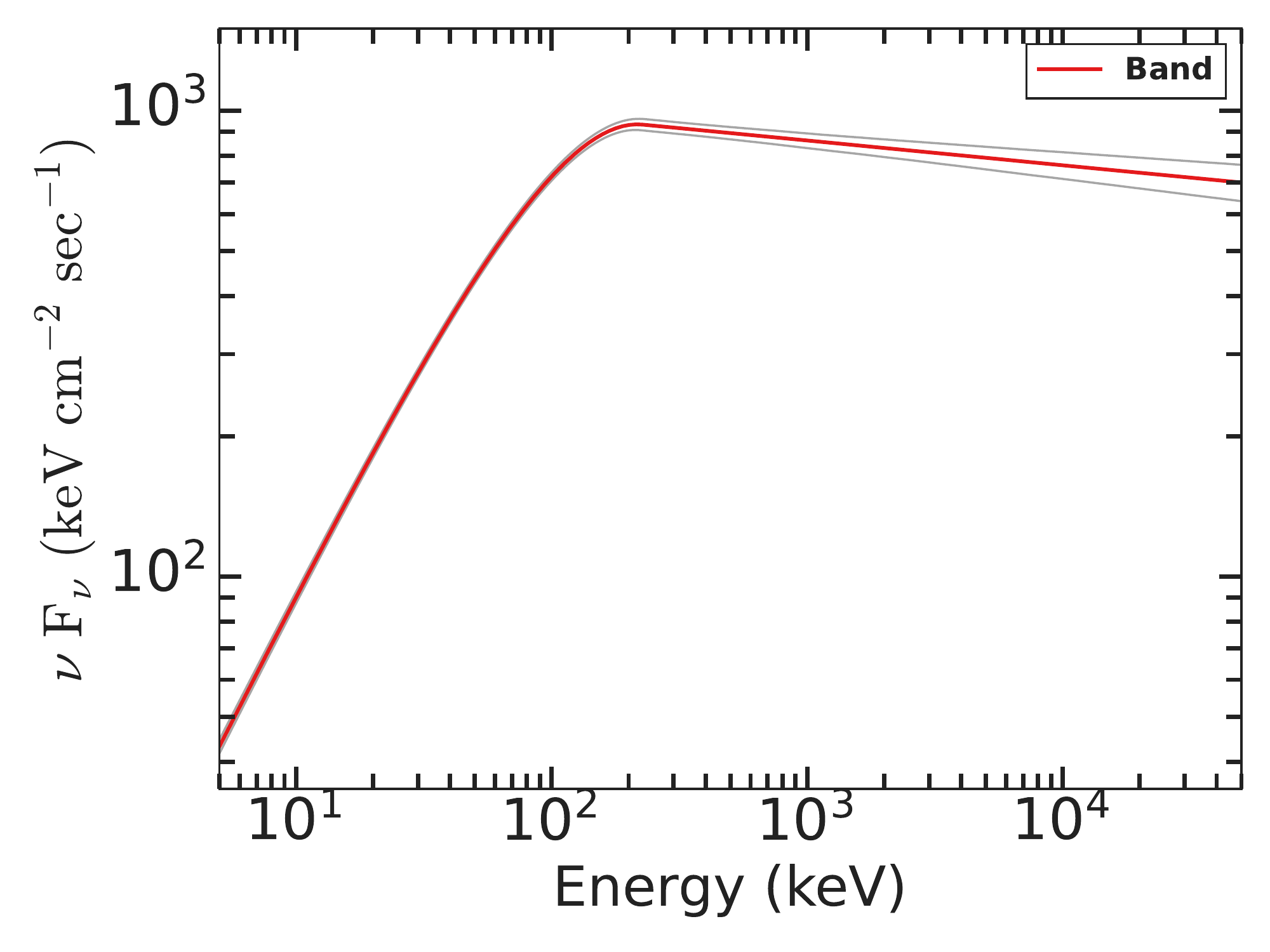}
\includegraphics[scale=0.38]{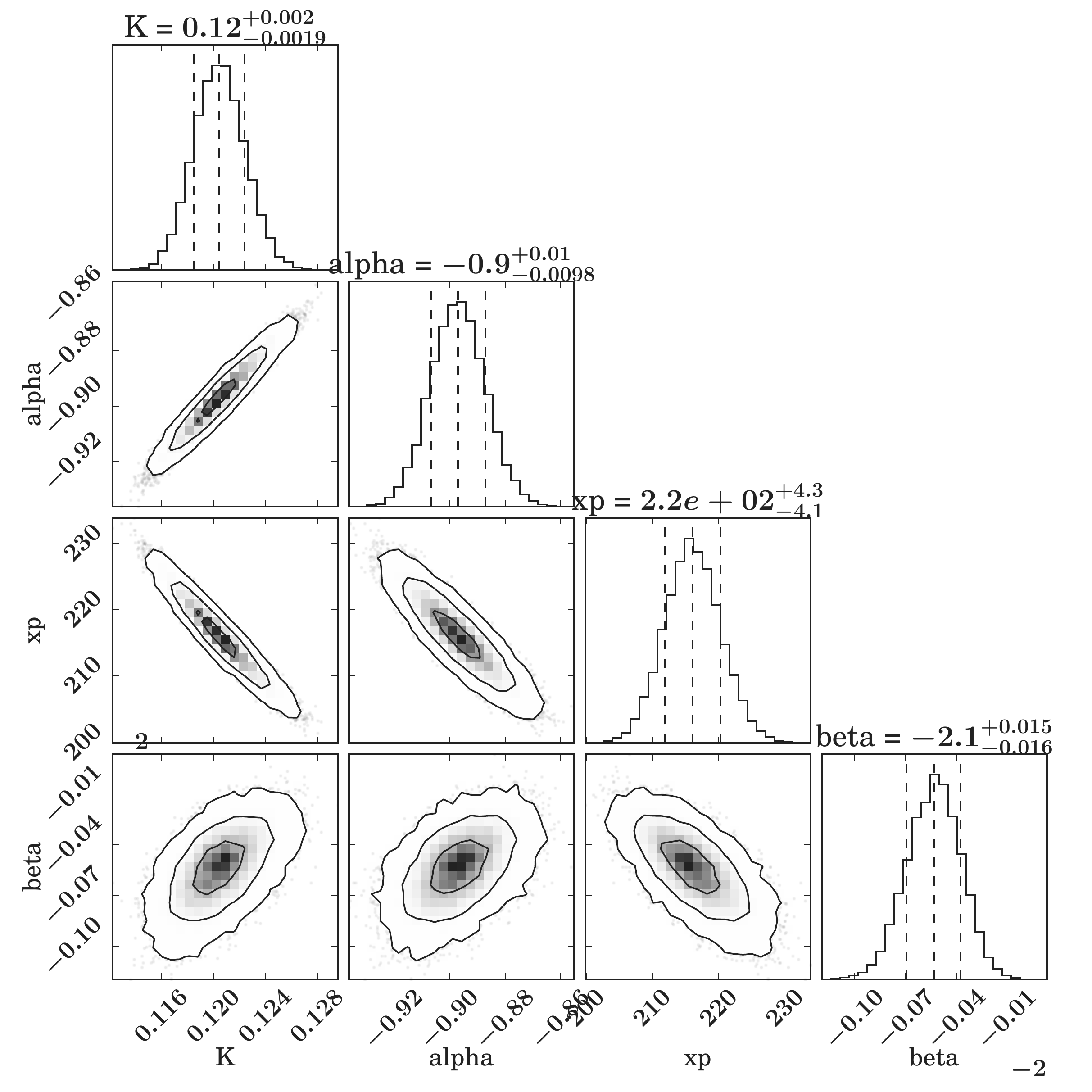}
\caption{{Top panel:} The best fit time-averaged (\fermiT to \fermiT+ 67.38\,s) spectrum (\sw{Band}) of \thisgrb in model space. The white shaded regions denote the 2 $\sigma$ uncertainty regions associated with the spectral model parameters. {Bottom panel:} The corresponding corner plot for 5000 number of simulations.}
\label{TAS}
\end{figure}

The prompt \fermi GBM and \swift BAT light curve of \thisgrb consists of a very bright main pulse followed by a long fainter emission activities up to $\sim$ 70\,s post trigger time. On the other hand, ASIM observed only the main bright pulse, displaying a Fast-Rise Exponential Decay (FRED) type structure (see Figure~\ref{ASIM_LC_SED}). Based on the DIC condition given in section \ref{GBM}, we find that the time-averaged spectra (\fermiT to \fermiT+67.38\,s) of \thisgrb in the energy range from 8 \keV to 40 MeV could be best fitted using the \sw{Band} model. We obtained the following best fit spectral parameters: $\alpha_{\rm pt}$=  -0.90$^{+0.01}_{-0.01}$, $\beta_{\rm pt}$= -2.05$^{+0.02}_{-0.02}$, and the spectral peak energy \Ep = 216.08$^{+4.20}_{-4.20}$ \keV. The best fit time-averaged spectrum of \thisgrb along with the corresponding corner plot is shown in Figure~\ref{TAS}. 

\subsubsection{Empirical correlations: Amati and Yonetoku}

\begin{figure}
\centering
\includegraphics[scale=0.35]{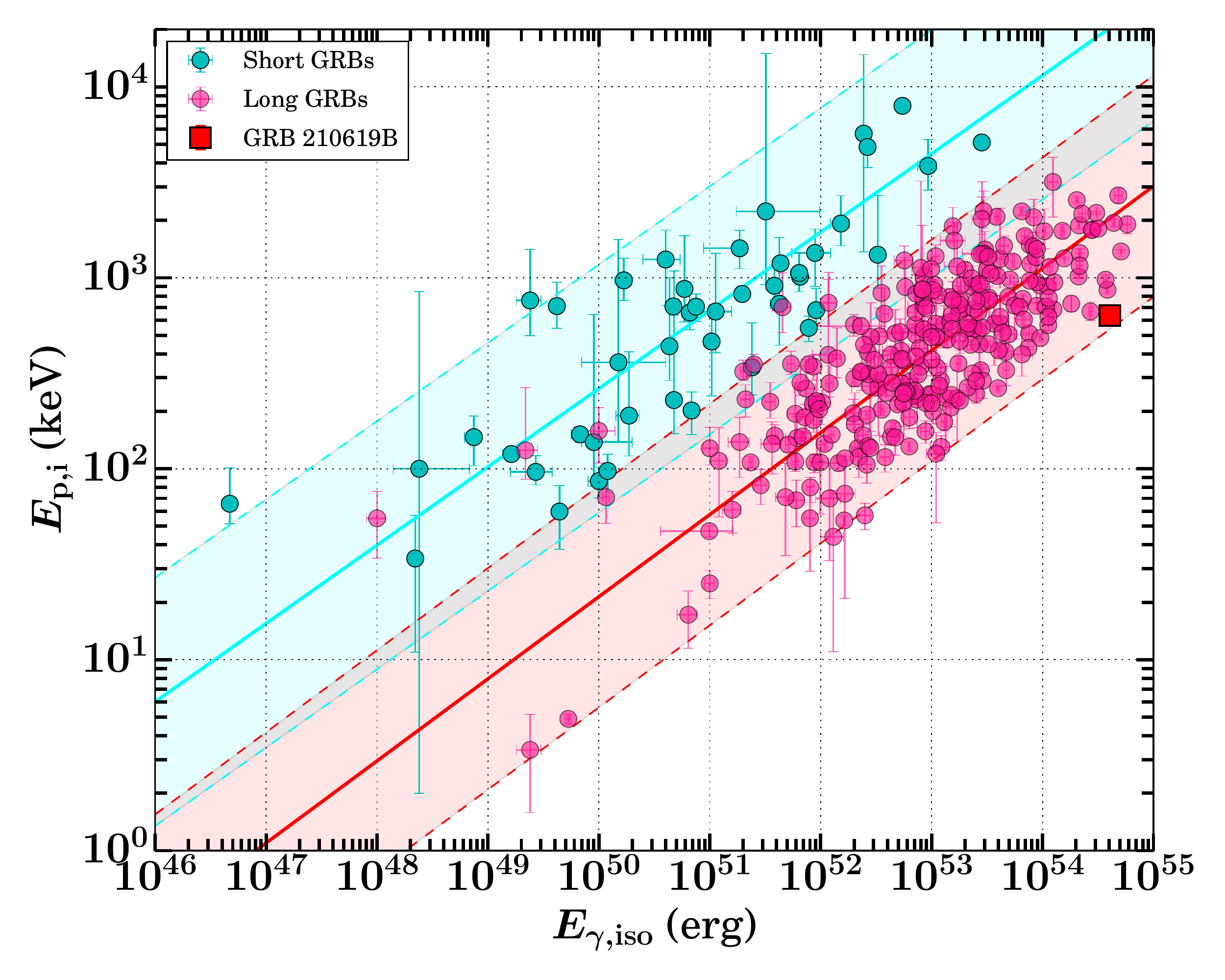}
\includegraphics[scale=0.35]{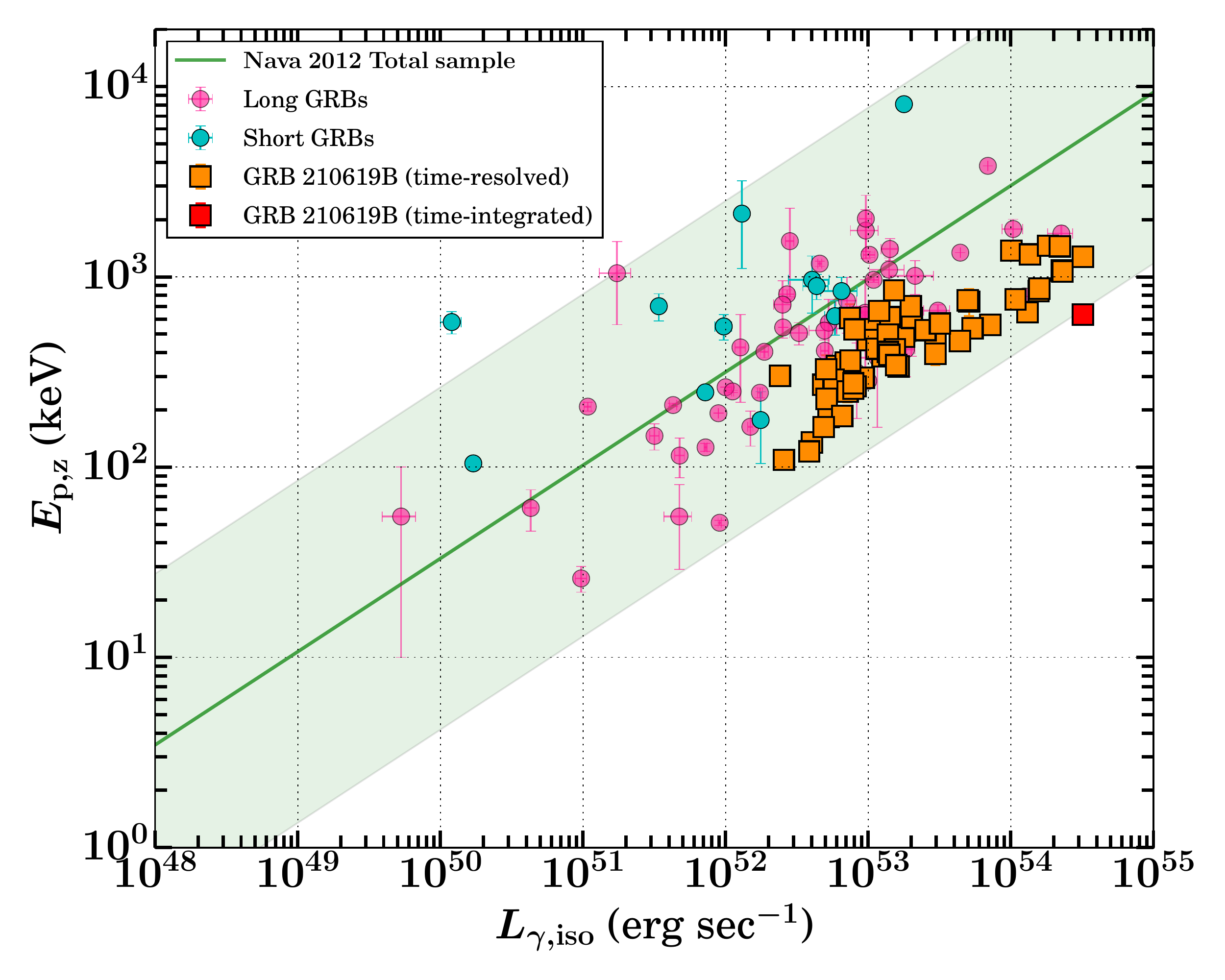}
\caption{{Top panel:} \thisgrb in Amati correlation plane (marked with a red square). The cyan and pink circles denote the long and short GRBs taken from \protect\cite{2020MNRAS.492.1919M}. The solid cyan and pink lines show the linear fit lines for short and long GRBs, respectively. The parallel shaded regions show the 3$\sigma$ scatter of the correlations. {Bottom panel:} \thisgrb in Yonetoku correlation plane (marked with a red square for the time-integrated duration and marked with orange squares for each of 59 spectra obtained using the time-resolved Bayesian bins with statistical significance $\geq$ 30). The cyan and pink circles denote the long and short GRBs taken from \protect\cite{2012MNRAS.421.1256N}. The parallel shaded regions show the 3$\sigma$ scatter of the correlations.}
\label{fig:prompt_properties_amati_yonetoku}
\end{figure}

The prompt properties of GRBs follow a few global correlations. These correlations are also used characterizing and classifying them into long and short bursts \citep{2020MNRAS.492.1919M}. For example, the time-integrated spectral peak energy (in the source frame) of the burst is positively correlated with the isotropic equivalent gamma-ray energy ($E_{\rm \gamma, iso}$), known as Amati correlation \citep{2006MNRAS.372..233A}. We calculated $E_{\rm \gamma, iso}$ and restframe peak energy for \thisgrb using the best fit time-integrated model parameters and compared the results with a larger sample of GRBs.
For the time-integrated interval, we calculated the energy flux equal to 6.27 $\rm \times ~10^{-6} erg ~cm^{-2} ~s^{-1}$ in the source frame (0.34 \keV to 3404.8 \keV) which is equivalent to $E_{\rm \gamma, iso}$= 4.05 $\rm \times ~10^{54} erg$. The $E_{\rm p, z}$-$E_{\rm \gamma, iso}$ correlation for \thisgrb along with other data points taken from \cite{2020MNRAS.492.1919M} are shown in Figure~\ref{fig:prompt_properties_amati_yonetoku}. \thisgrb is one of the brightest bursts ever detected by \fermi GBM and marginally satisfies the Amati correlation. In the literature, there is no consensus on the physical explanation for the Amati correlation. However, some studies indicate that the Amati correlation could be explained by the viewing angle effect within the framework of optically thin synchrotron emission \citep{2004ApJ...606L..33Y, 2004ApJ...614L..13E, 2005ApJ...629L..13L}. More recently, \cite{2021ApJ...908....9V} shows that the back-scattering dominated prompt emission model can naturally explain the Amati correlation. Furthermore, we also placed \thisgrb in the $E_{\rm \gamma, iso}$ and redshift distribution plane. The comparison with other well-studied samples \protect\citep{2020MNRAS.492.1919M, 2021ApJ...908L...2S} indicates that \thisgrb is one of the brightest bursts at its measured redshift.

\begin{figure}
\centering
\includegraphics[scale=0.43]{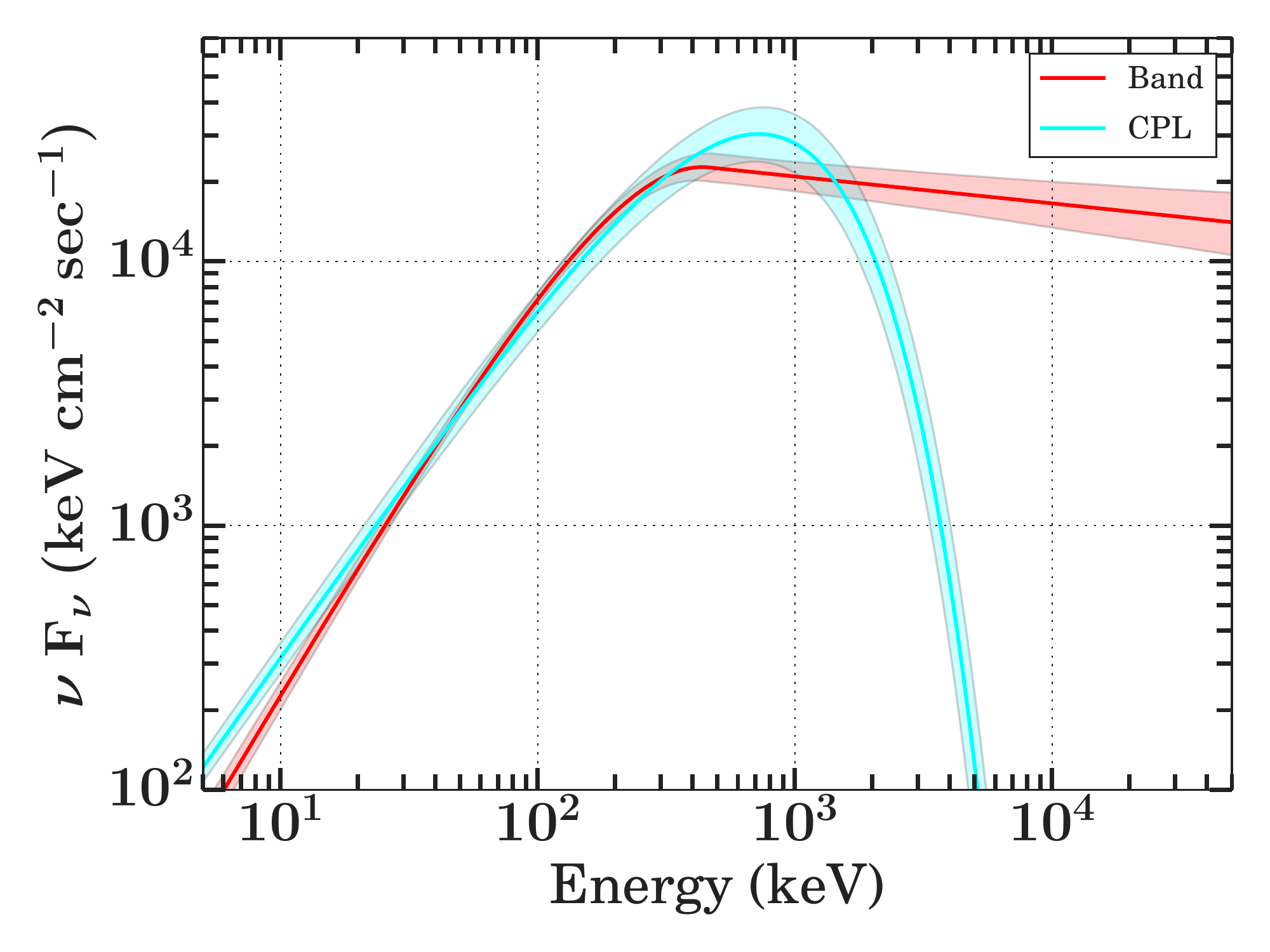}
\caption{The comparison between different empirical models (the \sw{Band} function shown in red, and the \sw{CPL} function shown in cyan) used to fit the peak spectra (time-interval from \fermiT+0.50\,s to \fermiT+ 1.01\,s, and energy range from 8 \keV to 40\,MeV) of \thisgrb in model space. The corresponding shaded colors show the 95 \% confidence levels.}
\label{Peak_spectra}
\end{figure}

Similarly, the time-integrated peak energy (in the source frame) of the burst is also correlated with the isotropic gamma-ray luminosity $L_{\rm \gamma, iso}$, known as Yonetoku correlation \citep{2010PASJ...62.1495Y}. The empirical Yonetoku correlation could be explained using photospheric dissipation model with the consideration that the subphotospheric dissipation take place far above the central engine \citep{2005ApJ...628..847R, 2007ApJ...666.1012T}. Recently, \cite{2019NatCo..10.1504I} proposed that the empirical Yonetoku correlation can also be interpreted as a natural consequence of viewing angle. We modelled the peak spectrum (\fermiT+0.50\,s to \fermiT+1.01\,s; see Figure \ref{Peak_spectra}) of \thisgrb to calculate the $L_{\rm \gamma, iso}$. For the peak time-interval, we calculated the energy flux equal to 1.14 $\rm \times ~10^{-4} erg ~cm^{-2} ~s^{-1}$ in the source frame (0.34 \keV to 3404.8 \keV) and it is equivalent to $L_{\rm \gamma, iso}$= 3.20 $\rm \times ~10^{54} erg ~s^{-1}$. The Yonetoku correlation for \thisgrb along with other data points taken from \citet{2012MNRAS.421.1256N} are shown in Figure~\ref{fig:prompt_properties_amati_yonetoku}. \thisgrb is one of the most luminous bursts ever detected by \fermi GBM and marginally satisfied the Yonetoku correlation. In addition to time-integrated $E_{\rm p, z}$-$L_{\rm \gamma, iso}$ correlation, we also examined the time-resolved $E_{\rm p, z}$-$L_{\rm \gamma, iso}$ relation for \thisgrb. For this purpose, we calculated the $L_{\rm \gamma, iso}$ values for each of the Bayesian bins with statistical significance (S) $\geq$ 30 used for the time-resolved spectral analysis. Figure~\ref{fig:prompt_properties_amati_yonetoku} shows the time-resolved $E_{\rm p, z}$-$L_{\rm \gamma, iso}$ correlation for \thisgrb. We noticed that the time-resolved Yonetoku correlation for \thisgrb is well consistent with that from the time-integrated Yonetoku correlation studied by \cite{2012MNRAS.421.1256N}. It confirmed this would indicate that the Yonetoku relation is intrinsic and not due to observational biases.

\subsubsection{\tninty-hardness distribution}

\begin{figure}
\centering
\includegraphics[scale=0.35]{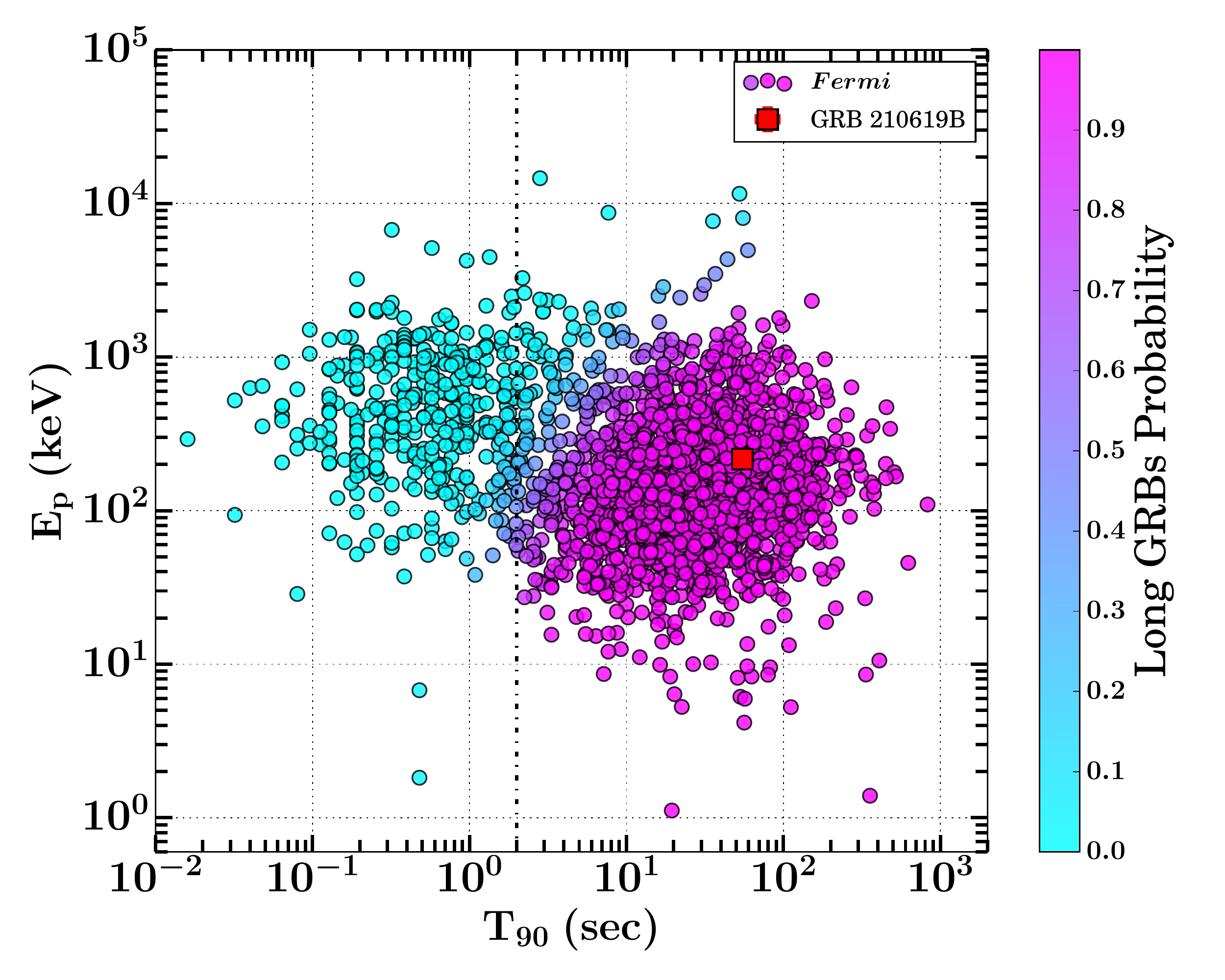}
\includegraphics[scale=0.35]{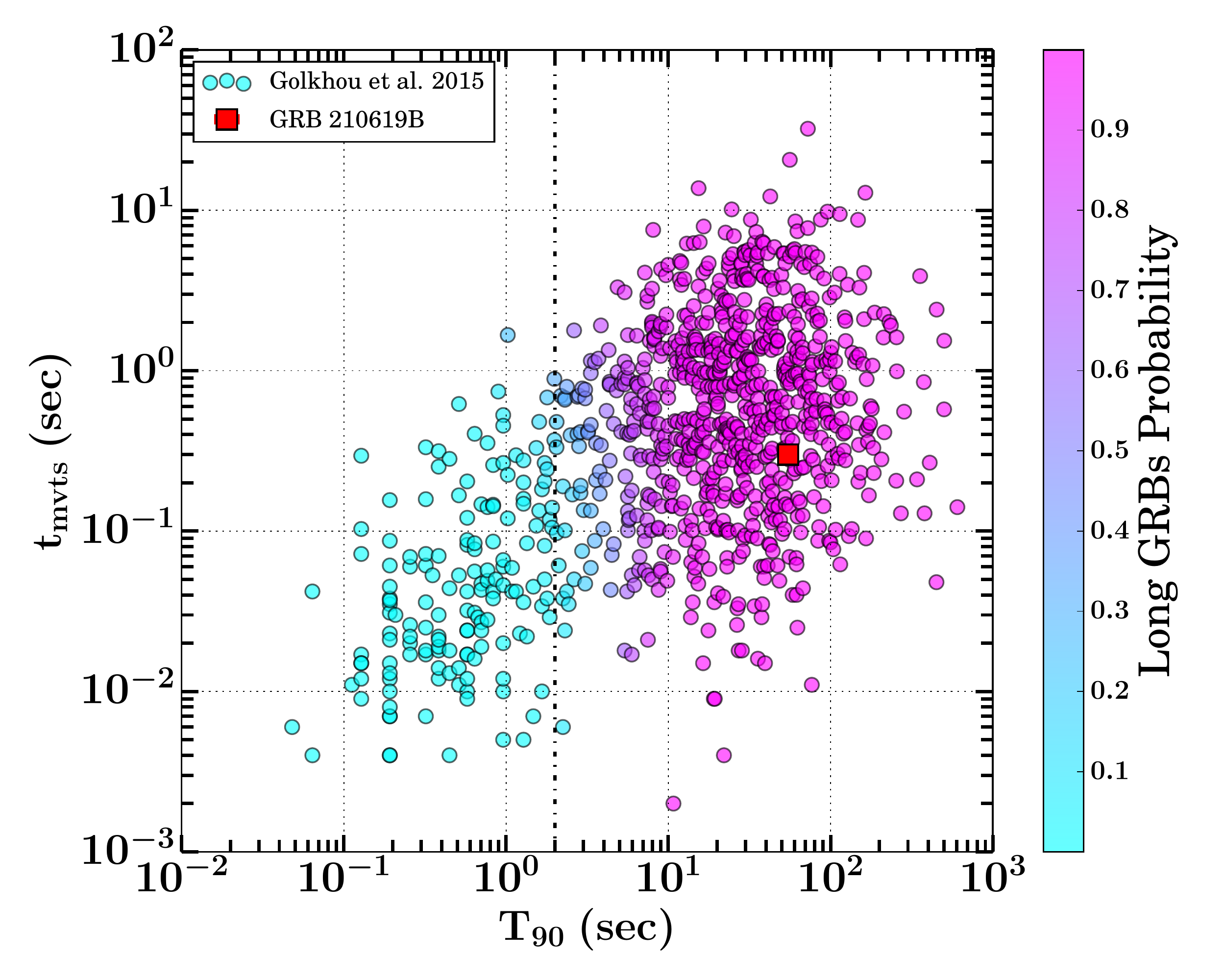}
\caption{{Top panel:} Spectral peak energy as a function of \tninty duration for \thisgrb along with all the bursts detected by \fermi GBM.
{Bottom panel:} Minimum variability time scale as a function of \tninty duration for \thisgrb along with other long and short duration GRBs taken from \protect\cite{2015ApJ...811...93G}. The right side color bar (Y-scale) shows the probability of long GRBs in respective plots. The vertical dashed-dotted black line denotes the boundary of two GRBs families.}
\label{promptproperties}
\end{figure}

GRBs are mainly classified based on the prompt emission properties such as \tninty duration and hardness ratio. Long duration GRBs have a softer spectrum in contrast to  short duration GRBs; therefore, long and short GRBs are positioned at different places in the \tninty-hardness distribution plane of GRBs. We collected the \tninty duration and peak energy values of all the bursts detected by \fermi GBM from the GBM burst catalog\footnote{\url{https://heasarc.gsfc.nasa.gov/W3Browse/fermi/fermigbrst.html}} \citep{2014ApJS..211...12G,2016yCat..22230028B,2020ApJ...893...46V}. The time-integrated \tninty-hardness distribution for \fermi GBM detected GRBs is shown in Figure~\ref{promptproperties}. The position of \thisgrb is shown with a red square, and \thisgrb follows the distribution of long GRBs. Furthermore, we also calculated the time-integrated hardness ratio (HR) by taking the ratio of counts in hard (50-300 \keV) and soft (10-50 \keV) energy channels. We find HR= 1.16 for \thisgrb. Figure~\ref{promptproperties} shows the \tninty-HR distribution for \thisgrb along with other long and short duration GRBs taken from \citet{2017ApJ...848L..14G}.

\subsubsection{\tninty-Minimum variability time}

The prompt emission light curve of GRBs is variable in nature due to being originated from internal shocks \citep{2015AdAst2015E..22P}. The minimum variability time for long (less variable) and short (more variable) GRBs follows a different distribution due to diverse compact central sources. The minimum time variability timescale, \mvts, is useful to constrain the minimum value of the bulk Lorentz factor ($\rm \Gamma_{\rm min}$) and the emission radius ($\rm R_{\rm c}$). We calculated the \mvts ($\sim$ 0.3\,s) for \thisgrb following the continuous wavelet transforms\footnote{\url{https://github.com/giacomov/mvts}} methodology given in \cite{2018ApJ...864..163V}. The \tninty-\mvts distribution for long and short GRBs (taken from \citealt{2015ApJ...811...93G}) along with \thisgrb are shown in Figure~\ref{promptproperties} (bottom panel).

Furthermore, we estimated $\rm \Gamma_{\rm min}$ and $\rm R_{\rm c}$ using the following equations from \citet{2015ApJ...811...93G}:

\begin{equation}
\rm \Gamma_{\rm min} \gtrsim 110 \, \left (\frac{L_{\rm \gamma, iso}}{10^{51} \, \rm erg/s} \, \frac{1+z}{\rm t_{\rm mvts} / 0.1 \, \rm s } \right )^{1/5}
\label{gamma_min}
\end{equation}
\begin{equation}
\rm R_c \simeq 7.3 {\times} 10^{13} \, \left (\frac{L_{\rm \gamma, iso}}{10^{51} \, \rm erg/s} \right )^{2/5} \left (\frac{ t_{\rm mvts} / 0.1 \, \rm s }{1+z} \right )^{3/5} \, \rm cm.
\label{minimum_source}
\end{equation}

We calculated the minimum value of the bulk Lorentz factor and the emission radius, giving $\gtrsim$ 550 and $\simeq$ 1.87 $\times$ $10^{15}$\,cm respectively, for \thisgrb. We also calculated the bulk Lorentz factor value ($\rm \Gamma$ = 817) for \thisgrb using the $\Gamma_{0}$-$E_{\gamma, \rm iso}$\footnote{$\Gamma_{0}$ $\approx$ 182 $\times$ $E_{\gamma, \rm iso, 52}^{0.25 \pm 0.03}$} correlation \citep{2010ApJ...725.2209L} and noticed that the minimum value of the bulk Lorentz factor constrained using the minimum variability time scale is consistent with the value of the Lorentz factor found using the correlation. The calculated radius for \thisgrb is much larger than the typical emission radius of the photosphere, indicating that the emission took place in an optically thin region away from the central engine \citep{2016ApJ...825...97U, 2019A&A...625A..60R, 2020NatAs...4..174B}.

\subsubsection{Spectral lag}
\label{lag}

\begin{figure}
\centering
\includegraphics[scale=0.33]{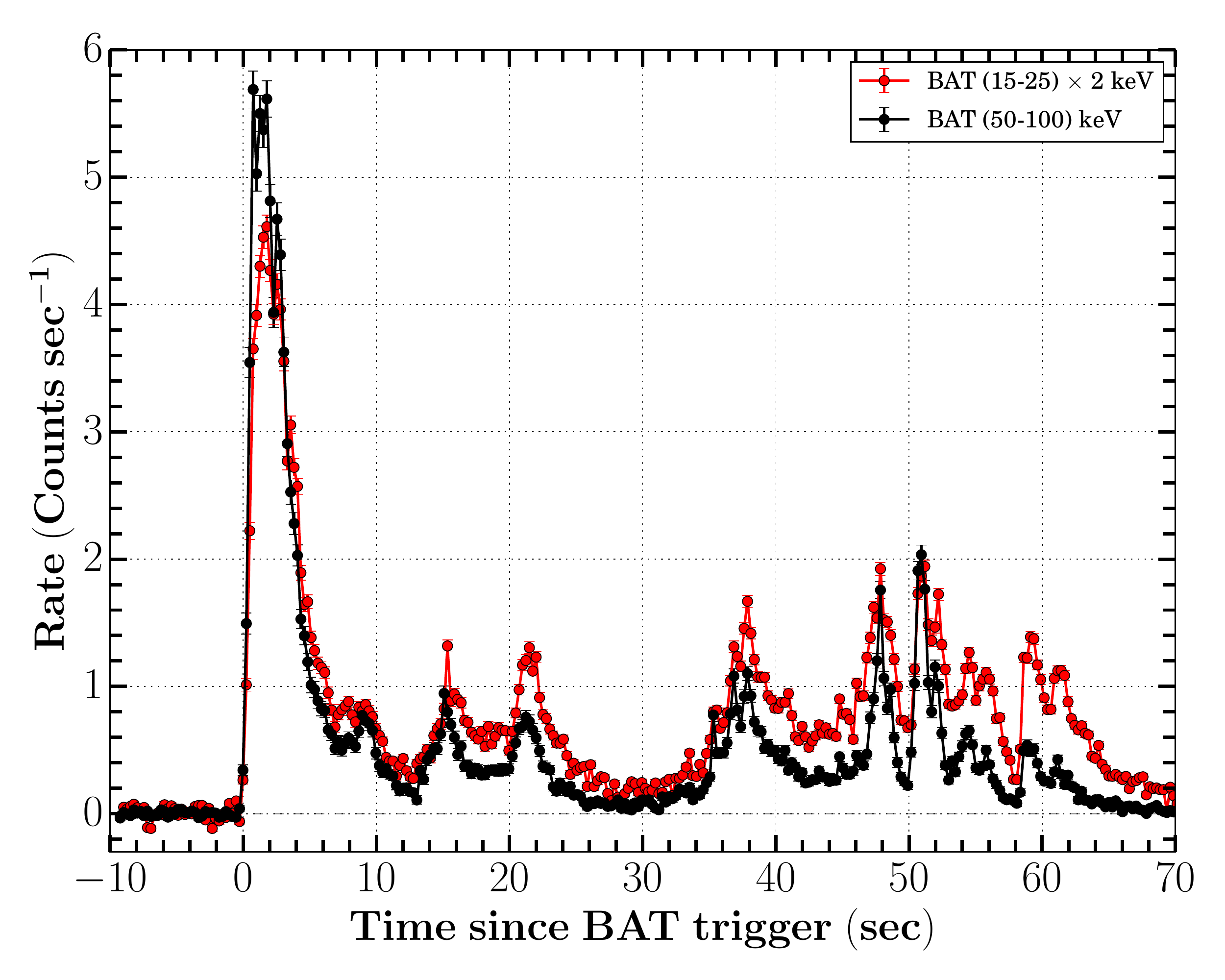}
\includegraphics[scale=0.35]{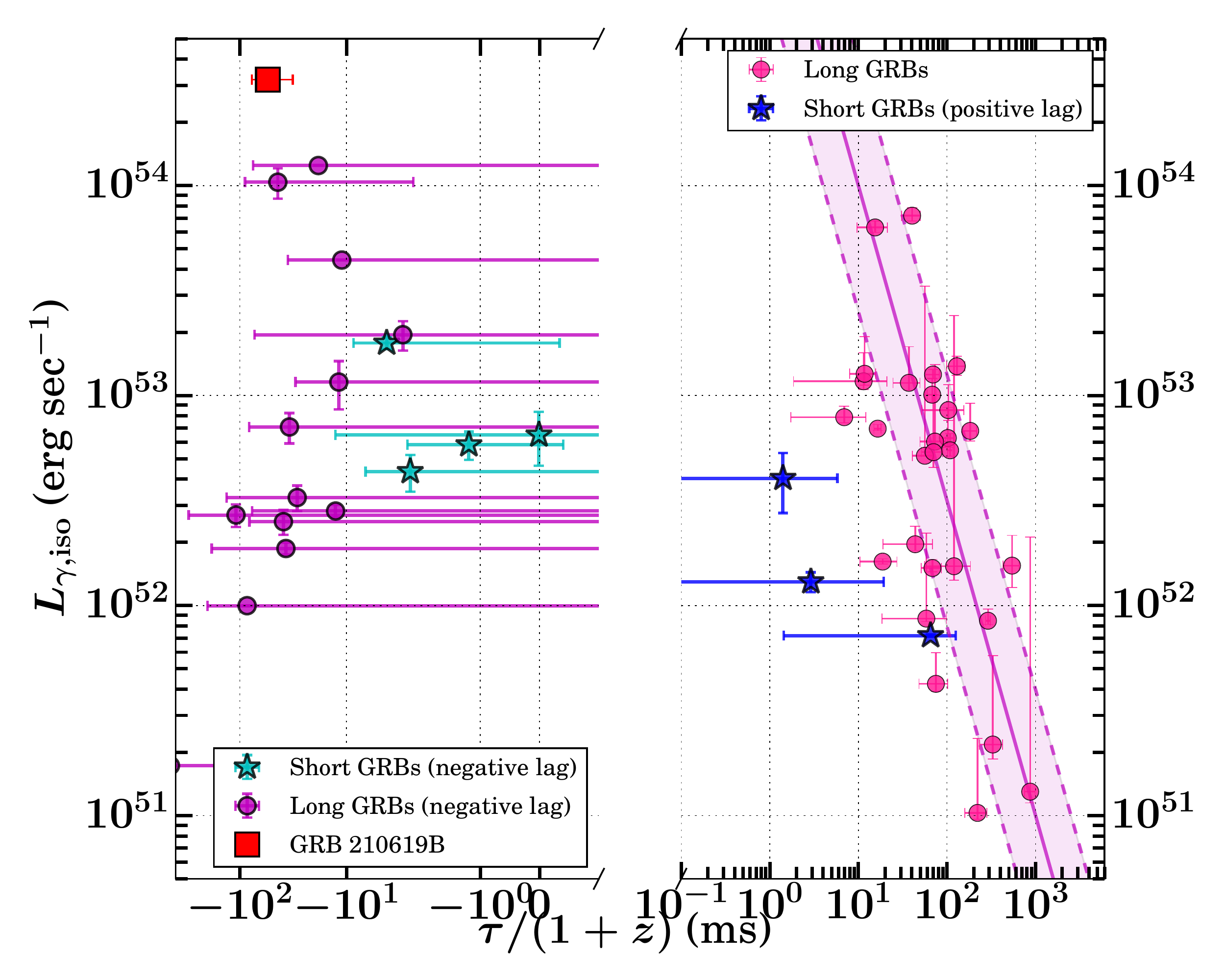}
\includegraphics[scale=0.33]{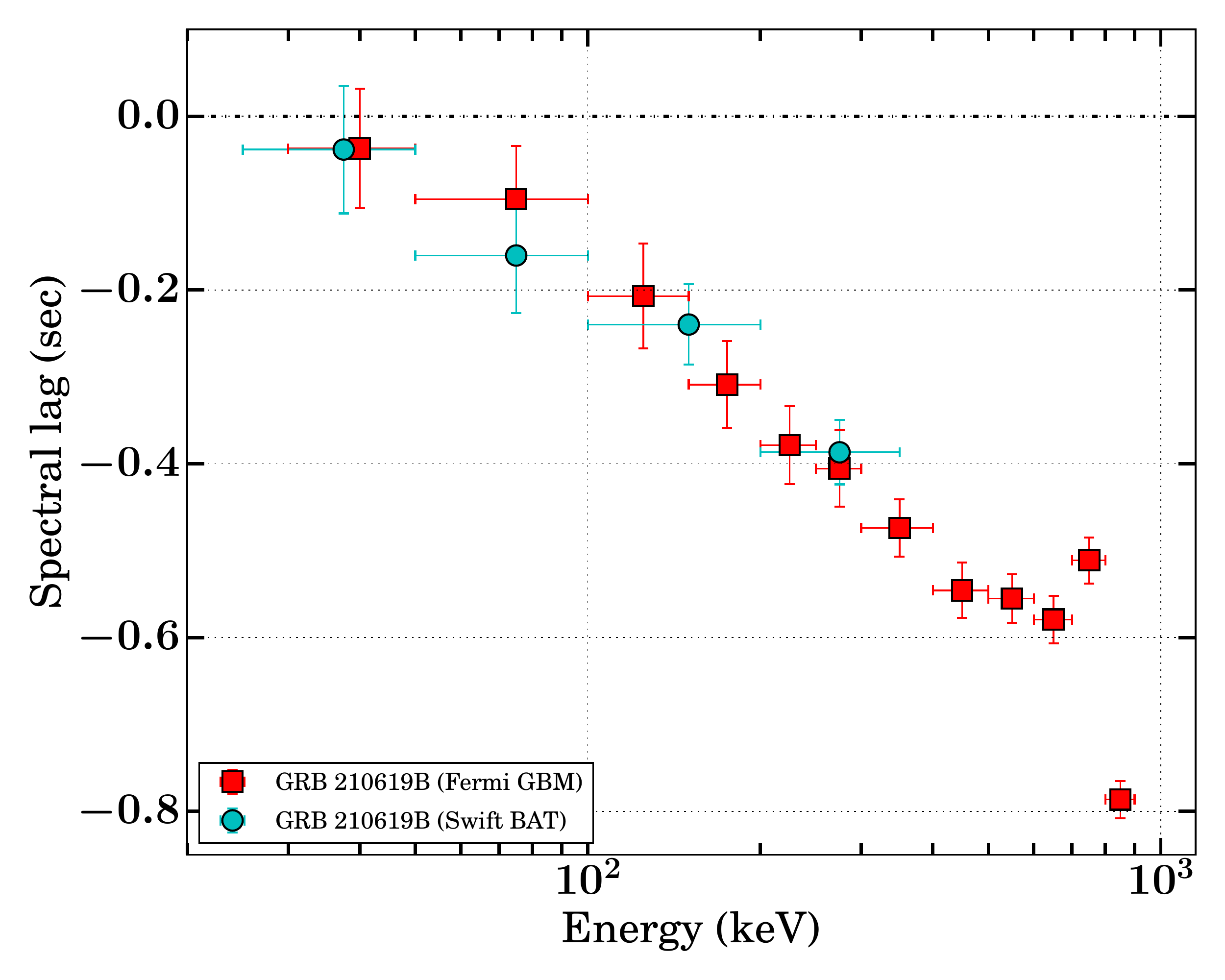}
\caption{{\bf Spectral lag of \thisgrb:} {Top panel:} \swift BAT count rate light curves in 15-25 \keV (shown with red) and 50-100 \keV (shown with black) energy range, with bin size of 256 ms. {Middle panel:} Lag-luminosity correlation for \thisgrb in \swift BAT 50-100 \keV and 15-25 \keV energy ranges along with other data points taken from \protect\cite{Ukwatta_2010, 2015MNRAS.446.1129B}. {Bottom panel:} Energy-resolved spectral lag analysis of \thisgrb using \swift BAT and \fermi GBM observations.}
\label{fig:lag}
\end{figure}

The prompt emission light curves of long GRBs (short GRBs) show significant (zero) delays in two different energy ranges, and this characteristic is known as spectral lag. If the high-energy photons of GRBs come before the low-energy photons, it is defined as positive spectral lag. On the other hand, we conventionally defined the lag as negative if the low-energy photons precede the high-energy photons. The observed spectral lag is usually explained in terms of the prompt intrinsic spectral evolution (mainly temporal evolution of \Ep) or due to the curvature effect of relativistic moving shocked shells \citep{2004ApJ...614..284D, 2016ApJ...825...97U}. \cite{2000ApJ...534..248N} reported the anti-correlation between the spectral lag and isotropic peak luminosity of GRBs using a limited long bursts sample (with known redshift) observed using the {\it BATSE} mission.

In the case of \thisgrb, we initially estimated the time-integrated (\fermiT to \fermiT+67.38\,s) spectral/timing lag for the \swift BAT light curves (see Figure~\ref{fig:lag}) in two different energy ranges (15-25 \keV and 50-100 \keV) using the cross-correlation function \citep[CCF;][]{2000ApJ...534..248N}. We followed the detailed methodology presented in \cite{2015MNRAS.446.1129B}. We have used \sw{emcee} package \citep{2013PASP..125..306F} to fit the cross-correlation function using an asymmetric Gaussian function. We obtained a negative spectral lag = -160$^{+67}_{-66}$\,ms and placed it in the anti-correlation relationship between the spectral lag and isotropic peak luminosity of GRBs (data taken from \citealt{Ukwatta_2010} in the same energy range). We noticed that \thisgrb is an outlier of the lag-luminosity anti-correlation relationship (see middle panel of Figure~\ref{fig:lag}). Furthermore, we examined the literature and found that many long GRBs (GRB 060814, GRB 061021, GRB 070306, GRB 080721, GRB 080804, GRB 090426C, GRB 100728B, GRB 110205A, GRB 140102A, GRB 150213A, and many others) are reported with such a large value ($>$ -100\,ms, although with large error bars) for the negative spectral lag \citep{2015MNRAS.446.1129B, 2018JHEAp..18...15C, 2021MNRAS.505.4086G}. \cite{2015MNRAS.446.1129B} studied a larger sample of long and short GRBs and noticed that the bursts with high luminosity might have smaller lags. \cite{2014AstL...40..235M} proposed that the GRBs independent emission pulses exhibit hard-to-soft spectral evolution of the peak energy, suggesting a positive spectral lag \citep{2018ApJ...869..100U}. However, a strong/complex spectral evolution of the peak energy could result in superposition effects, and this superposition effect could explain the negative spectral lag observed for \thisgrb.

In addition, we also performed the energy-resolved spectral lag analysis (\fermiT to \fermiT+67.38\,s) of \thisgrb using \swift BAT and \fermi GBM observations. We considered the 15-25 \keV and 8-30 \keV light curves as reference to calculate the lag using \swift BAT and \fermi GBM respectively. We noticed that the energy-resolved lag analysis also shows the negative lag, suggesting that the low-energy photons precede the high-energy photons for \thisgrb. Figure~\ref{fig:lag} (bottom panel) shows the spectral evolution of the lags for \thisgrb. The energy-resolved spectral lags obtained using BAT, and GBM data are tabulated in Table B3 of the appendix.

\subsubsection{Spectral Evolution and correlations}

\begin{figure}
\centering
\includegraphics[scale=0.37]{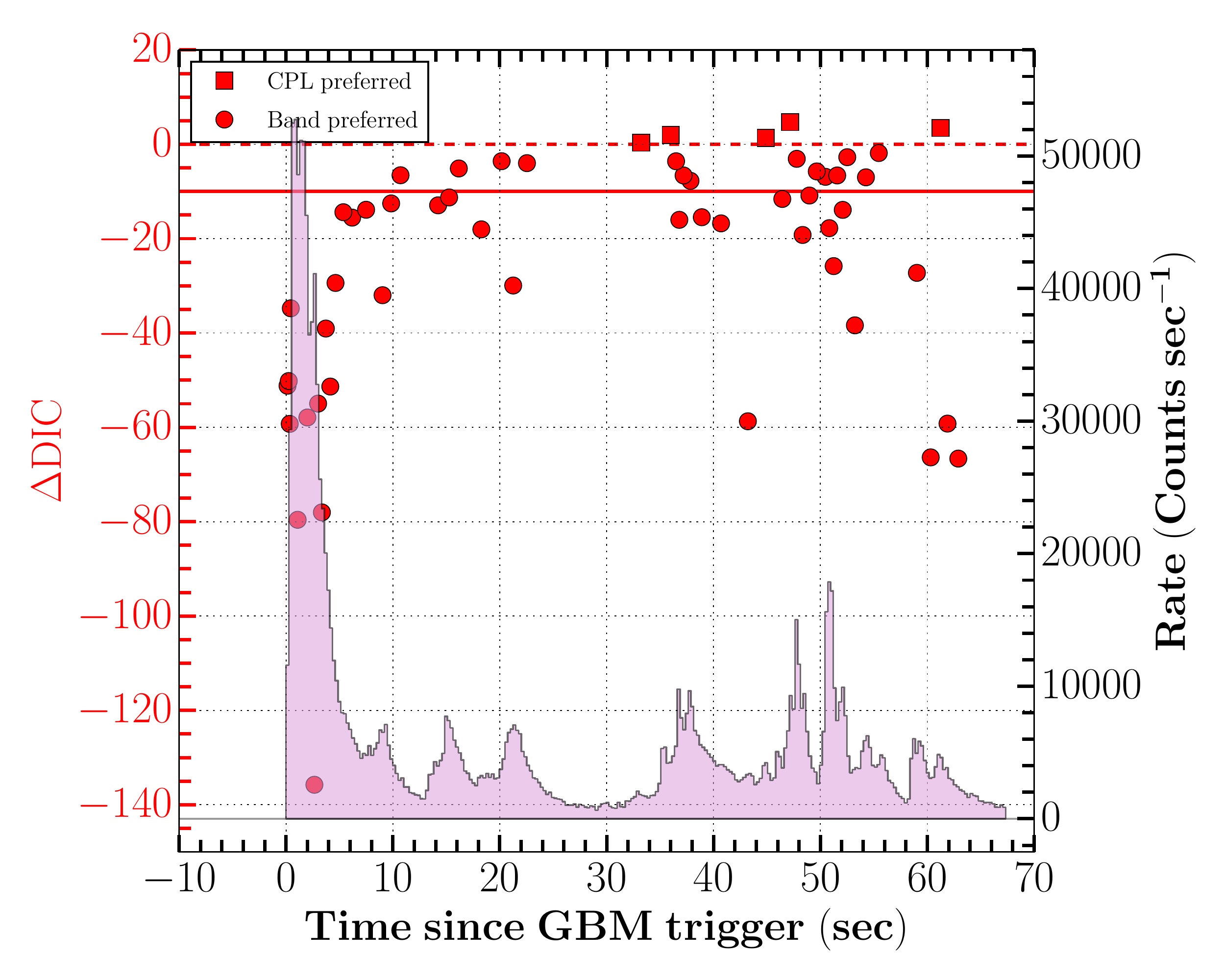}
\caption{The difference between DIC values of \sw{Band} and \sw{Cutoff-power law} models for each of the time-resolved bins obtained using Bayesian block analysis (with S $\geq$ 30). The horizontal red dashed, and solid red lines show the difference between DIC values equal zero and -10, respectively.}
\label{delta_DIC}
\end{figure}

\begin{figure*}
    \centering
    \includegraphics[scale=0.45]{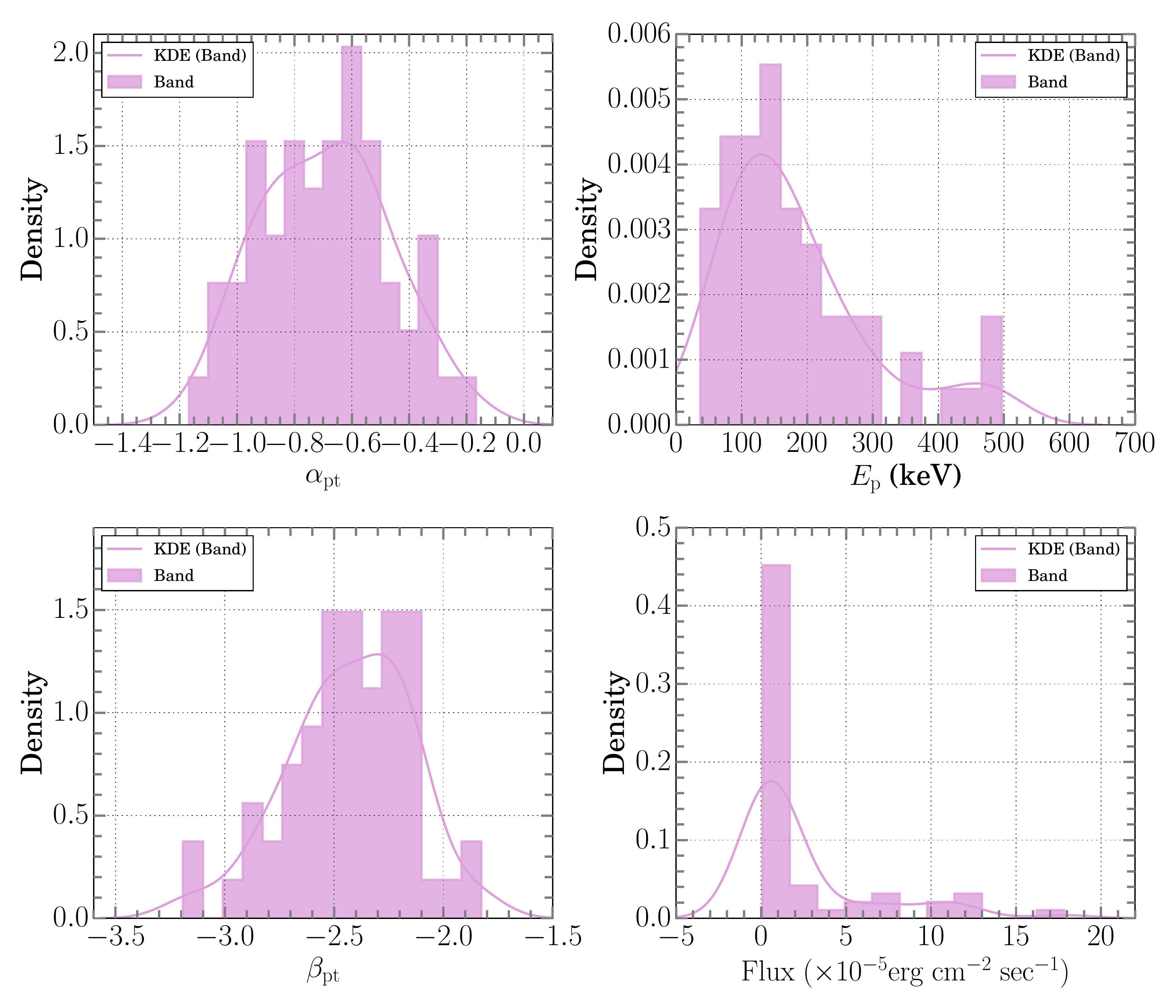}
    \caption{The spectral parameters ($\alpha_{\rm pt}$, \Ep, $\beta_{\rm pt}$, and flux) distributions for \thisgrb obtained using the time-resolved spectral analysis of \fermi GBM data using the \sw{Band} model. The plum curves show the kernel density estimation (KDE) of the respective parameters distributions fitted using the \sw{Band} model.}
    \label{parameter_dist}
\end{figure*}

The prompt emission is believed to consist of various spectral components, the combination of which provides the shape of the observed spectrum. To constrain these components for \thisgrb, we performed the time-resolved spectral analysis (see section \ref{GBM}). Figure~\ref{delta_DIC} shows the difference of deviation information criterion values from \sw{Band} and \sw{Cutoff-power law} functions for each of the time-resolved bins. The comparison of differences of the DIC values for each bin suggests that most of the bins (except five) support the traditional \sw{Band} model. Figure \ref{parameter_dist} shows the distribution of spectral parameters (\Ep, $\alpha_{\rm pt}$, $\beta_{\rm pt}$, and flux) obtained using the \sw{Band} model. We have also shown the kernel density estimation (KDE) for each of the parameters. The averaged and standard deviation values of individual spectral parameter distributions are given in Table \ref{tab:mean_std}.

\begin{table*}
\begin{scriptsize}
\begin{center}
\caption{The averaged and standard deviation of individual spectral parameters distributions for \thisgrb.}
\label{tab:mean_std}
\begin{tabular}{ccccc} \hline 
\textbf{Model} & \multicolumn{4}{c}{\textbf{Spectral parameters}}  \\ \hline
 & $\bf \alpha_{\rm \bf pt}$ & \bf \Ep /$\bf E_{\it c}$ (\keV) & $\bf \beta_{\rm \bf pt}$ & Flux ($\times 10^{-05}$\,${\rm erg}\,{\rm s}^{-1}$)\\ \hline
\sw{Band}&$-0.69 \pm 0.22$&$183.47 \pm 113.90$&$-2.43 \pm 0.28$& 2.37 $\pm$ 3.88 \\
\sw{CPL}&$-0.88 \pm 0.20$&$214.52 \pm 128.63$&-& 1.45 $\pm$ 2.34 \\\hline
\end{tabular}
\end{center}
\end{scriptsize}
\end{table*}

\begin{figure}
\centering
\includegraphics[scale=0.37]{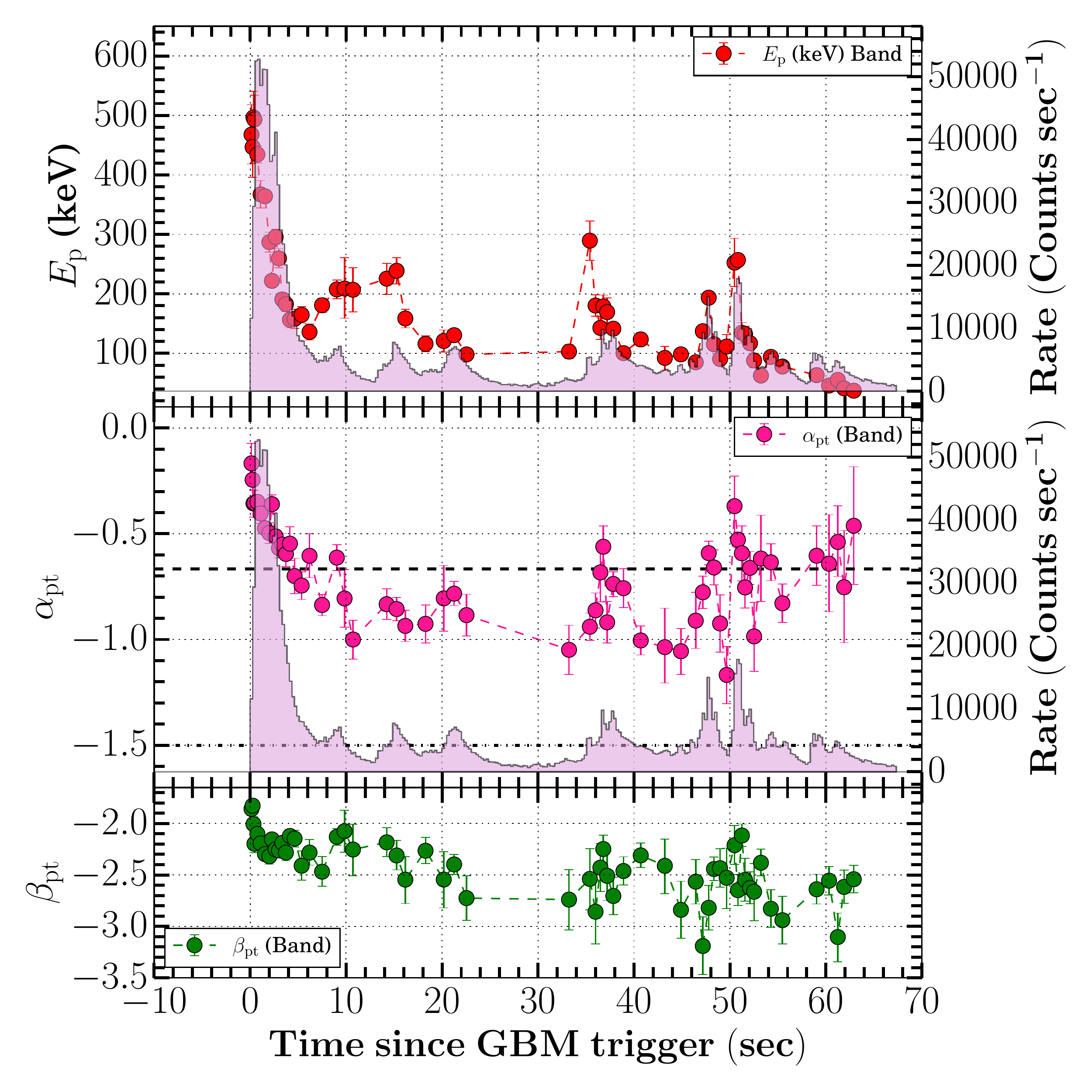}
\caption{Temporal evolution of spectral parameters obtained using time-resolved spectral modelling with the \sw{Band} model: {Top panel:} Evolution of the peak energy and it tracks the intensity of the burst. {Medium panel:}  Evolution of low-energy spectral index. The horizontal dashed and dotted-dashed lines represent the synchrotron line of death ($\alpha_{\rm pt}$ = -2/3) and the synchrotron fast cooling line ($\alpha_{\rm pt}$ = -3/2). {Bottom panel:} Evolution of high-energy spectral index obtained using the \sw{Band} spectral model.}
\label{TRS_evolution_band}
\end{figure}

The prompt emission spectral parameters of GRBs exhibit a strong spectral evolution and are useful to understand the radiation mechanisms and jet composition of GRBs. Figure~\ref{TRS_evolution_band} shows the evolution of spectral parameters (\Ep, $\alpha_{\rm pt}$, $\beta_{\rm pt}$, and flux) for \thisgrb obtained using the 
the \sw{Band} model. In the Figure, the evolution of spectral parameters is over-plotted on the count-rate prompt emission light curve of \thisgrb. We noticed that the evolution of observed \Ep exhibits an intensity tracking pattern for \thisgrb. The observed negative spectral lag (see section \ref{lag}) and intensity tracking behaviour of \Ep for \thisgrb is consistent with the prediction of a connection between the spectral lag and spectral evolution \citep{2018ApJ...869..100U}. Moreover, the low-energy index also exhibits an intensity tracking trend, suggesting the double-tracking characteristics seen in a few cases of GRBs. Further, we examined the overall spectral parameters correlations. 

\begin{table*}
\caption{The correlation between the spectral parameters obtained using the time-resolved spectral analysis of \thisgrb. We have used Pearson correlation to calculate the correlation strengths (Pearson correlation coefficient, r) and the probability of null hypothesis (p) for individual parameters correlation.}
\label{tab:correlation}
\centering
\begin{tabular}{ccccccc} \hline 
\textbf{Model} & \multicolumn{2}{c|}{\textbf{log (Flux)-log (\Ep)}} & \multicolumn{2}{c|}{\textbf{log (Flux)-$\alpha_{\rm pt}$}} & \multicolumn{2}{c}{\textbf{log (\Ep)-$\alpha_{\rm pt}$}} \\ \hline
 & \bf r & \bf p &  \bf r & \bf p &  \bf r & \bf p  \\ \hline
\sw{Band}& 0.87 & 2.07 $\times 10^{-19}$& 0.75 & 1.10 $\times 10^{-11}$& 0.47 & 1.85 $\times 10^{-4}$ \\ \hline
\end{tabular}
\end{table*}

\begin{figure}
\centering
\includegraphics[scale=0.34]{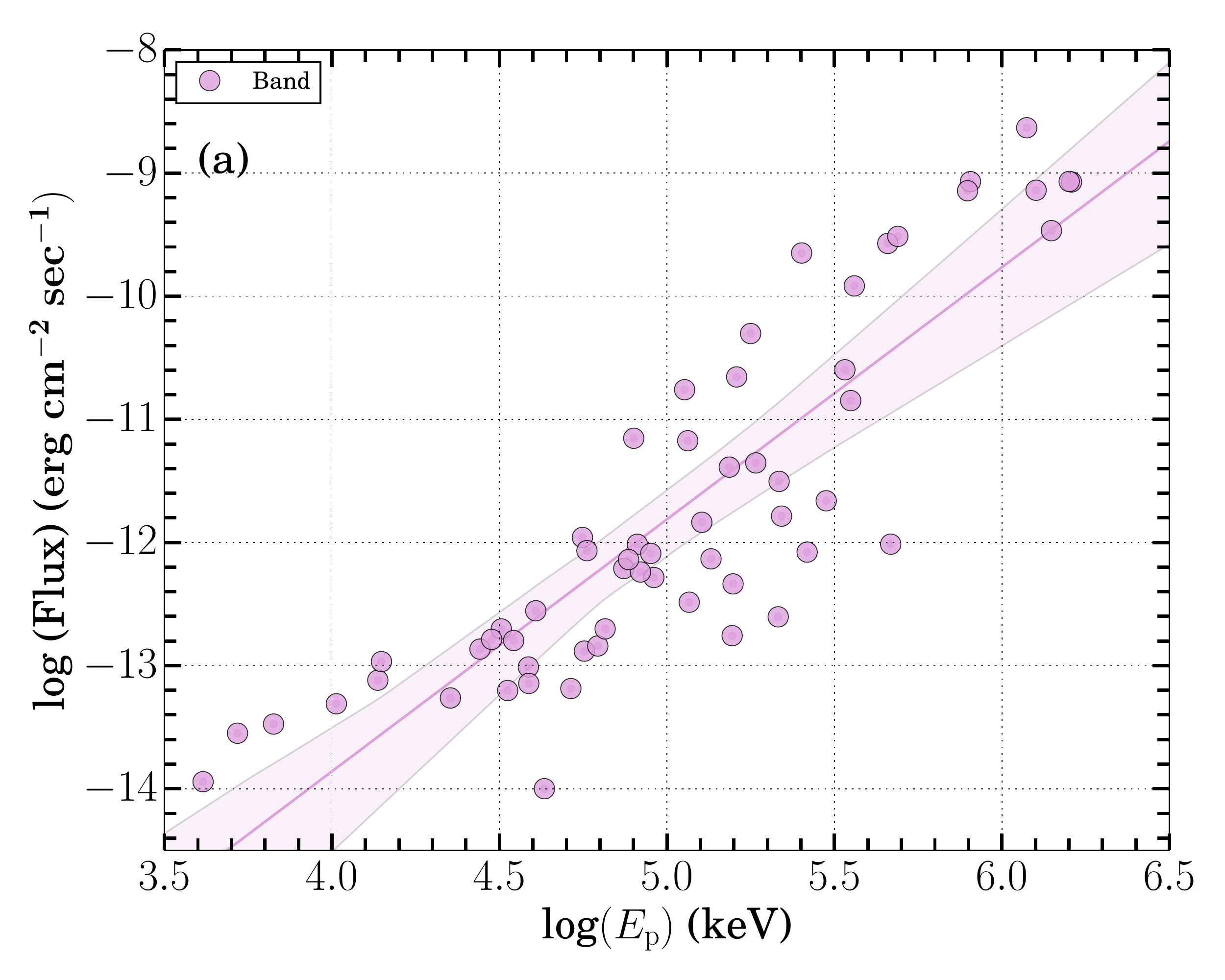}
\includegraphics[scale=0.34]{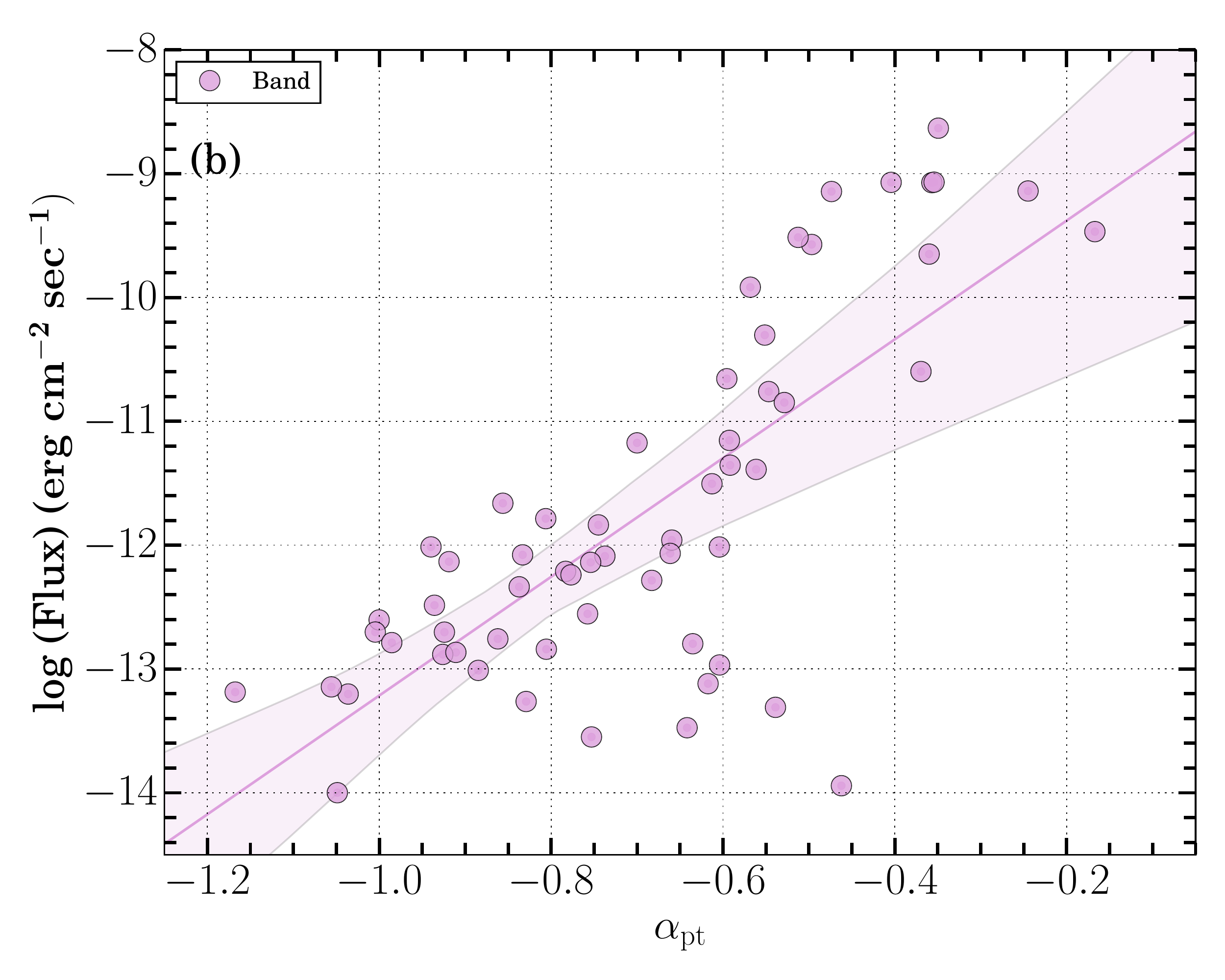}
\includegraphics[scale=0.34]{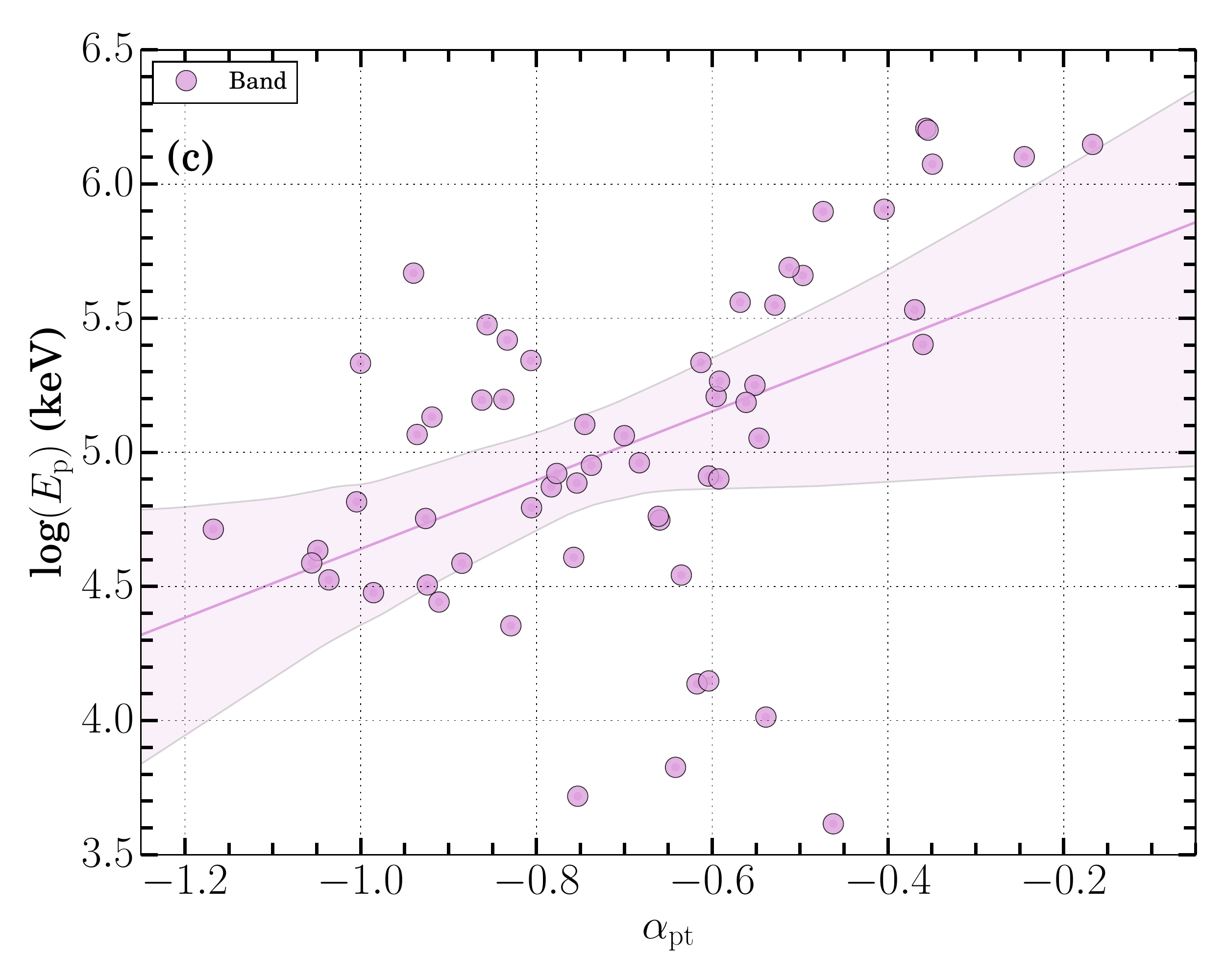}
\caption{{\bf Correlations between the spectral parameters obtained using time-resolved spectral modelling with the \sw{Band} model:} (a) Correlation between the peak energy and flux. (b) Correlation between the low-energy spectral index and flux. (c) Correlation between the peak energy and the low-energy spectral index.}
\label{TRS_correlation_band}
\end{figure}

We studied the correlation between the following spectral parameters: (a) log (Flux)-log (\Ep), (b) log (Flux)-$\alpha_{\rm pt}$, and (c) log (\Ep)-$\alpha_{\rm pt}$. We have utilized Pearson correlation to calculate the correlation strengths (Pearson correlation coefficient, r) and the probability of a null hypothesis (p) for individual parameters correlation \footnote{The null hypothesis probability (p-value) we mention here is the probability of the correlation occurring by chance, i.e. with respect to a constant relationship. If the p-value of a statistical test is small enough we can say that the null hypothesis (i.e. constant) is false and the correlation is a good model.}. We have noticed strong correlations of (a) log (Flux)-log (\Ep) and (b) log (Flux)-$\alpha_{\rm pt}$. On the other hand, log (\Ep)-$\alpha_{\rm pt}$ shows a moderate degree of correlation (see Table \ref{tab:correlation}). Figure \ref{TRS_correlation_band} shows the correlations between the spectral parameters obtained using the time-resolved spectral modelling with the \sw{Band} model. The time-resolved spectral analysis (using ASIM data) of the main bright pulse observed with ASIM shows evidence for hard-to-soft spectral evolution (softening when brightening behaviour), which is significant at 95\% confidence level for the first two (three) time intervals (see Table \ref{tab:asim} in the appendix). This characteristic is consistent with \fermi time-resolved spectral analysis results during the main bright pulse; therefore it shows agreement between the two observations.

\subsection{Discussion}
\label{discussion}
Based on the above mentioned results using the high energy analysis of the prompt emission properties of \thisgrb, our major findings are discussed below in the following sub-sections. 

\subsubsection{Prompt emission mechanism and jet composition}

Regarding the observed prompt emission, the spectrum was explored to find the possible radiation mechanism of GRBs. There are mainly two possibilities, non-thermal thin shell synchrotron emission or photospheric thermal emission \citep{2015AdAst2015E..22P, 2020NatAs...4..210Z}. The non-thermal thin shell synchrotron emission suggests that the low-energy photon indices could not be harder than the -2/3 value, also known as ``synchrotron line of death (LOD)”. However, in the case of \thisgrb, the evolution of the low-energy photon indices exceeds the synchrotron LOD during the main bright/hard pulse, and during the softer/longer pulses, the low-energy photon indices become softer and remain consistent with this limit (see Figure~\ref{TRS_evolution_band}). The observed hard values of the low-energy photon indices (during the main pulse emission) suggest that the observed prompt emission spectrum of \thisgrb is inconsistent with the non-thermal thin shell synchrotron emission model (both in slow and fast cooling cases). This hard $\alpha_{\rm pt}$ also suggests that our viewing angle for the burst is not significantly off the burst emission axis and that the central engine has been long-time active. The hard $\alpha_{\rm pt}$ could be explained using thermal ``photospheric emission" \citep{2015AdAst2015E..22P, 2018JApA...39...75I}. 
\par
On the other hand, the softer values of the low-energy photon indices during the longer emission phase are consistent with the non-thermal thin shell synchrotron emission model. This suggests that the radiation process responsible for \thisgrb shows a transition between photospheric thermal emission (hard $\alpha_{\rm pt}$) and non-thermal synchrotron emission (soft $\alpha_{\rm pt}$). This further supports a transition in the jet composition of \thisgrb, between a matter-dominated hot fireball \citep{1992Natur.357..472U, 2015AdAst2015E..22P} to Poynting flux dominated outflow \citep{1997ApJ...482L..29M, 2001MNRAS.321..177L}. Such a transition has been observed only for a few bursts \citep{2018NatAs...2...69Z, 2019ApJS..242...16L}. However, this is not the only possible scenario to explain the above observations in spectral evolution of \thisgrb. A few complex theoretical models have been proposed that can produce a hard $\alpha_{\rm pt}$ (as observed during the main emission phase of \thisgrb) within the framework of synchrotron emission model, such as synchrotron emission in decaying magnetic field \citep{2006ApJ...653..454P, 2014NatPh..10..351U}, time-dependent cooling of synchrotron electrons \citep{2018MNRAS.476.1785B, 2020NatAs...4..174B}, etc.
\par
In addition, the non-thermal synchrotron emission model might have a narrow spectral width (alike the Planck function) similar the case of the photosphere model as for example GRB 081110A \citep{2018JApA...39...75I}, results in a harder $\alpha_{\rm pt}$. On the other hand, a soft $\alpha_{\rm pt}$ (as observed during the longer emission phase of \thisgrb) can be produce within the framework of photospheric emission model. The photosphere model does not always look like the Planck function (a narrow spectral width) but may also have a similar shape to the case of non-thermal synchrotron emission from the region above the photosphere \citep{2005ApJ...633.1018P, 2005ApJ...628..847R, 2015MNRAS.454L..31A, 2019MNRAS.485..474A}, for example GRB 090902B \citep{2010ApJ...709L.172R, 2011MNRAS.415.3693R} and GRB 110920A \citep{2015MNRAS.450.1651I}. In some cases (for example GRB110920A, \citealt{2018JApA...39...75I}), it is shown that two different models having different shapes equally fit the same observed spectrum. These arguments suggest that many times there is a degeneracy between the spectral models. Therefore spectroscopy alone may not be sufficient to test the viability of the proposed models. The limitation of spectral fitting can be resolved using prompt emission polarization measurements \citep{2022JApA...43...37I, 2022MNRAS.511.1694G}. Nevertheless, such a work is out from the scope of this paper.

\subsubsection{Comparison of spectral parameters with a larger sample}

\begin{figure}
\centering
\includegraphics[scale=0.32]{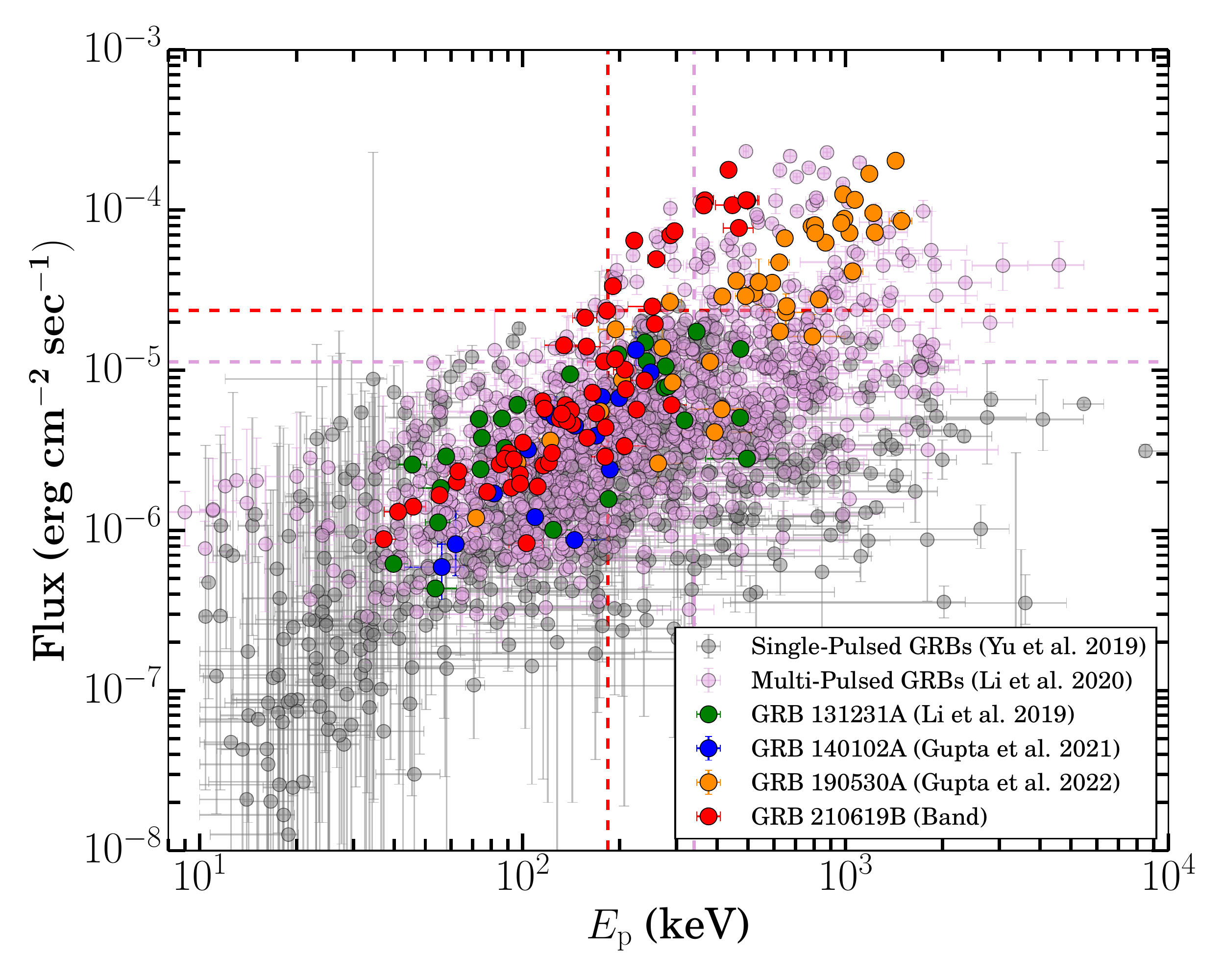}
\includegraphics[scale=0.32]{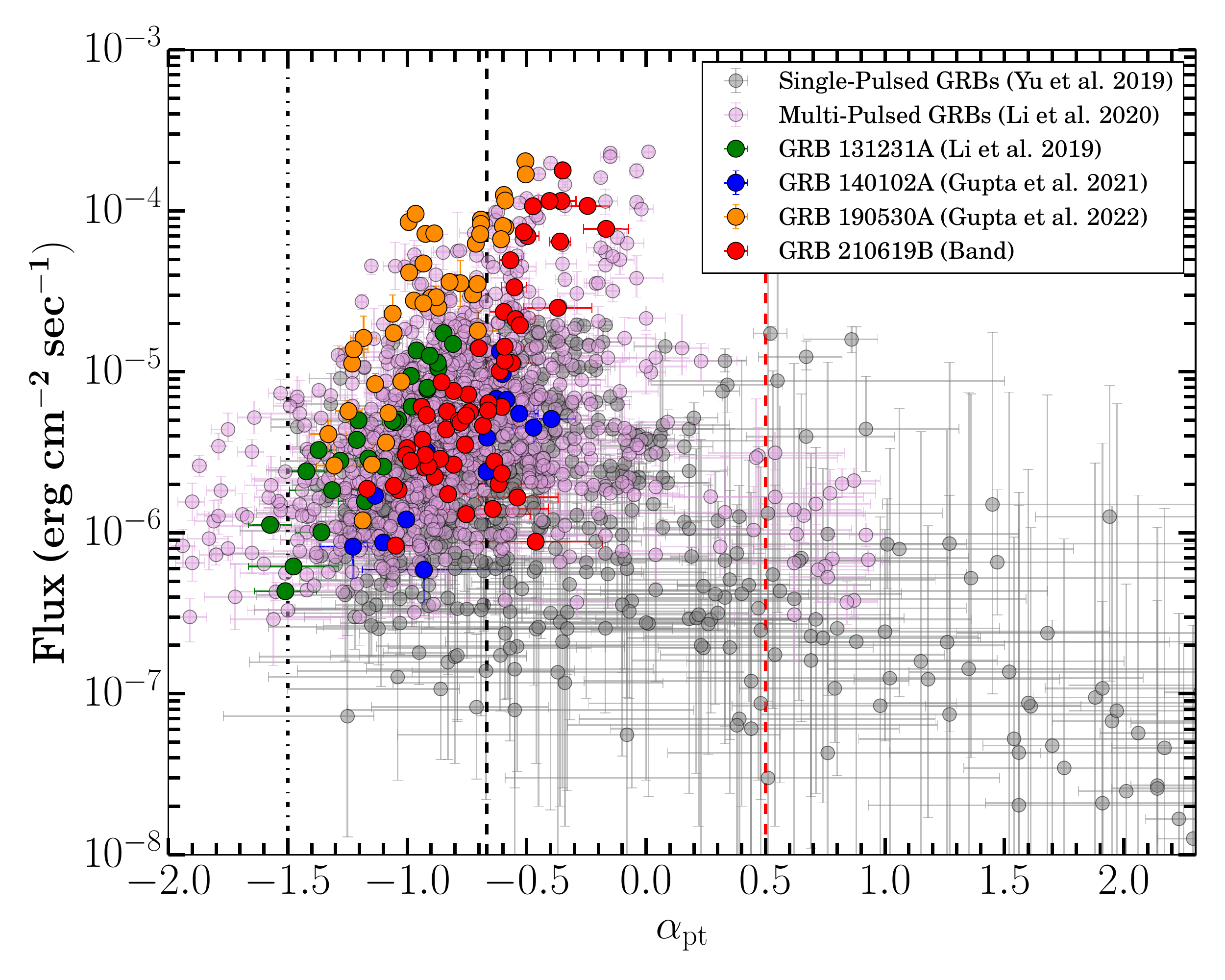}
\includegraphics[scale=0.32]{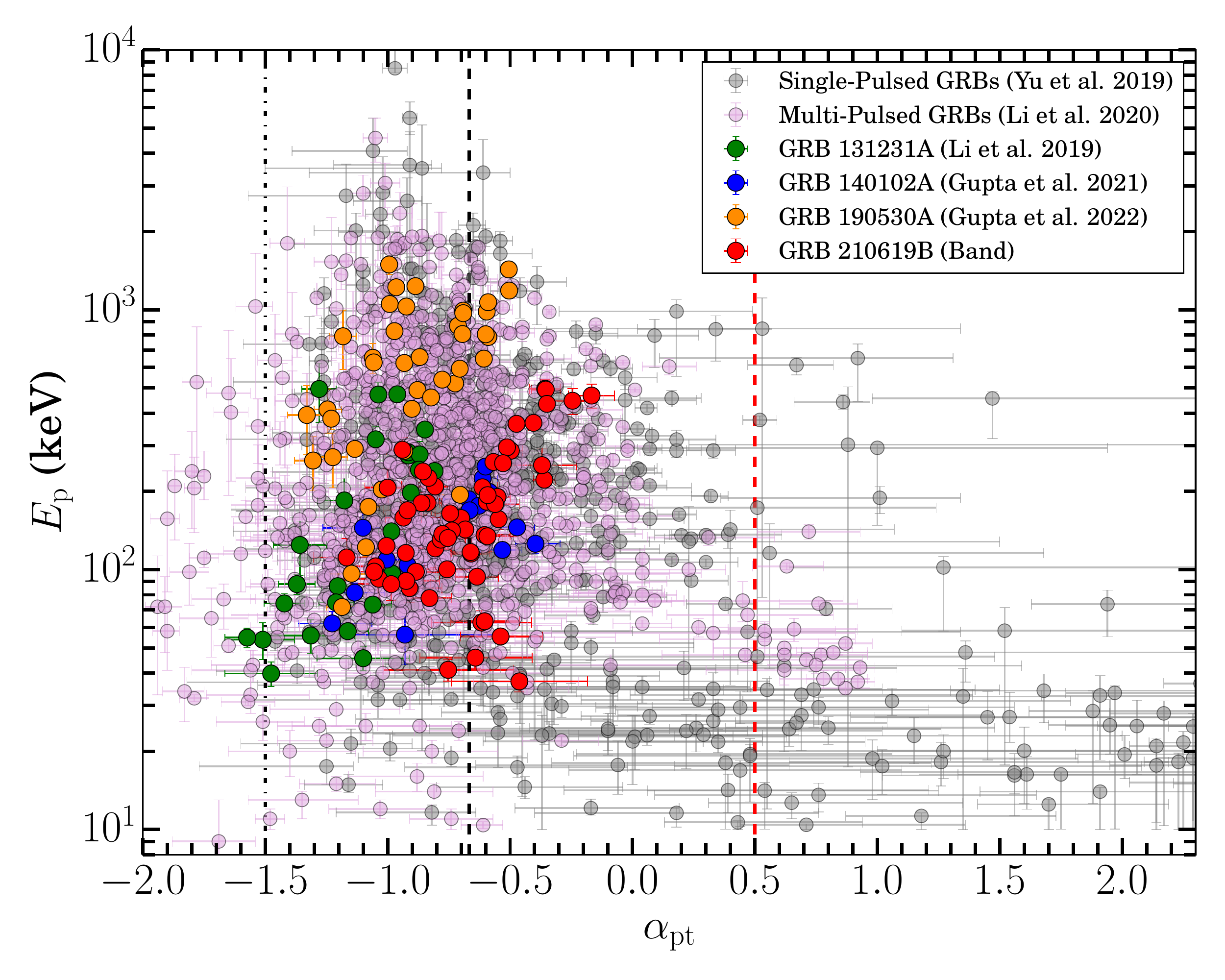}
\caption{{\bf Comparison between the spectral parameters:} {Top panel:} Comparison between the peak energy and energy flux obtained using time-resolved spectral modelling with the 
\sw{Band} model for \thisgrb with well-studied sample of single, multi-pulsed, and Double tracking bursts. The vertical/horizontal pink and plum lines show the mean values of \Ep/flux of \thisgrb and sample of multi-pulsed GRBs, respectively. {Middle panel:} Comparison between the low-energy power-law index and energy flux. {Bottom panel:} Comparison between the low energy power-law index and peak energy. The vertical black dotted-dashed, black dashed, and red dashed lines shows the synchrotron fast cooling, slow cooling, and jitter radiation.}
\label{TRS_comparison}
\end{figure}

We collected the spectral parameters of single and multi-episodic bursts and compared them with those of \thisgrb. Figure~\ref{TRS_comparison} (top panel) shows the distribution of time-resolved peak energy as a function of energy flux of \thisgrb along with a large sample of GRBs. We noticed that the mean values of energy flux from \thisgrb are significantly larger than the mean value of similar multi-pulsed GRBs; however, the mean values of the peak energy from \thisgrb are softer in comparison with the mean values of multi-pulsed GRBs. Figure~\ref{TRS_comparison} (middle panel) shows the distribution of time-resolved low-energy photon indices as a function of energy flux of \thisgrb along with a large sample of GRBs. The $\alpha_{\rm pt}$ values of \thisgrb exceed the synchrotron LOD during the brighter phase. We also found that for some of the single and multi-pulsed GRBs even exceed the low-energy photon index, predicted from jitter radiation ($\alpha_{\rm pt}$ = +0.5; \citealt{2000ApJ...540..704M}). Figure~\ref{TRS_comparison} (bottom panel) shows the distribution of time-resolved peak energy as a function of $\alpha_{\rm pt}$ of \thisgrb along with a large sample of GRBs. 
\par
Furthermore, we also compared the spectral parameters of \thisgrb with those bursts (GRB 131231A, GRB 140102A, GRB 190530A, and others) having ``Double-tracking" evolution pattern of \Ep and $\alpha_{\rm pt}$ \citep{2019ApJ...884..109L, 2021MNRAS.505.4086G, 2022MNRAS.511.1694G}. We noticed that some bursts are consistent with the non-thermal thin shell synchrotron emission model, however, some of them required a hybrid non-thermal synchrotron plus thermal photospheric emission. This suggests that the observed ``Double-tracking" evolution characteristics could be independent of the emission mechanisms of GRBs, though many more such bright bursts are needed to be analysed and physically modelled to confirm his.   

\section{Broadband afterglow emission of \thisgrb}
\label{afterglow}

\subsection{Observations and data analysis}
\label{afterglow_observations}

Followed by the \fermi and \swift detection of \thisgrb with larger error circles; further localization improvements were made with the X-ray Telescope (XRT) and the optical-UV telescope (UVOT) on-board \swift \citep{2021GCN.30267....1B, 2021GCN.30269....1P}. This triggered several facilities on-ground and further in space. This section presents the broadband afterglow observational and data analysis details carried out by our optical follow-up programs using various ground-based telescopes, including the 10.4\,m GTC.  

\subsubsection{\fermi LAT}

At the \fermi GBM trigger time, the burst was outside the \fermi Large Area Telescope (LAT) FoV (the boresight angle was 109 deg.) and it came into the FoV $\sim$ 200 s post GBM trigger time. \fermi LAT detected high energy emission associated from \thisgrb at the position RA= 319.7, DEC = 33.9 deg. (J2000) with an error radius of 0.13 deg. \citep{2021GCN.30270....1A}.

We extracted the LAT high energy data of \thisgrb for a time interval of $\sim$ 10\,ks (from \fermiT+285\,s to \fermiT+10\,ks) using the LAT data server web-page \footnote{\url{https://fermi.gsfc.nasa.gov/cgi-bin/ssc/LAT/LATDataQuery.cgi}} and performed the unbinned likelihood analysis using the official LAT data analysis GUI (python) based software known as \sw{gtburst} \footnote{\url{https://fermi.gsfc.nasa.gov/ssc/data/analysis/scitools/gtburst.html}}. After downloading the LAT data, we used the \sw{Make navigation plots} tool to check when the source was within the FoV of the instrument. We noticed that the source was outside the FoV (source off-axis angle $\> 65^{\circ}$) at the time of \fermiT . It comes within the FoV at $\sim$ 285\,s post \fermi trigger and remains till $\sim$ 2850\,s post \fermiT. \fermi LAT observed the source multiple times between \fermiT+285\,s to \fermiT+10\,ks. For the unbinned likelihood analysis of LAT data, we have used the following parameters: region of interest (ROI) = $\rm 10^{\circ}$, energy range = 100\,MeV - 100\,GeV, and angular cut = 105$^{\circ}$. For the total time intervals, the source (\sw{P8R3\_SOURCE\_V2}) response has been used as the data class, and for short temporal bins, we used the transient (\sw{P8R2\_TRANSIENT020E\_V6}) response. Further, we estimated the probability of the LAT photons being associated with the source using the \sw{gtsrcprob} tool. A more detailed methodology for \fermi LAT data analysis is presented in \cite{2021MNRAS.505.4086G}. We also compared the characteristics of LAT detected \thisgrb with \fermi LAT second GRB catalog (see \S~\ref{Comparison with LAT catalogue}).

\subsubsection{\swift XRT}

The soft X-ray observations of \thisgrb were started by the \swift X-ray Telescope (XRT; \citealt{2005SSRv..120..165B}) in search of the X-ray afterglow at 00:04:53.4 UT on 20 June 2021 ($\sim$ 328.1\,s post \swift BAT detection). The XRT discovered a bright and fading uncatalogued X-ray source at the position RA= 319.7161, DEC= 33.8495 deg. (J2000) with an error radius of 11.1\,arcseconds \citep{2021GCN.30261....1D}. This position was within the \swift BAT uncertainty region, 38\,arcseconds from the \swift BAT onboard location. We obtained the \swift XRT temporal and spectral data for the X-ray afterglow of \thisgrb from the \swift XRT GRB lightcurve repository\footnote{\url{https://www.swift.ac.uk/xrt_curves/}} and \swift XRT GRB spectrum repository\footnote{\url{https://www.swift.ac.uk/xrt_spectra/}}, respectively. These repositories are provided by UK Swift Science Data Centre and web pages \citep{2007A&A...469..379E,2009MNRAS.397.1177E}. For the spectral analysis of \swift XRT data, we have used the {\tt XSPEC} software. We have used the XRT spectrum in the 0.3-10 \keV energy range and fitted the spectrum using an absorbed power-law model (one model with absorption fixed to the contribution of our Galaxy \sw{phabs} and another with free absorption to constrain this parameter from the contribution of the host galaxy \sw{zphabs}).

\subsubsection{BOOTES}

The Burst Observer and Optical Transient Exploring System (BOOTES) followed the \thisgrb trigger with the 60\,cm BOOTES-4/MET robotic telescope located at Lijiang Astronomical Observatory, China \citep{2012ASInC...7..313C}. The observation was performed on 20 June 2021 at 14:20:16 UT, which is $\sim 0.6$ day after the burst trigger. A series of images were taken using the clear filter with 60 s exposures. Due to the poor weather condition, the optical afterglow was not detected in the stacked image after the standard correction procedure of bias and flat. Finally, an upper limit was obtained, calibrated with nearby stars in the USNO-B1.0 catalog, as mentioned in Table B4 of the appendix.

\subsubsection{OSN}

Eight hours after the burst trigger, we used the 1.5\,m telescope at {\it Observatorio de Sierra Nevada} (OSN, Granada, Spain) to monitor the optical afterglow evolution of \thisgrb \citep{2021GCN.30293....1H}. Four epochs observations were executed on 20-25 June 2021. In the first epoch, images in BVRI-filters with 90 s exposure each were obtained. The afterglow is clearly seen in the single-frame except for the B-band even after stacked. Therefore, the second and third epochs were obtained using 300\,s exposures in RI-filters, but the afterglow was barely seen in single images. Even though the I-band images with 300\,s exposures  taken in order to get deep monitoring of it, it was indeed weakly seen in the stacked image. We analysed these imaging data with IRAF software using aperture photometry performed with standard procedures after bias and flat-field corrections. The magnitudes were calibrated using the reference stars in the same field of view from the Pan-STARRS catalog \citep{2020ApJS..251....7F} with equations from \citet{2004PASP..116..133L} for the transformation between them. Results are shown in Table B4 of the appendix.

\subsubsection{CAHA}

We also triggered the 2.2\,m telescope at Calar Alto Observatory located at Almeria, Spain. Using the onboard instrument Calar Alto Faint Object Spectrograph (CAFOS), images in Sloan-r filter (with 180\,s exposure each) were obtained on 27 June 2021 starting at 02:59 UT, which is already about 8 days after the burst trigger. The optical afterglow is still detected in the stacked image. The reduction process followed standard procedures using the IRAF routine after bias and flat correction. Then the photometric results were calibrated with nearby stars from Pan-STARRS catalog \citep{2020ApJS..251....7F}. A full observation log is shown in Table B4 of the appendix.

\subsubsection{GTC/OSIRIS: spectroscopic}

We performed optical spectroscopy observations of \thisgrb afterglow with the Optical System for Imaging and low-Resolution Integrated Spectroscopy (OSIRIS), mounted on the 10.4\,m GTC, under the program GTCMULTIPLE2F-21A (PI: A. J. Castro-Tirado) on 25 June 2021. The observation consisted of 3x1200\,s frames using the grism R2000B, which covers a spectral range from 3950\,{\AA} to 5700\,{\AA} as tabulated in Table B4 of the appendix.

We reduced the GTC/OSIRIS data using \emph{PypeIt} v1.8.1\footnote{\url{https://github.com/pypeit/PypeIt/releases/tag/1.8.1}} \citep{2022zndo...3506872P}. This is an open-source Python-based spectroscopic data reduction pipeline \citep[see][]{2020JOSS....5.2308P}.
Basic image processing (overscan, bias, flat-fielding, cosmic removal, etc.) followed standard techniques. Wavelength solutions (in vacuum, heliocentric frame) for individual images were computed from HgAr, Xe, and Na arc lamps. Flux calibration was derived from the G24-9 standard star. Given that the afterglow trace is not detected in the individual frames, we combined only the 2D frames with different spectral binnings (from 1 to 6 pixels) in order to find the best compromise between SNR and spectral resolution for the final 1D spectrum. However, we only detect (at best) a featureless, very low significance continuum at the position of the GRB afterglow. For the analysis reported hereafter, we have used the redshift value ($z$ = 1.937) for \thisgrb reported by \cite{2021GCN.30272....1D} using an earlier GTC spectrum, obtained only 2.5\,hr after the GRB detection.

\subsubsection{GTC/OSIRIS: optical imaging}

In order to get the deep image of the burst field, a late time follow-up observation was executed with the 10.4\,m GTC under the same program on 9 July, i.e., 20 days after the trigger. During this epoch, the seeing was 1.6" which is not as good as expected, therefore, we carried out a second epoch deep imaging on 30 July, i.e. 41 days after the trigger. During this epoch, the seeing was 0.9", better than the first epoch observations. 
 
A series of images were taken using Sloan-griz filters with exposures of 90\,s, 60\,s, and 50\,s, respectively. These images were analysed using the aperture photometry method after bias and flat correction with IRAF routines and calibrated with reference stars in the nearby field from the Pan-STARRS catalog. In our first epoch observations, the optical counterpart of \thisgrb located in the UVOT/\swift error region was detected in the r band with a magnitude of 24.75 mag. However, in our second epoch of observations, we did not detect the optical counterpart of \thisgrb, and we obtained an upper limit of r $\sim$ 25.7 mag.

In addition to our late-time photometric observations of the optical afterglow of \thisgrb (see Table B4 of the appendix), for completeness, we also utilized those reported in Gamma-ray Coordination Network (see Table B5 of the appendix) in our analysis.

\subsubsection{Host galaxy Search}

\begin{figure*}
	\centering
	\includegraphics[angle=0,scale=0.27]{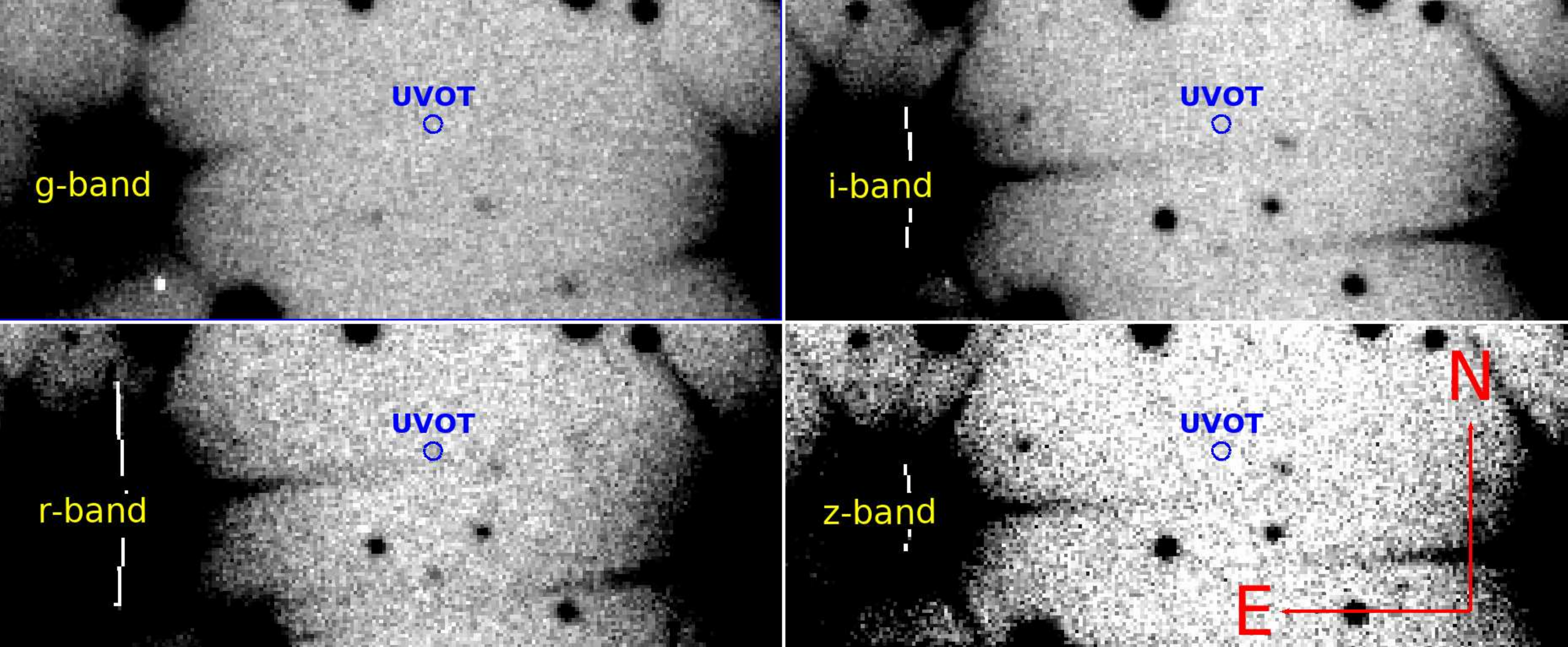}
	\caption{Late deep observation obtained with Sloan griz-filters by the 10.4\,m GTC telescope on 29 July 2021 (i.e., $\sim$ 40 days after trigger) at the \thisgrb field. The UVOT position is marked with the blue circle. There is no possibility of an existing underlying host galaxy detected within the UVOT error region. East is leftwards, and North is upwards.} 
	\label{GRB210619Bhostgalaxy}
\end{figure*}

On 10 July 2021, late-time observations of the burst took place with the 10.4\,m GTC telescope. Then at 20 days post-trigger, a point-like source i.e., the afterglow was still clearly detected in the stacked images of each gri and z-bands at the UVOT position \citep{2021GCN.30278....1K}. Later on, second epoch observations were taken on July 30, showing that the source faded beyond detection and no underlying object was found within the error region (see Figure~\ref{GRB210619Bhostgalaxy}). An excess of counts/signal in the late time r-band stacked image can be seen aligned in the direction of a nearby star's spark. Furthermore, the source position in redder filters, i.e., iz-bands, was out of the spark direction, and still, no credible detection was found. Therefore, we conclude that the host galaxy has not been detected during our two epochs' observations and that the host's brightness is fainter than our estimated upper limits (see Table~B4 of the appendix). For this burst, the i-band absolute magnitude of $<$-19.5 mag, points to a fainter and perhaps dwarf galaxy host as discussed in \citet{2016ApJ...817....8P, 2017MNRAS.467.1795L}.  

\subsection{Results}

The present section highlights the broadband late-time afterglow features of \thisgrb using our observations (see \S~\ref{afterglow_observations}) along with those reported in the public domain (see Table~B5 of the appendix). The broadband afterglow behaviour of GRBs could be understood either indirectly through empirical fitting constraining the temporal and spectral indices (closure relationship) or directly through external shock fireball modelling. Here, we have explored the empirical fitting method to understand the nature of the late-time broadband afterglow emission of \thisgrb and have compared it with the afterglow results reported by \cite{2021arXiv210900010O}. 

\subsubsection{Temporal and spectral evolution of X-ray and optical afterglows}

\begin{figure}
\centering
\includegraphics[scale=0.35]{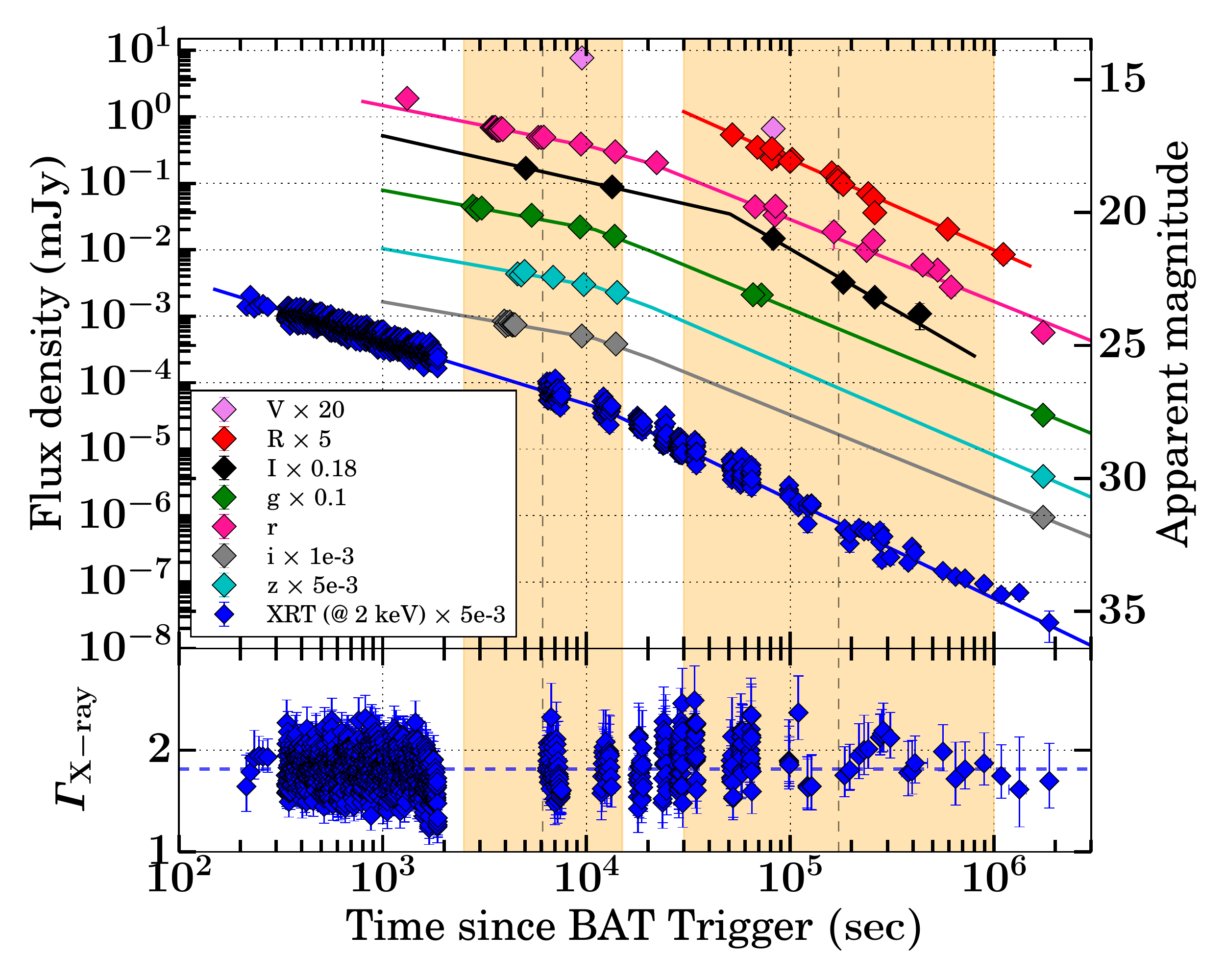}
\caption{{\bf Multi-wavelength afterglow of \thisgrb.} {Top panel:} The various colored solid lines shows the empirical power-law or broken power-law model fitting to the observed data points. The optical data sets are corrected for the foreground extinction. {Bottom panel:} The temporal evolution of X-ray photon indices. The horizontal blue dashed line shows the mean value of the photon indices. The vertical orange shaded regions indicate the epochs of spectral energy distributions. The vertical black dashed lines show the corresponding centered epochs of both the SEDs.}
\label{Afterglowlc}
\end{figure}

The multi-wavelength afterglow light curve of \thisgrb is shown in Figure~\ref{Afterglowlc} (top panel). The X-ray afterglow light curve of \thisgrb does not show any flaring activity and could be fitted using a broken power-law function with the following temporal parameters: temporal index before the break $\alpha_{\rm x,1}$= $0.95 \pm 0.01$, temporal index after the break $\alpha_{\rm x,2}$= $1.49 \pm 0.03$, and break time t$_{\rm x, b}$= $12241 \pm 1141$\,s ($\chi^2/d.o.f=1397/1519$). 
We modelled the late time time-averaged (from \fermiT+ 6394 to \fermiT+ 34829\,s) X-ray PC mode spectrum (early WT mode observations may not provide accurate column density due to the fluctuations) to constrain the intrinsic hydrogen column density and other X-ray spectral parameters of \thisgrb. We obtained the following spectral parameters using time-integrated XRT observations: X-ray photon indices $\Gamma_{\rm X-ray}$ = $1.90^{+0.06}_{-0.06}$, and intrinsic hydrogen column density $N_\mathrm{H,z}$ = $9.25^{+3.01}_{-2.86} \times 10^{21}~\mathrm{cm}^{-2}$. The bottom panel of Figure~\ref{Afterglowlc} shows the temporal evolution of X-ray photon indices during the entire X-ray afterglow emission phase. We noticed that the X-ray photon indices do not change during the entire XRT observations epochs (mean value of $\Gamma_{\rm X-ray}$ = 1.81). The observed temporal break in the X-ray afterglow light curve could be explained either due to the crossing of break frequencies or due to a possible jet break. However, as the spectral indices do not change, it supports the later scenario (jet break). Based on the above and in the light of applicable closure relationships for the ISM forward shock model, the X-ray afterglow of \thisgrb seems to be consisted with normal decay phase followed by a jet break phase.

The late-time optical afterglow light curve of \thisgrb using data from the present analysis and those published ones could be well described either by a single power-law (only the R band light curve due to unavailability of early, i.e. before the jet break, data in the same filter) or broken power-law functions. We have used the r-band light curve (longest and densest temporal coverage) to understand the temporal features of the optical emission of \thisgrb. The foreground extinction corrected r-band light curve could be fitted using a broken power-law function with the following temporal parameters: temporal index before the break $\alpha_{\rm r,1}$= $0.61 \pm 0.04$, temporal index after the break $\alpha_{\rm r,2}$= $1.23 \pm 0.05$, and break time t$_{\rm r, b}$= $16076 \pm 3332$\,s. We noticed that the break time in the optical light curve is consistent with the break time in the X-ray afterglow light curve. 

\subsubsection{Joint X-ray and optical SED evolution}
\label{SED}

We created joint spectral energy distributions (SEDs) to constrain the intrinsic host extinction and the X-ray and optical spectral indices (both before and after the jet break phase), as described below. Then, we utilized the closure relationships using temporal and spectral indices (both for optical and X-ray) in the ISM as well as WIND mediums (without energy injection from the central engine) to constrain the location of the cooling-break frequency ($\nu_{c}$) and other afterglow properties.

\begin{figure}
\centering
\includegraphics[scale=0.55]{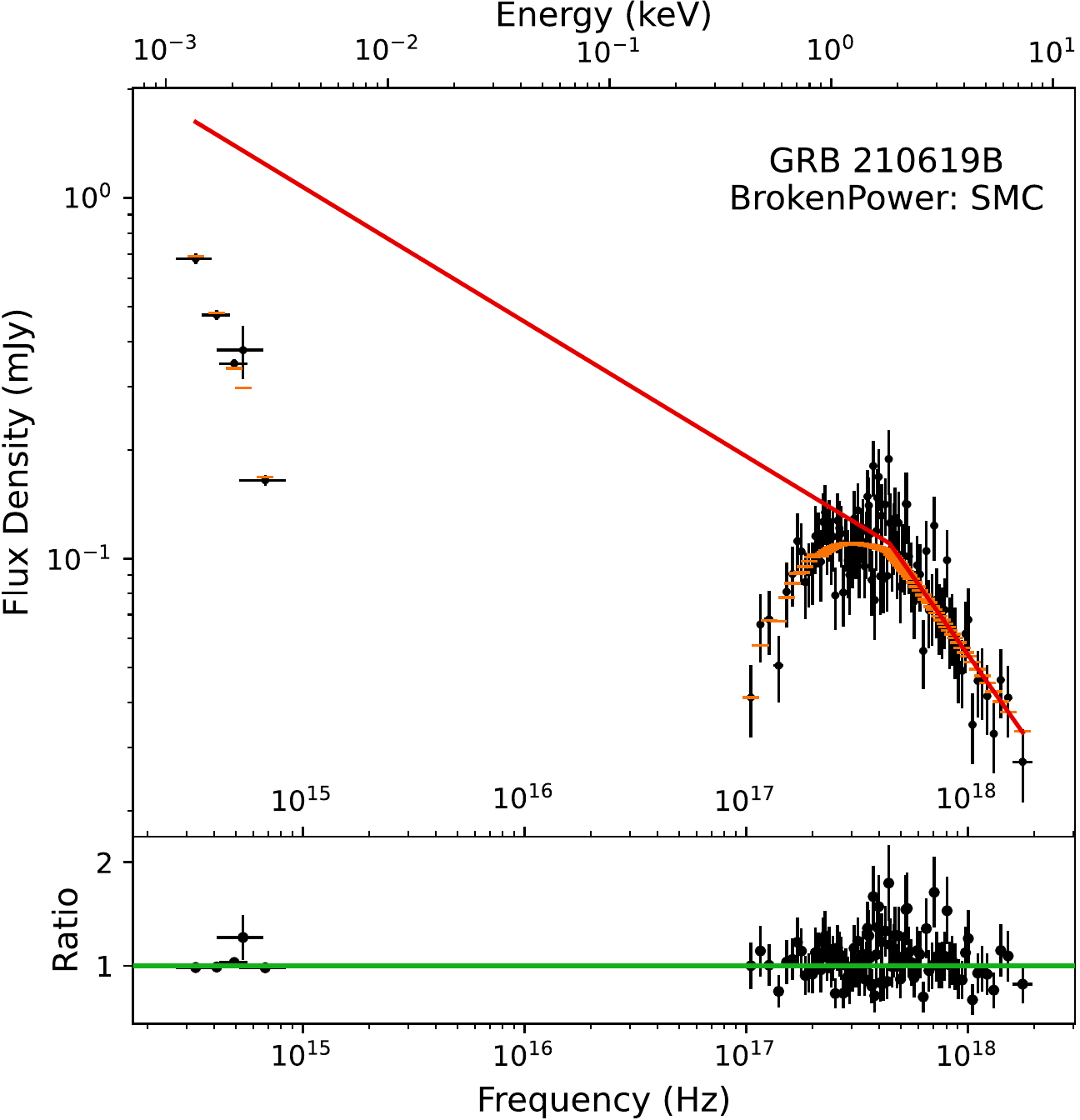}
\includegraphics[scale=0.55]{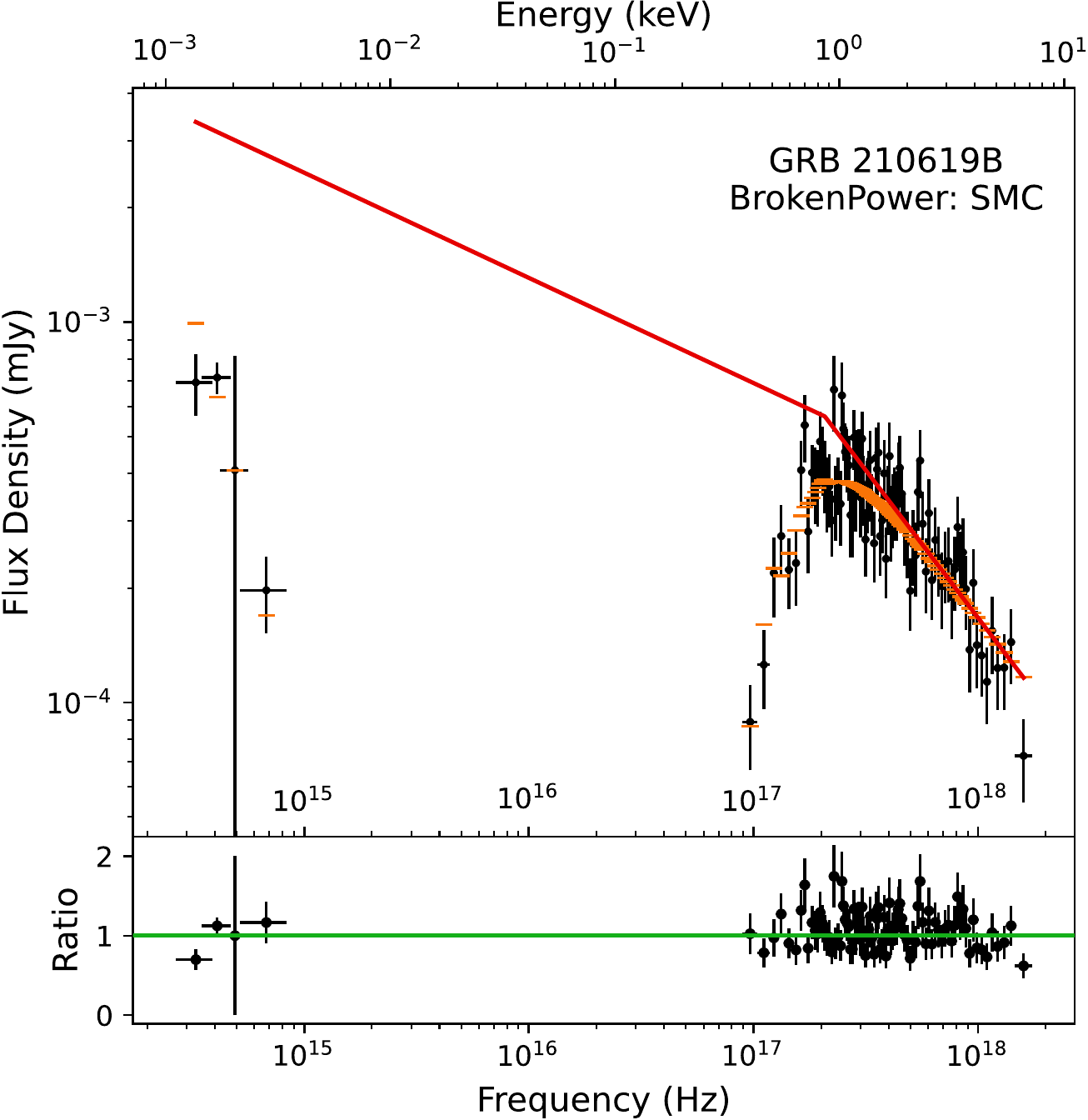}
\caption{{\bf Joint X-ray and optical SED of \thisgrb.} {Top panel:} SED 1 centered at 6100\,s (\fermiT+2500  to \fermiT+15\,ks) (plus optical data in Vrigz filters; data from Figure~\ref{Afterglowlc}) and {Bottom panel:} SED 2 centered at 173\,ks using XRT data from (\fermiT+$30\,{\rm ks}$  to \fermiT+ $10^{6}\,{\rm s}$) plus optical data in griz filters from GTC. Both SEDs were the best fit using the broken power-law model with the SMC as the best fit host extinction law. The host galaxy absorbed and extinguished spectral model points are given in orange and the best-fit absorption and extinction corrected spectral model is given by the solid red line. The ratio of the data divided by the folded model is given in the sub-panel of each panel.}
\label{AfterglowSED}
\end{figure}

We constructed two optical- X-ray SEDs centered at 6100\,s and 173\,ks. The first SED at 6100\,s was constructed with the methodology of \cite{sch10}, using data within a time range of $2500-15\,{\rm ks}$. To construct the pha files for the different optical filters at $6100\,{\rm s}$, we first constructed a single filter light curve by normalising the individual filter light curves together. Then using this single filter light curve, we determined the temporal slope within the $2500-15\,{\rm ks}$ interval. We then fitted a power-law to the individual filter light curves within this same time interval, fixing the slope to the value determined from the single filter light curve. We used the derived normalisations to compute the count rate and count rate error at $6100\,{\rm s}$, which was then applied to the relevant spectral file. For the XRT data, we have taken the PC mode spectral files built using the time-sliced spectra option on the Leicester \textit{Swift} XRT webpage \footnote{https://www.swift.ac.uk/xrt\_spectra} \citep{2009MNRAS.397.1177E}. The spectral files were normalized to correspond to the $0.3-10\,{\rm keV}$ flux of the afterglow at the mid point of the selected time range. The flux, used to normalize a given spectrum, was determined by fitting a power-law to the temporal data within the SED selected time range. The best-fit decay index was used to compute the flux at the mid-point of the SED, in the same way as was done for the optical data. The optical pha files for the second SED centered at 173\,ks, were constructed directly from the GTC magnitudes taken at that time, assuming the data are quasi-simultaneous. The X-ray data for the second SED was constructed using the same methodology as the 6100\,s SED, except that the data were extracted from $30000-10^6\,{\rm s}$ and normalised to 173\,ks. 

Both SEDs were fitted using {\textsc XSPEC} (version 12.12.0), following the procedure outlined in \cite{sch10,schady07}. We tested two different models for the continuum: a power-law and a broken power-law, with the change in spectral slope fixed to be $\Delta \beta = 0.5$ for the broken-power law component, which corresponds to the expected change in spectral slope caused by the synchrotron cooling frequency \citep{1998ApJ...497L..17S}. In each of these models, we also included two dust and gas components to account for the Galaxy and host galaxy dust extinction and photoelectric absorption (\textit{phabs}, \textit{zphabs}, \textit{zdust}; we used \textit{zdust} for both the Galaxy and host galaxy dust components, but with the redshift set to zero for the Galaxy dust component). The Galactic components were frozen to the reddening and column density values from \cite{schlegel} and \cite{kalberla}, respectively. For the host galaxy dust extinction, we tested the dependence of dust extinction on wavelength for three different scenarios: the Small and Large Magellanic clouds (SMC and LMC, respectively) and on the Milky Way (MW). In addition, we also accounted for the absorption due to the Lyman series in the $912-1215$\, \AA \, restframe wavelength range, with the XSPEC \textit{zigm} component. We set this component to used the prescription of \cite{mad95} to estimate the effective optical depth from the Lyman series as a function of wavelength and redshift, and we set this component to also include attenuation due to photoelectric absorption.

\begin{table*} 
\centering 
\caption{Results from simultaneous UV/optical and X-ray spectral fits for the SMC, LMC, and MW dust-extinction law models, for both a power-law (POW) and a broken power-law (BKP) continuum. 
For the broken power-law model, the second spectral index is fixed as $\beta_2= \beta + 0.5$. The fourth and fifth columns give the host galaxy equivalent column density and E(B-V), the sixth column gives the break energy for the broken power-law spectral models, the $\chi ^2$ and degree of freedom (dof) of the fit are given in the seventh column, and the eighth gives the null hypothesis probability.
\label{tab:sedfits}} 
\begin{tabular}{@{}cccccccc} 
\hline 
SED & Model & $\beta$ & $N_H$ & E(B-V) & $E_{bk}$ & $\chi ^2$ (dof) & Null Hypothesis\\ 
 & & & $10^{21}$~cm$^{-2}$ & (mag) & (\keV) & & Probability\\ 
\hline
\hline
6100\,s &  MW/POW  & $ 0.47 ^{+ 0.01 }_{ -0.01 } $ & $ 0.60 ^{+ 1.70 }_{ -0.60 } $ & $ 0.24 ^{+ 0.01 }_{ -0.01 } $ & $ -$ & 378 (95)  &  <1e-05 \\
6100\,s &  LMC/POW  & $ 0.45 ^{+ 0.01 }_{ -0.01 } $ & $ 0.27 ^{+ 1.65 }_{ -0.27 } $ & $ 0.19 ^{+ 0.01 }_{ -0.01 } $ & $ -$ & 184 (95)  &  <1e-05 \\
6100\,s &  SMC/POW  & $ 0.41 ^{+ 0.00 }_{ -0.01 } $ & $ <3.95$ & $ 0.15 ^{+ 0.01 }_{ -0.01 } $ & $ -$ & 170 (95)  &  <1e-05 \\
6100\,s &  MW/BKP  & $ 0.42 ^{+ 0.02 }_{ -0.02 } $ & $ 4.80 ^{+ 2.13 }_{ -1.98 } $ & $ 0.21 ^{+ 0.01 }_{ -0.01 } $ & $ 2.02 ^{+0.20}_{-0.18} $ & 319 (94)  &  <1e-05 \\
6100\,s &  LMC/BKP  & $ 0.42 ^{+ 0.01 }_{ -0.01 } $ & $ 5.42 ^{+ 2.15 }_{ -1.98 } $ & $ 0.18 ^{+ 0.01 }_{ -0.01 } $ & $ 1.93 ^{+0.19}_{-0.16} $ & 116 (94)  &  0.06 \\
6100\,s &  SMC/BKP  & $ 0.37 ^{+ 0.01 }_{ -0.01 } $ & $ 4.24 ^{+ 2.12 }_{ -1.94 } $ & $ 0.14 ^{+ 0.01 }_{ -0.01 } $ & $ 1.84 ^{+0.17}_{-0.17} $ & 87 (94)  &  0.695 \\
\hline
173\,ks &  MW/POW  & $ 0.72 ^{+ 0.05 }_{ -0.05 } $ & $ 7.90 ^{+ 2.45 }_{ -2.23 } $ & $ 0.73 ^{+ 0.08 }_{ -0.07 } $ & $ - $ & 146 (90)  &  1.86e-04 \\
173\,ks &  LMC/POW  & $ 0.74 ^{+ 0.06 }_{ -0.05 } $ & $ 8.63 ^{+ 2.62 }_{ -2.41 } $ & $ 0.72 ^{+ 0.08 }_{ -0.08 } $ & $ - $ & 146 (90)  &  1.84e-04 \\
173\,ks &  SMC/POW  & $ 0.76 ^{+ 0.05 }_{ -0.05 } $ & $ 9.52 ^{+ 2.61 }_{ -2.41 } $ & $ 0.79 ^{+ 0.08 }_{ -0.08 } $ & $ - $ & 147 (90)  &  1.37e-04 \\
173\,ks &  MW/BKP  & $ 0.34 ^{+ 0.04 }_{ -0.04 } $ & $ 8.79 ^{+ 3.10 }_{ -3.77 } $ & $ 0.32 ^{+ 0.05 }_{ -0.05 } $ & $ 0.82 ^{+0.27}_{-0.16} $ & 100 (89)  &  0.205 \\
173\,ks &  LMC/BKP  & $ 0.31 ^{+ 0.04 }_{ -0.03 } $ & $ 7.50 ^{+ 2.78 }_{ -3.37 } $ & $ 0.25 ^{+ 0.04 }_{ -0.04 } $ & $ 0.84 ^{+0.25}_{-0.14} $ & 91 (89)  &  0.413 \\
173\,ks &  SMC/BKP  & $ 0.28 ^{+ 0.04 }_{ -0.03 } $ & $ 5.95 ^{+ 2.36 }_{ -3.08 } $ & $ 0.21 ^{+ 0.04 }_{ -0.04 } $ & $ 0.86 ^{+0.23}_{-0.13} $ & 92 (89)  &  0.403 \\
\hline
\end{tabular} 
 \end{table*}

The results of the SED fits are shown in Figure \ref{AfterglowSED} and tabulated in Table~\ref{tab:sedfits}. For the $6100\,{\rm s}$, for both the power-law and broken power-law, the SMC type extinction gives the best fit, for the power-law model the fit gives $\chi^2/d.o.f=170/95$, while for the broken power-law model the fit gives $\chi^2/d.o.f=87/94$. An F-test gives a p-value of $<<10^{-5}$, suggesting the break is statistically required. For the $173\,{\rm ks}$ SED, for the power-law model the fits with the 3 different extinction laws give similar $\chi^2/d.o.f$, however, the MW extinction law is preferred. The broken power-law model gives a smaller $\chi^2/d.o.f$ compared to the power-law for each of the three extinction curves. In this case, the $\chi^2/d.o.f$ for the LMC and SMC fits are very similar and slightly preferred over the MW. Given the preference for the SMC in the first SED, we assume this as the best extinction curve for the 173\,ks SED. The F-test on SMC power-law ($\chi^2/d.o.f=147/90$) and SMC broken power-law ($\chi^2/d.o.f=92/89$) gives a p-value of $<<10^{-5}$, suggesting the break is statistically required. Overall, for both the SEDs, we take the best fit model as the broken power-law, with SMC type extinction.

\subsubsection{GeV afterglow and Comparison with LAT catalog}
\label{Comparison with LAT catalogue}

\begin{figure}
\centering
\includegraphics[scale=0.35]{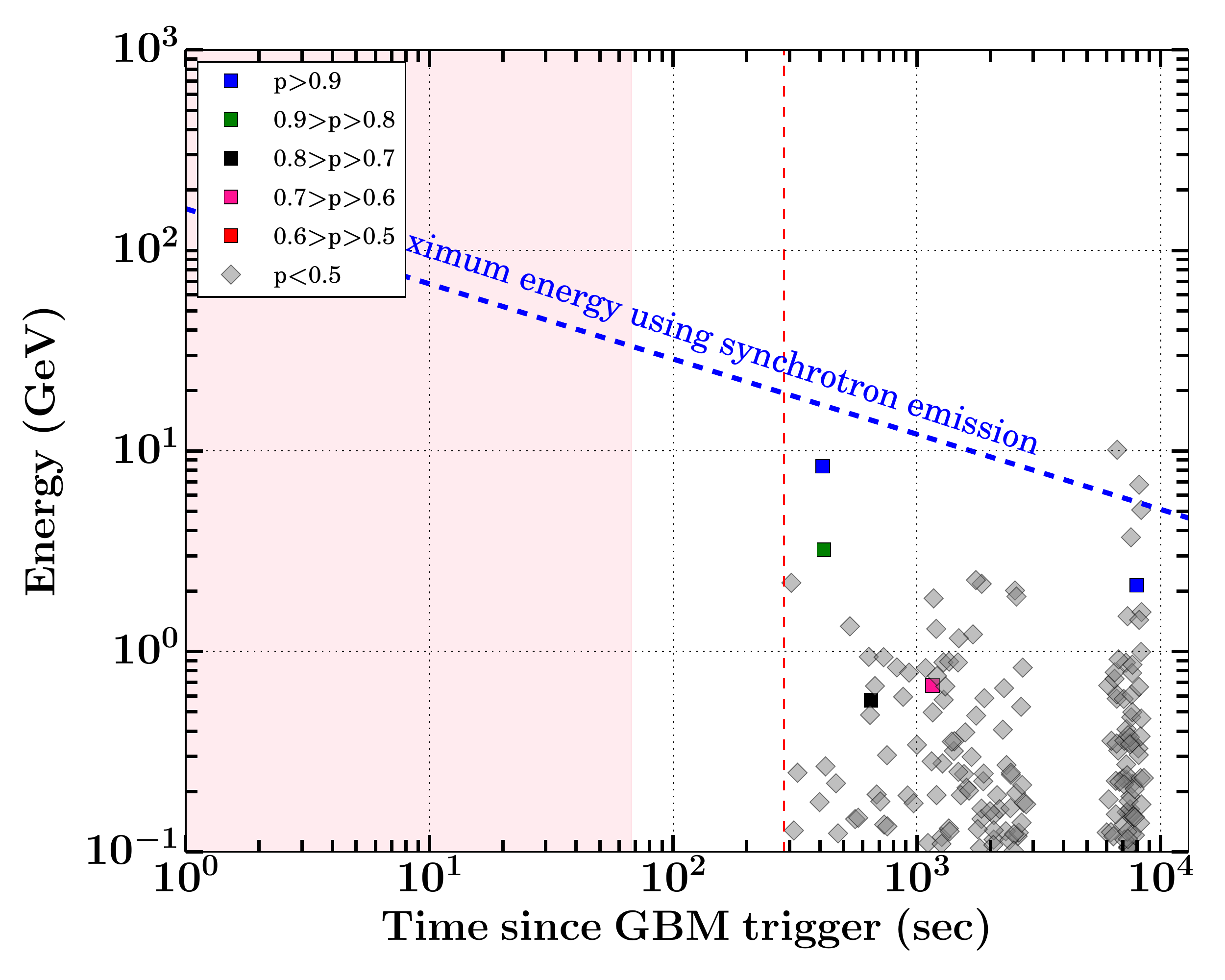}
\caption{{\bf Delayed LAT emission:} Distribution of high energy GeV photons observed by \fermi LAT from \thisgrb. The colored squares and diamond shows the probability of individual photons to be associated with \thisgrb. The blue dashed line represents the maximum possible high energy GeV energy using the synchrotron process.}
\label{GeV_LAT_GRB210619B}
\end{figure}

Figure~\ref{GeV_LAT_GRB210619B} shows the distribution of the high energy photons and their association with \thisgrb, detected by \fermi LAT. \fermi LAT observed the highest-energy photon at $\sim$ 410\,s post \fermi 
trigger with an energy of 8.38 GeV \citep{2021GCN.30270....1A}. During the time window of our LAT analysis (\fermiT+ 285\,s to \fermiT+ 10\,ks), we found the following spectral parameters: energy 
flux = $(4.35 \pm 2.15) ~\times 10^{-10}$  $\rm ~erg ~cm^{-2} ~s^{-1}$ (in 100\,MeV - 10\,GeV energy range), photon flux = $(2.38 \pm 1.52) ~ \times 10^{-7}$  $\rm ~ph. ~cm^{-2} ~ s^{-1}$ (in 100\,MeV - 10\,GeV energy range), the LAT photon index ($\it \Gamma_{\rm LAT}$) = $-1.42 \pm 0.28$ with the test-statistic (TS) of detection = 30. During this window, we noticed that there were only a few GeV photons ($>$ 100\,MeV) with probability larger than 90 \% to be associated with \thisgrb. To constrain the possible emission process of GeV photons, we used equation 9 from \cite{2010ApJ...718L..63P} to estimate the maximum energy emitted by the synchrotron process during the decelerating phase in the case of ISM-like ambient medium. We noticed that all the GeV photons observed from \thisgrb with associated probability greater than 50 \% could be described using synchrotron emission.

In the top panel of Figure~B2 of the appendix, we have plotted the distribution of the maximum energy of the highest-energy LAT photons with respect to the arrival time for \thisgrb. We have also shown the other LAT detected GRBs data points obtained from the second \fermi LAT GRB Catalog (2FLGC). In the case of \thisgrb, the highest-energy GeV photon detected by \fermi LAT arrived after the \fermi GBM \tninty duration, in agreement with a large fraction of \fermi LAT detected bursts. Furthermore, we also measured the restframe isotropic $\gamma$-ray energy in the 100\,MeV -10\,GeV  ($E_{\rm LAT, iso}$) using \fermi LAT observations for \thisgrb. We have shown the distribution of $E_{\rm LAT, iso}$ with respect to redshift ($z$) for \thisgrb and have compared 34 other LAT detected GRBs with their measured redshift from the 2FLGC (see bottom panel of Figure~B2 in the appendix). We notice that \thisgrb is a moderately energetic event amongst the \fermi LAT detected GRB at $z$ $\sim$ 2, with the highest-energy photon at 16.76\,GeV in the source restframe. 

\subsection{Discussion}

Based on the above mentioned results using the multi-wavelength analysis of the afterglow properties of \thisgrb, our major findings are discussed below in the following sub-sections.

\subsubsection{Closure relations and cooling break frequency}

We implemented closure relations before and after the jet break phase of \thisgrb, following the predictions of the fireball model without energy injection from the central engine \citep{2009ApJ...698...43R, 2013NewAR..57..141G, 2018ApJ...859..160W}. These ``closure relations'' depend on properties like jet geometry, electron energy index ($p$), spectral regimes (slow/fast) and the surrounding environment \citep{1998ApJ...497L..17S, 2004RvMP...76.1143P}. Our joint optical and SED fitting (see \S~\ref{SED}) before the jet break phase clearly indicates a break (1.84 $\pm$ 0.17 \keV) in the SED within the X-ray band (see the top panel of Figure~\ref{AfterglowSED}). Assuming the break due to cooling break frequency and the slow cooling case for an ISM like ambient medium for $p >$ 2 (a softer power-law (non-thermal) distribution of the shock-accelerated electrons), we have also explored various other possible spectral regimes; however, we noticed that the first SED could be well explained using $\nu_m$ $< \nu_c$ $< \nu_{x}$ (where $\nu_m$, $\nu_c$, and $\nu_{x}$ are the synchrotron peak, cooling-break, and X-ray frequencies, respectively) spectral regime for the ISM slow cooling case. 
For this spectral regime, we calculated the X-ray, i.e. $\alpha_{x}$ = 3($\beta_{x}$-1)/2, and optical ($\alpha_{o}$ = 3$\beta_{o}$/2) temporal indices using their observed values of spectral indices and found $\alpha_{x}$ = 0.81 $\pm$ 0.02 and $\alpha_{o}$ = 0.56 $\pm$ 0.02. These predicted values of X-ray and optical temporal indices are smaller than the observed ones. Using the calculated value of the X-ray spectral index, we estimated the $p$ value of $p$ = 1.74 $\pm$ 0.02 (a smaller value with respect to normally observed cases, i.e. $p >$ 2). \par

Therefore we also examined the ``closure relation'' for the same spectral regime but for harder electron energy index case (with 1 $<~p~<$ 2; \citealt{2008MNRAS.388..144R, 2009ApJ...698...43R}). In this case, we calculated $\alpha_{x}$ = 0.95 $\pm$ 0.01 and $\alpha_{o}$ = 0.70 $\pm$ 0.01 using the observed values of X-ray and optical spectral indices, respectively. We noticed that these values are well in agreement with the observed values ($\alpha_{x}$ = 0.95 $\pm$ 0.01 and $\alpha_{o}$ = 0.61 $\pm$ 0.04). However \cite{2021arXiv210900010O} considered similar values (without host galaxy reddening correction) of optical/X-ray spectral indices ($\beta_{x}$=$\beta_{o}$=0.9) and claimed that at least before the jet break phase the observed $\nu_c$ is above the X-ray band, which is inconsistent with our findings and are ruled out by our SED fitting (see \S~\ref{SED}). Additionally we also implemented the ``closure relations" post jet break phase and noticed that $\nu_c$ follows the typical evolution predicted post jet break phase (i.e. t$^{0}$) which remains close to the XRT spectral window ($\beta_{x}$=$\beta_{o}$+0.5) throughout the afterglow emission phase. The derived value of $\beta_{o}$ using near-simultaneous griz observations with the 10.4\,m GTC at very late epochs, i.e., $\sim$ 20 days post burst, are also consistent with post jet break forward shock model predictions.  

\section{Summary and Conclusions}
\label{summaryandconclusions}

In this work, we present a detailed analysis of the prompt emission plus thermal and spectral characteristics of a long, luminous, and peculiar multi-structured burst \thisgrb (which was also observed by the Atmosphere-Space Interactions Monitor installed on the International Space Station). The prompt light curve shows a very bright and brief pulse ($\sim$ 4\,s) followed by softer and rather long emission episodes up to $\sim$ 65\,s. A detailed time-resolved spectral analysis indicates hard low-energy photon indices, exceeding the synchrotron LOD in most of the Bayesian bins with high significance during brighter pulses, and it becomes softer (within the error bar) during the longer/softer emission phase. Harder $\alpha_{\rm pt}$ values observed during brighter emission pulses could be attributed to thermal emission arising from photospheric regions. On the other hand, the softer values of the low-energy photon indices during the longer emission phase are consistent with the non-thermal thin shell synchrotron emission model. This suggests that the radiation process responsible for \thisgrb shows a transition between photospheric thermal emission (hard $\alpha_{\rm pt}$) and non-thermal synchrotron emission (soft $\alpha_{\rm pt}$). \par

In addition, we noticed a peculiar spectral evolution of \Ep and $\alpha_{\rm pt}$, where both of these spectral parameters exhibit the `flux tracking' pattern. We found strong/moderate positive correlations among various parameters: log (Flux)-log (\Ep), log (Flux)-$\alpha_{\rm pt}$, and log (\Ep)-$\alpha_{\rm pt}$, supporting the observed tracking pattern of the \Ep and $\alpha_{\rm pt}$. The flux tracking behaviour of \Ep could be understood in terms of cooling and expansion of the fireball. In such scenario, during the cooling of relativistic electrons, the magnetic field will decrease, resulting in a lower intensity and \Ep values; however, during the expansion of the fireball, the magnetic field will increase, giving rise to higher intensity and \Ep values \citep{2021MNRAS.505.4086G}. Recently, \cite{2021NatCo..12.4040R} found a relation between the spectral index and the flux by examining the spectral evolution during the steep decay phase (observed with \swift XRT) of bright \swift BAT GRB pulses. They found that all the cases of their sample show a strong spectral softening, which is consistent with a shift of the peak energy. They demonstrated that such behavior is well explained assuming an adiabatic cooling process of the emitting particles and a decay of the magnetic field. \par

Furthermore, \cite{2021A&A...656A.134G} performed numerical simulations and tracked the evolution of electrons to examine the observed evolution patterns of \Ep from GRBs. They suggested that the observed strong correlation between \Ep and flux can be explained if electrons are dominated by adiabatic cooling (considering a constant Lorentz factor). These studies support that adiabatic cooling could play a major role to understand the observed intensity tracking pattern of the peak energy. On the other hand, \cite{2019MNRAS.484.1912R} presented the possible physical interpretation of the flux tracking behaviour of $\alpha_{\rm pt}$ for the bursts having photospheric signatures (for example GRB 090719, GRB 100707A, GRB 160910A, plus many more) in their observed spectra. However, the possible physical interpretation of the flux tracking behavior of $\alpha_{\rm pt}$ for the bursts having synchrotron signature (for example GRB 131231A, and GRB 140102A) in their observed spectra is still an open question. In the case of GRB 210619B, hard $\alpha_{\rm pt}$ values during the main pulse mean that the spectrum has a different origin than synchrotron at the beginning (although, due to limitation of spectroscopy, it could be confirmed using additional prompt emission polarisation measurements). Therefore, assuming photoshperic signatures in the early spectra of GRB 210619B, the flux tracking behaviour of $\alpha_{\rm pt}$ could be understood in terms of photospheric heating in a flow with a varying entropy \citep{2019MNRAS.484.1912R}. \par

We also observed a negative spectral lag for this burst, thus being an outlier of the typical known lag-luminosity anti-correlation of long bursts. The observed negative spectral lag could be explained in terms of a superposition of effects. Recently, some authors explained the observed spectral lag using emission mechanism models such as the photosphere \citep{2021MNRAS.505L..26J} and the optically-thin synchrotron model \citep{2016ApJ...825...97U} of GRBs. However, according to the photosphere jet model, only positive spectral lags of long GRBs can be explained. The observed negative lag for \thisgrb might be explained using optically-thin synchrotron model with the following assumptions: a curved spectrum, an emission radius, a decaying magnetic field (expanding jet), and the acceleration of the emitting regions \citep{2016ApJ...825...97U}. Moreover, the spectral lag is believed to be connected with the spectral evolution (correlation between \Ep and flux): (1) only the positive lags are possible for the hard-to-soft evolution of \Ep, (2) both the positive and negative lags can occur in the
case of intensity tracking pattern of \Ep. Thus, the observed negative lag and intensity tracking behaviour of \Ep for \thisgrb are consistent with each other \citep{2018ApJ...869..100U}.  \par

Further, we explored the empirical fitting method to understand the nature of late-time multi-wavelength afterglow (from GeV to Optical frequencies) of \thisgrb and found that the broadband afterglow emission including delayed GeV afterglow emission detected by \fermi LAT could be explained by the synchrotron process in the case of an ISM-like ambient medium. We produced the joint SED using XRT and optical photometric data and constrained the location of cooling frequency and host extinction of \thisgrb. We observed that the SED could be described within the framework of an external forward shock model with $\nu_m$ $< \nu_c$ $< \nu_{x}$ spectral regime supporting the rarely observed hard electron energy index (with $p<$ 2) in an ISM-like ambient medium. However, \cite{2021arXiv210900010O} claimed that at least before the jet break phase the observed $\nu_c$ is above the X-ray band, which is inconsistent with our findings and are ruled out by our SED fitting (see \S~\ref{SED}). \par

The hard electron energy index calculated from our optical observations and analysis indicates a flat spectrum (a harder power-law distribution of the shock-accelerated electrons) at higher frequencies and can be explained in terms of two possible scenarios. The first case is a single power-law distribution of electrons (1 $<$ $p$ $<$ 2) having an exponential cutoff corresponding to maximum electron Lorentz factor \citep{2001BASI...29..107B}; and the second scenario being a double power-law distribution of electrons (1 $<$ $p1$ $<$ 2 and $p2$ $>$ 2) having an ``injection break'' \citep{2001ApJ...554..667P,2004ASPC..312..411B,2008MNRAS.388..144R}. The other alternative scenario which can explain the multi-wavelength afterglow emission is that a very narrow jet propagating into an unusually rarefied interstellar medium \citep{2021arXiv210900010O}. More recently, \cite{2022arXiv220813821M} reported the polarization measurement of the optical afterglow of \thisgrb in the time window \fermiT+5967 -- \fermiT +8245\,s and found changes in polarization properties such as polarization degree and angle. Such changes in the polarization properties have been reported for a few cases during the prompt emission indicating some evidence for a change in the order of magnetic field \citep{2019ApJ...882L..10S}. However, this has been the first case where a change in afterglow polarization properties has been observed. The temporal, spectral, and polarimetric observations of more such bright GRBs using space (\fermi, {\it INTEGRAL}, ASIM, {\it AstroSat})/ground-based instruments) will be useful to decipher the evolution of spectral properties, effects of hard electron energy index $p$ (giving rise to additional pieces of information to understand particle acceleration and shock physics in more detail) and changes in polarization properties during both prompt and afterglow phases.
 
\section*{Acknowledgements}
Authors acknowledge comments by the anonymous referee for his/her constructive comments that improved the contents and shape of this paper. 
MCG acknowledges support from the Ram\'on y Cajal Fellowship RYC2019-026465-I (funded by the MCIN/AEI /10.13039/501100011033 and the European Social Funding). YDH acknowledges support under the additional funding from the RYC2019-026465-I. RS-R acknowledges support under the CSIC-MURALES project with reference 20215AT009. MCG and AJCT acknowledge financial support from the State Agency for Research of the Spanish MCIU through the ``Center of Excellence Severo Ochoa" award to the Instituto de Astrofísica de Andalucía (SEV-2017-0709). RG and SBP acknowledge BRICS grant {DST/IMRCD/BRICS /PilotCall1 /ProFCheap/2017(G)} for the financial support. RG and SBP also acknowledge the financial support of ISRO under AstroSat archival Data utilization program (DS$\_$2B-13013(2)/1/2021-Sec.2). RG thanks to Dr. Liang Li for sharing the data files of GRB 131231A presented in Figure~13. RG is also thankful to Dr. S. Iyyani, and Dr. V. Sharma for the fruitful discussion. AA acknowledges funds and assistance provided by the Council of Scientific \& Industrial Research (CSIR), India with file no. 09/948(0003)/2020-EMR-I. AJCT acknowledges support from the Spanish Ministry Project PID2020$-$118491GB$-$I00 and Junta de Andalucia Project P20$\_$01068. MM, AL, DS,AM, and N{\O} acknowledge final support from the Research Council of Norway under Contracts 208028/F50 and 223252/F50 (CoE). Based on observations collected at the Centro Astron\'omico Hispano-Alem\'an (CAHA) at Calar Alto, operated jointly by Junta de Andalucia and Consejo Superior de Investigaciones Cientificas (IAA-CSIC). Based on observations made with the Gran Telescopio Canarias (GTC), installed at the Spanish Observatorio del Roque de los Muchachos of the Instituto de Astrofisica de Canarias, on the island of La Palma. Based on observations collected at the Observatorio de Sierra Nevada, operated by the Instituto de Astrofisica de Andalucia (IAA-CSIC). This research has used data obtained through the HEASARC Online Service, provided by the NASA-GSFC, in support of NASA High Energy Astrophysics Programs. ASIM is a mission of ESA’s SciSpace programme for scientific utilization of the {\it ISS} and non-{\it ISS} space exploration platforms and space environment analogues. ASIM and the ASIM Science Data Centre are funded by ESA and by national grants of Denmark, Norway and Spain.

\section*{Data Availability}

The data presented in this work can be made available based on the individual request to the corresponding authors.

\bibliographystyle{mnras}
\bibliography{GRB210619B}

\appendix
\section{Detectors/telescopes description}

\subsection{\fermi GBM and LAT instruments}

The \fermi Gamma-ray Burst Monitor (GBM; \citealt{2009ApJ...702..791M}) has twelve sodium iodide (NaI) and bismuth germanate (BGO) detectors, covering an energy range approximately from 8 \keV to 1\,MeV and 200 \keV to 40\,MeV, respectively. These detectors are arranged in such a way that GBM covers approximately all-sky, not hidden by the Earth, making it a crucial GRB detecting instrument with a large energy coverage. On the other hand, \fermi Large Area Telescope (LAT; \citealt{2009ApJ...697.1071A}) is a Gigaelectron-Volts (GeV) instrument working from 20\,MeV to 300\,GeV with a wide field of view (FoV). 

\subsection{ASIM instrument}

The Atmosphere-Space Interactions Monitor (ASIM; \citealt{2019arXiv190612178N}) is a mission of the European Space Agency (ESA) mounted on the starboard side of the Columbus module of the International Space Station ({\it ISS}) and operational since 2018. The main science objectives of the ASIM mission are Terrestrial Gamma-ray Flashes (TGF) and Transient Luminous Events (TLE), atmospheric phenomena associated with lightning activity. However, ASIM proved effective also in the detection of cosmic GRBs and similar phenomena \citep{2021Natur.600..621C}. The ASIM payload consists of two nadir-viewing instruments: the Modular X- and Gamma-ray Sensor (MXGS; \citealt{2019SSRv..215...23O}) and the Modular Multispectral Imaging Assembly (MMIA; \citealt{2019SSRv..215...28C}). MXGS consists of two detection planes: the Low Energy Detector (LED) based on Cadmium-Zinc Telluride (CZT) crystals, sensitive in the 50-400 \keV range, and the High-Energy Detector (HED) made of 12 crystals of Bismuth Germanium Oxide (BGO) with photomultiplier readout. HED is sensitive to photons between 300\,keV and >30\,MeV. 
Data download is enabled by a trigger logic active on time scales spanning from $300~\mu s$ to $20~ms$, tailored for the detection of sub-millisecond transients such as TGFs. When a trigger is issued, event data for a two-second time frame centered at trigger time are sent to telemetry for subsequent download.

\subsection{\swift BAT instrument}

The Burst Alert Telescope (BAT; \citealt{2005SSRv..120..143B}) on-board the Neil Gehrels Swift Observatory (\swift; \citealt{2004ApJ...611.1005G}) mission is a highly sensitive coded aperture telescope (with a large FoV) designed to detect GRBs. It works in the energy range from 15-350 \keV.

\section{FIGURES and TABLES}

\renewcommand{\thefigure}{B\arabic{figure}}
\setcounter{figure}{0}

\begin{figure}
\centering
\includegraphics[scale=0.35]{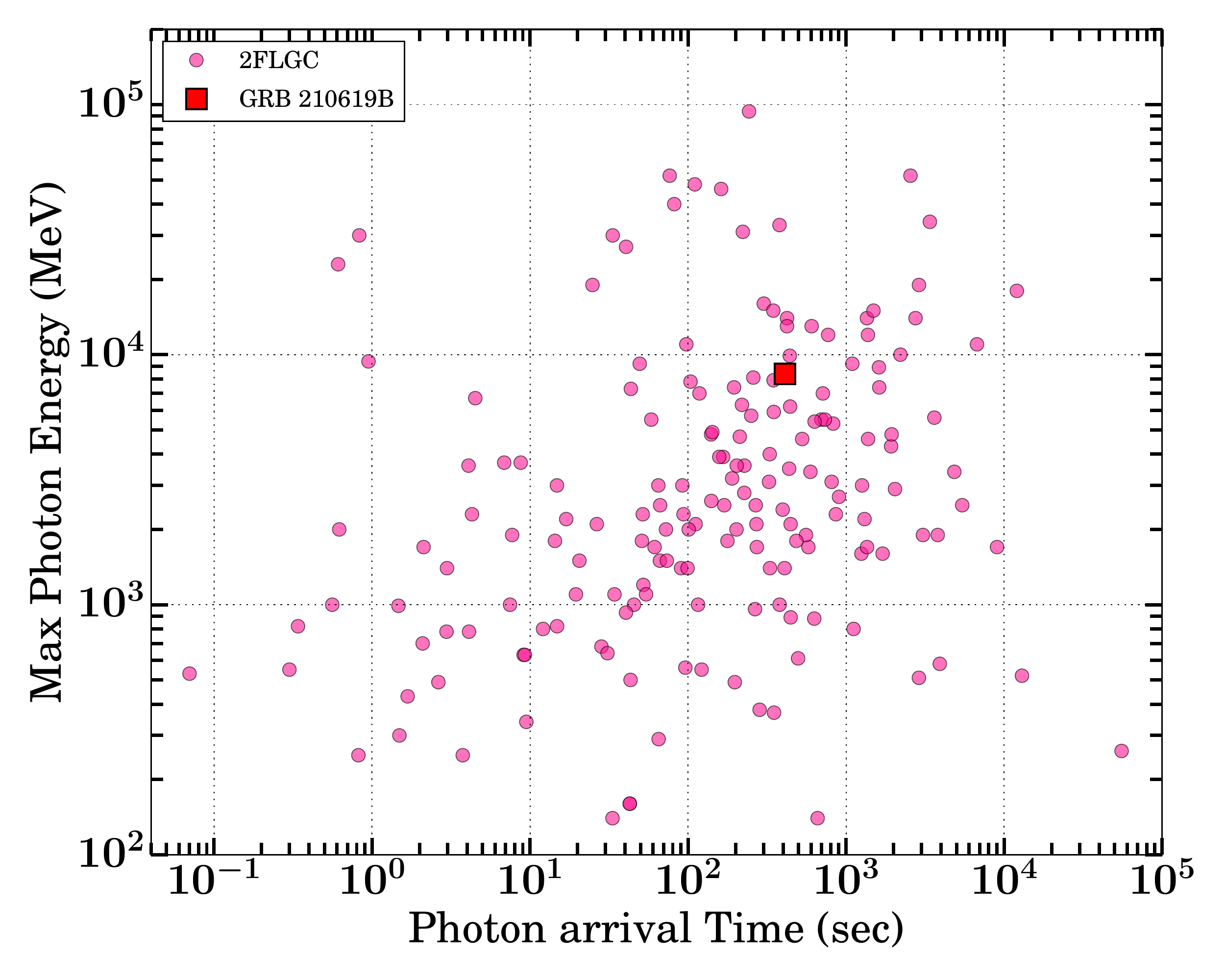}
\includegraphics[scale=0.35]{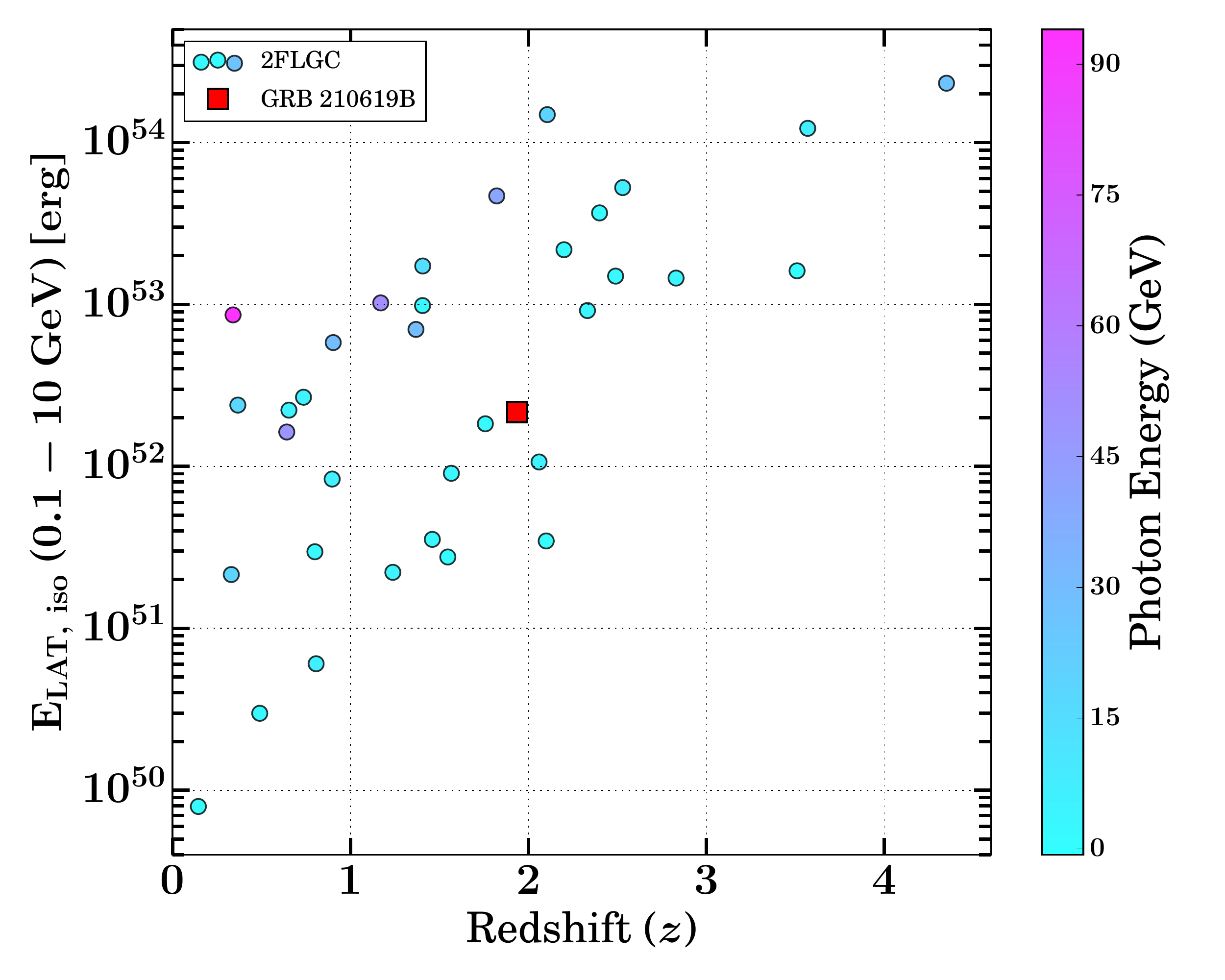}
\caption{The comparison of \thisgrb (shown with red squares) with second \fermi LAT GRB catalog (2FLGC). {Top panel:} GeV photons arrival time as a function of maximum energy. {Bottom panel:} Restframe isotropic energy in LAT energy range as a function of redshift.}
\label{GeV_LAT}
\end{figure}

\onecolumn

\begin{landscape}
\begin{scriptsize}
\begin{center}
\begin{longtable}{|c|c|c|ccccc|cccc|c|} 
\caption{Results of the time-resolved prompt emission spectral analysis of \thisgrb using \sw{Band} and \sw{CPL} functions and \fermi GBM data. The energy flux values (in erg $\rm cm^{-2}$ $\rm s^{-1}$) are calculated in 8 \keV-40\,MeV energy range. The notation S denotes the significance of the spectrum of each time bin.}
\label{TRS_Table_Bayesian} \\ \hline
T$_{\rm start}$ (s) & T$_{\rm stop}$ (s) & S & \boldmath $\it \alpha_{\rm pt}$ & \boldmath $\it \beta_{\rm pt}$ & \boldmath \Ep (\keV) &  \bf (Flux $\times 10^{-06}$)  & \bf DIC$_{\rm Band}$ & \boldmath $\it \Gamma_{\rm CPL}$ & \boldmath $E_{\rm c}$ (\keV) &  \bf (Flux $\times 10^{-06}$)  & \bf DIC$_{\rm CPL}$ & \rm \bf $\Delta$ DIC\\ \hline
0.068 & 0.213 & 39.61 & $-0.17_{-0.10}^{0.09}$ & $-1.86_{-0.06}^{+0.06}$ & $467.71_{-49.46}^{+50.36}$ & 77.32 & 100.11 & $-0.45_{-0.05}^{+0.05}$ & $530.00_{-49.55}^{+49.80}$ & 32.6 & 151.29 & -51.18 \\
0.213 & 0.302 & 41.06 & $-0.24_{-0.09}^{+0.09}$ & $-1.83_{-0.06}^{+0.06}$ & $446.77_{-50.90}^{+51.20}$ & 107.36 & -453.79 & $-0.55_{-0.04}^{+0.04}$ & $603.21_{-56.25}^{+56.72}$ & 46.82 & -403.6 & -50.19 \\
0.302 & 0.397 & 55.82 & $-0.36_{-0.07}^{+0.06}$ & $-2.01_{-0.07}^{+0.07}$ & $496.58_{-43.73}^{+44.12}$ & 114.85 & -290.18 & $-0.57_{-0.03}^{+0.03}$ & $554.48_{-42.29}^{+42.52}$ & 60.7 & -230.91 & -59.27 \\
0.397 & 0.502 & 73.81 & $-0.35_{-0.06}^{+0.06}$ & $-2.20_{-0.09}^{+0.09}$ & $493.20_{-39.94}^{+40.49}$ & 115.28 & -67.78 & $-0.58_{-0.03}^{+0.03}$ & $544.08_{-38.24}^{+37.99}$ & 78.1 & -33.02 & -34.76 \\
0.502 & 1.008 & 213.07 & $-0.35_{-0.02}^{+0.02}$ & $-2.10_{-0.03}^{+0.03}$ & $434.48_{-14.15}^{+14.07}$ & 178.35 & 2468.1 & $-0.61_{-0.01}^{+0.01}$ & $524.01_{-16.24}^{+16.30}$ & 110.24 & 3082.47 & -614.37 \\
1.008 & 1.172 & 110.46 & $-0.40_{-0.05}^{+0.05}$ & $-2.19_{-0.06}^{+0.06}$ & $367.26_{-22.66}^{+23.03}$ & 115.13 & 649.43 & $-0.61_{-0.03}^{+0.03}$ & $387.09_{-20.13}^{+20.27}$ & 72.5 & 729.02 & -79.59 \\
1.172 & 1.921 & 250.46 & $-0.47_{-0.02}^{+0.02}$ & $-2.30_{-0.03}^{+0.03}$ & $364.12_{-9.40}^{+9.48}$ & 106.9 & 2915.46 & $-0.64_{-0.01}^{+0.01}$ & $367.53_{-8.83}^{+8.93}$ & 73.2 & 3248.94 & -333.48 \\
1.921 & 2.078 & 104.55 & $-0.50_{-0.05}^{+0.05}$ & $-2.32_{-0.08}^{+0.08}$ & $286.95_{-16.52}^{+16.71}$ & 69.7 & 419.03 & $-0.68_{-0.03}^{+0.03}$ & $297.70_{-17.37}^{+17.22}$ & 47.5 & 476.94 & -57.91 \\
2.078 & 2.463 & 143.64 & $-0.36_{-0.04}^{+0.04}$ & $-2.15_{-0.04}^{+0.04}$ & $221.87_{-10.09}^{+10.13}$ & 64.49 & 1722.38 & $-0.66_{-0.02}^{+0.02}$ & $260.37_{-10.09}^{+10.33}$ & 37.8 & 1875.05 & -152.67 \\
2.463 & 2.849 & 157.93 & $-0.51_{-0.03}^{+0.03}$ & $-2.25_{-0.05}^{+0.05}$ & $295.65_{-12.54}^{+12.67}$ & 73.82 & 1813.37 & $-0.69_{-0.02}^{+0.02}$ & $314.59_{-12.51}^{+12.76}$ & 48.2 & 1949.14 & -135.77 \\
2.849 & 3.159 & 119.25 & $-0.57_{-0.05}^{+0.05}$ & $-2.26_{-0.07}^{+0.07}$ & $259.66_{-15.65}^{+16.17}$ & 49.37 & 1300.49 & $-0.75_{-0.02}^{+0.03}$ & $293.17_{-15.22}^{+15.12}$ & 32.5 & 1355.46 & -54.97 \\
3.159 & 3.556 & 114.24 & $-0.55_{-0.05}^{+0.05}$ & $-2.19_{-0.06}^{+0.06}$ & $190.54_{-11.22}^{+11.27}$ & 33.53 & 1573.62 & $-0.80_{-0.03}^{+0.03}$ & $238.38_{-13.52}^{+13.62}$ & 20.0 & 1651.65 & -78.03 \\
3.556 & 3.905 & 93.9 & $-0.60_{-0.06}^{+0.06}$ & $-2.28_{-0.08}^{+0.08}$ & $182.74_{-12.13}^{+11.97}$ & 23.57 & 1246.25 & $-0.81_{-0.03}^{+0.03}$ & $217.83_{-14.53}^{+14.30}$ & 15.3 & 1285.31 & -39.06 \\
3.905 & 4.386 & 89.92 & $-0.55_{-0.08}^{+0.08}$ & $-2.12_{-0.06}^{+0.06}$ & $156.45_{-13.59}^{+13.47}$ & 21.25 & 1685.38 & $-0.86_{-0.03}^{+0.04}$ & $224.93_{-16.10}^{+16.59}$ & 11.8 & 1736.74 & -51.36 \\
4.386 & 4.878 & 72.66 & $-0.70_{-0.08}^{+0.08}$ & $-2.15_{-0.08}^{+0.08}$ & $157.93_{-15.92}^{+15.93}$ & 14.04 & 1650.0 & $-0.97_{-0.04}^{+0.04}$ & $244.94_{-23.72}^{+23.45}$ & 8.05 & 1679.4 & -29.4 \\
4.878 & 5.855 & 78.09 & $-0.75_{-0.07}^{+0.07}$ & $-2.41_{-0.14}^{+0.14}$ & $164.70_{-13.68}^{+13.73}$ & 7.24 & 2533.13 & $-0.88_{-0.04}^{+0.04}$ & $183.12_{-14.66}^{+14.69}$ & 5.06 & 2547.54 & -14.41 \\
5.855 & 6.519 & 53.16 & $-0.60_{-0.10}^{+0.10}$ & $-2.28_{-0.12}^{+0.13}$ & $135.87_{-13.49}^{+13.28}$ & 6.06 & 1836.63 & $-0.85_{-0.06}^{+0.06}$ & $163.89_{-18.66}^{+18.28}$ & 3.75 & 1852.18 & -15.55 \\
6.519 & 8.47 & 73.61 & $-0.84_{-0.05}^{+0.05}$ & $-2.47_{-0.14}^{+0.15}$ & $180.81_{-11.83}^{+11.74}$ & 4.39 & 3434.79 & $-0.92_{-0.04}^{+0.04}$ & $198.40_{-17.00}^{+16.87}$ & 3.19 & 3448.67 & -13.88 \\
8.47 & 9.594 & 67.88 & $-0.61_{-0.06}^{+0.06}$ & $-2.13_{-0.08}^{+0.08}$ & $207.45_{-15.92}^{+15.68}$ & 10.09 & 2693.48 & $-0.80_{-0.04}^{+0.04}$ & $247.71_{-21.32}^{+21.19}$ & 5.42 & 2725.47 & -31.99 \\
9.594 & 10.083 & 35.24 & $-0.81_{-0.14}^{+0.14}$ & $-2.07_{-0.21}^{+0.20}$ & $208.97_{-49.91}^{+51.72}$ & 7.61 & 1424.88 & $-0.96_{-0.06}^{+0.06}$ & $291.76_{-46.09}^{+45.79}$ & 3.61 & 1437.42 & -12.54 \\
10.083 & 11.346 & 37.26 & $-1.00_{-0.09}^{+0.09}$ & $-2.25_{-0.26}^{+0.25}$ & $206.89_{-37.22}^{+37.08}$ & 3.36 & 2684.86 & $-1.08_{-0.06}^{+0.06}$ & $277.93_{-44.76}^{+44.64}$ & 1.99 & 2691.42 & -6.56 \\
13.652 & 14.807 & 47.61 & $-0.83_{-0.07}^{+0.07}$ & $-2.18_{-0.14}^{+0.14}$ & $225.54_{-26.34}^{+25.82}$ & 5.68 & 2628.19 & $-0.96_{-0.05}^{+0.05}$ & $292.08_{-38.63}^{+38.30}$ & 3.3 & 2641.13 & -12.94 \\
14.807 & 15.716 & 66.97 & $-0.86_{-0.05}^{+0.05}$ & $-2.31_{-0.14}^{+0.15}$ & $238.77_{-21.99}^{+22.26}$ & 8.62 & 2384.08 & $-0.95_{-0.04}^{+0.04}$ & $287.80_{-27.58}^{+27.73}$ & 5.71 & 2395.35 & -11.27 \\
15.716 & 16.624 & 50.21 & $-0.94_{-0.07}^{+0.07}$ & $-2.55_{-0.23}^{+0.23}$ & $158.62_{-15.74}^{+15.61}$ & 3.78 & 2314.14 & $-1.02_{-0.06}^{+0.06}$ & $187.58_{-24.39}^{+24.90}$ & 2.87 & 2319.26 & -5.12 \\
16.624 & 19.933 & 62.17 & $-0.93_{-0.09}^{+0.09}$ & $-2.26_{-0.13}^{+0.13}$ & $115.98_{-13.21}^{+13.11}$ & 2.54 & 4089.06 & $-1.09_{-0.05}^{+0.05}$ & $169.93_{-18.91}^{+19.06}$ & 1.54 & 4107.09 & -18.03 \\
19.933 & 20.414 & 30.45 & $-0.81_{-0.16}^{+0.16}$ & $-2.54_{-0.28}^{+0.27}$ & $120.73_{-18.61}^{+18.56}$ & 2.65 & 1302.6 & $-0.95_{-0.11}^{+0.11}$ & $139.61_{-27.72}^{+27.80}$ & 1.97 & 1306.19 & -3.59 \\
20.414 & 22.098 & 83.25 & $-0.78_{-0.06}^{+0.06}$ & $-2.40_{-0.10}^{+0.10}$ & $130.48_{-7.73}^{+7.68}$ & 4.97 & 3285.78 & $-0.93_{-0.04}^{+0.04}$ & $151.82_{-11.91}^{+11.72}$ & 3.46 & 3315.72 & -29.94 \\
22.098 & 22.991 & 44.62 & $-0.88_{-0.10}^{+0.10}$ & $-2.73_{-0.21}^{+0.22}$ & $98.14_{-7.82}^{+7.86}$ & 2.24 & 2266.88 & $-0.99_{-0.08}^{+0.08}$ & $109.19_{-15.68}^{+15.81}$ & 1.8 & 2270.88 & -4.0 \\
31.821 & 34.636 & 31.63 & $-1.05_{-0.12}^{+0.12}$ & $-2.74_{-0.29}^{+0.29}$ & $102.95_{-12.24}^{+12.10}$ & 0.83 & 3794.2 & $-1.13_{-0.09}^{+0.09}$ & $133.40_{-25.91}^{+25.80}$ & 0.68 & 3793.86 & 0.34 \\
35.169 & 35.633 & 38.92 & $-0.94_{-0.06}^{+0.06}$ & $-2.54_{-0.30}^{+0.29}$ & $289.46_{-32.92}^{+32.93}$ & 6.06 & 1492.65 & $-0.95_{-0.06}^{+0.06}$ & $298.59_{-42.07}^{+41.31}$ & 4.49 & 1494.31 & -1.66 \\
35.633 & 36.354 & 38.32 & $-0.86_{-0.08}^{+0.08}$ & $-2.86_{-0.31}^{+0.31}$ & $180.31_{-18.21}^{+17.92}$ & 2.88 & 1972.32 & $-0.90_{-0.07}^{+0.07}$ & $175.68_{-24.39}^{+24.20}$ & 2.48 & 1970.36 & 1.96 \\
36.354 & 36.638 & 31.9 & $-0.68_{-0.14}^{+0.14}$ & $-2.43_{-0.23}^{+0.24}$ & $142.74_{-19.41}^{+19.11}$ & 4.62 & 683.44 & $-0.83_{-0.10}^{+0.10}$ & $150.54_{-27.45}^{+27.34}$ & 3.06 & 687.04 & -3.6 \\
36.638 & 36.961 & 52.31 & $-0.56_{-0.10}^{+0.10}$ & $-2.25_{-0.13}^{+0.13}$ & $178.78_{-19.76}^{+19.23}$ & 11.33 & 1081.59 & $-0.79_{-0.06}^{+0.06}$ & $210.28_{-25.39}^{+25.90}$ & 6.96 & 1097.6 & -16.01 \\
36.961 & 37.44 & 47.35 & $-0.92_{-0.10}^{+0.10}$ & $-2.51_{-0.24}^{+0.24}$ & $169.29_{-24.24}^{+23.96}$ & 5.38 & 1552.48 & $-1.03_{-0.06}^{+0.06}$ & $212.09_{-30.73}^{+30.22}$ & 4.03 & 1559.1 & -6.62 \\
37.44 & 38.25 & 75.05 & $-0.74_{-0.06}^{+0.06}$ & $-2.70_{-0.18}^{+0.18}$ & $141.42_{-8.28}^{+8.41}$ & 5.62 & 2240.16 & $-0.82_{-0.05}^{+0.05}$ & $134.20_{-10.28}^{+10.14}$ & 4.52 & 2247.91 & -7.75 \\
38.25 & 39.556 & 66.59 & $-0.76_{-0.09}^{+0.09}$ & $-2.46_{-0.14}^{+0.14}$ & $100.32_{-7.94}^{+8.08}$ & 3.53 & 2905.42 & $-0.96_{-0.05}^{+0.05}$ & $121.76_{-11.60}^{+11.72}$ & 2.55 & 2920.88 & -15.46 \\
39.556 & 41.849 & 64.94 & $-1.00_{-0.07}^{+0.07}$ & $-2.31_{-0.12}^{+0.12}$ & $123.41_{-11.14}^{+11.25}$ & 3.05 & 3676.57 & $-1.14_{-0.05}^{+0.05}$ & $187.30_{-21.69}^{+21.62}$ & 1.98 & 3693.33 & -16.76 \\
41.849 & 44.578 & 54.62 & $-1.04_{-0.17}^{+0.18}$ & $-2.41_{-0.27}^{+0.26}$ & $92.25_{-20.53}^{+19.47}$ & 1.85 & 3815.0 & $-1.25_{-0.05}^{+0.05}$ & $168.82_{-23.01}^{+23.58}$ & 1.28 & 3873.71 & -58.71 \\
44.578 & 45.205 & 34.78 & $-1.06_{-0.11}^{+0.11}$ & $-2.84_{-0.28}^{+0.28}$ & $98.25_{-9.77}^{+9.71}$ & 1.96 & 1733.51 & $-1.11_{-0.10}^{+0.10}$ & $120.29_{-22.03}^{+21.62}$ & 1.66 & 1732.15 & 1.36 \\
45.869 & 46.996 & 56.31 & $-0.91_{-0.13}^{+0.13}$ & $-2.57_{-0.22}^{+0.21}$ & $84.94_{-10.13}^{+10.06}$ & 2.58 & 2607.69 & $-1.10_{-0.07}^{+0.07}$ & $115.68_{-14.22}^{+14.13}$ & 1.94 & 2619.28 & -11.59 \\
46.996 & 47.344 & 51.11 & $-0.78_{-0.08}^{+0.08}$ & $-3.19_{-0.28}^{+0.28}$ & $137.16_{-8.78}^{+8.71}$ & 4.83 & 1027.46 & $-0.81_{-0.07}^{+0.07}$ & $121.35_{-13.39}^{+13.26}$ & 4.4 & 1022.74 & 4.72 \\
47.6 & 47.974 & 76.9 & $-0.59_{-0.05}^{+0.06}$ & $-2.82_{-0.21}^{+0.22}$ & $193.50_{-10.68}^{+10.54}$ & 11.73 & 1421.85 & $-0.65_{-0.04}^{+0.04}$ & $158.08_{-10.98}^{+10.83}$ & 9.77 & 1424.93 & -3.08 \\
47.974 & 48.695 & 72.5 & $-0.66_{-0.08}^{+0.08}$ & $-2.44_{-0.11}^{+0.12}$ & $115.31_{-8.21}^{+8.12}$ & 6.4 & 2157.85 & $-0.88_{-0.05}^{+0.05}$ & $132.80_{-11.48}^{+11.45}$ & 4.58 & 2177.07 & -19.22 \\
48.695 & 49.243 & 41.06 & $-0.92_{-0.14}^{+0.14}$ & $-2.43_{-0.19}^{+0.19}$ & $90.52_{-10.98}^{+10.76}$ & 3.04 & 1640.68 & $-1.13_{-0.10}^{+0.10}$ & $134.34_{-26.72}^{+26.46}$ & 2.15 & 1651.54 & -10.86 \\
49.243 & 50.071 & 31.76 & $-1.17_{-0.14}^{+0.13}$ & $-2.53_{-0.30}^{+0.29}$ & $111.41_{-20.52}^{+20.29}$ & 1.88 & 2164.14 & $-1.26_{-0.09}^{+0.09}$ & $178.14_{-40.38}^{+40.32}$ & 1.37 & 2169.89 & -5.75 \\
50.431 & 50.517 & 33.99 & $-0.37_{-0.14}^{+0.14}$ & $-2.21_{-0.19}^{+0.19}$ & $252.46_{-40.36}^{+41.02}$ & 24.98 & -670.88 & $-0.62_{-0.08}^{+0.08}$ & $269.88_{-38.82}^{+39.72}$ & 14.6 & -663.99 & -6.89 \\
50.517 & 51.143 & 111.14 & $-0.53_{-0.04}^{+0.04}$ & $-2.65_{-0.15}^{+0.15}$ & $256.88_{-11.44}^{+11.30}$ & 19.45 & 2254.79 & $-0.59_{-0.03}^{+0.03}$ & $203.12_{-9.28}^{+9.53}$ & 15.1 & 2272.57 & -17.78 \\
51.143 & 51.339 & 48.03 & $-0.59_{-0.13}^{+0.13}$ & $-2.12_{-0.11}^{+0.11}$ & $134.42_{-17.31}^{+17.36}$ & 14.33 & 440.09 & $-0.87_{-0.07}^{+0.07}$ & $180.85_{-26.03}^{+25.98}$ & 7.31 & 465.91 & -25.82 \\
51.339 & 51.804 & 49.62 & $-0.75_{-0.10}^{+0.10}$ & $-2.55_{-0.21}^{+0.21}$ & $132.36_{-12.83}^{+12.90}$ & 5.35 & 1493.78 & $-0.88_{-0.07}^{+0.07}$ & $140.52_{-16.10}^{+16.14}$ & 4.02 & 1500.38 & -6.6 \\
51.804 & 52.376 & 68.71 & $-0.66_{-0.08}^{+0.08}$ & $-2.63_{-0.15}^{+0.16}$ & $116.88_{-7.24}^{+7.27}$ & 5.75 & 1853.42 & $-0.79_{-0.06}^{+0.06}$ & $112.55_{-9.99}^{+9.95}$ & 4.48 & 1867.32 & -13.9 \\
52.376 & 52.651 & 31.12 & $-0.99_{-0.17}^{+0.16}$ & $-2.66_{-0.28}^{+0.27}$ & $87.95_{-12.01}^{+11.95}$ & 2.8 & 630.89 & $-1.14_{-0.12}^{+0.12}$ & $122.50_{-27.16}^{+26.87}$ & 2.23 & 633.62 & -2.73 \\
52.651 & 53.803 & 42.75 & $-0.62_{-0.20}^{+0.21}$ & $-2.38_{-0.13}^{+0.13}$ & $62.61_{-7.19}^{+7.18}$ & 2.01 & 2655.99 & $-1.04_{-0.09}^{+0.09}$ & $92.16_{-13.91}^{+14.01}$ & 1.36 & 2694.36 & -38.37 \\
53.803 & 54.719 & 56.65 & $-0.64_{-0.09}^{+0.09}$ & $-2.83_{-0.18}^{+0.18}$ & $93.91_{-4.88}^{+4.75}$ & 2.77 & 2387.05 & $-0.74_{-0.07}^{+0.07}$ & $82.11_{-7.98}^{+7.84}$ & 2.34 & 2394.05 & -7.0 \\
54.719 & 56.207 & 55.52 & $-0.83_{-0.09}^{+0.09}$ & $-2.94_{-0.23}^{+0.23}$ & $77.76_{-4.39}^{+4.40}$ & 1.74 & 3054.06 & $-0.91_{-0.07}^{+0.07}$ & $77.23_{-7.70}^{+7.80}$ & 1.5 & 3055.92 & -1.86 \\
58.382 & 59.675 & 64.29 & $-0.60_{-0.14}^{+0.14}$ & $-2.64_{-0.14}^{+0.15}$ & $63.29_{-4.54}^{+4.54}$ & 2.34 & 2832.41 & $-0.90_{-0.07}^{+0.07}$ & $69.32_{-6.64}^{+6.58}$ & 1.82 & 2859.66 & -27.25 \\
59.675 & 60.966 & 45.17 & $-0.64_{-0.23}^{+0.23}$ & $-2.56_{-0.14}^{+0.14}$ & $45.87_{-4.27}^{+4.25}$ & 1.41 & 2723.52 & $-1.11_{-0.11}^{+0.11}$ & $66.40_{-10.29}^{+10.45}$ & 1.08 & 2789.89 & -66.37 \\
60.966 & 61.509 & 38.74 & $-0.54_{-0.16}^{+0.17}$ & $-3.10_{-0.24}^{+0.23}$ & $55.32_{-3.21}^{+3.23}$ & 1.66 & 1541.7 & $-0.72_{-0.13}^{+0.13}$ & $47.45_{-6.29}^{+6.22}$ & 1.47 & 1538.24 & 3.46 \\
61.509 & 62.273 & 35.2 & $-0.75_{-0.26}^{+0.27}$ & $-2.62_{-0.16}^{+0.16}$ & $41.18_{-3.87}^{+3.98}$ & 1.31 & 1948.82 & $-1.21_{-0.14}^{+0.14}$ & $66.40_{-13.28}^{+13.36}$ & 1.01 & 2008.03 & -59.21 \\
62.273 & 63.528 & 30.68 & $-0.46_{-0.28}^{+0.28}$ & $-2.54_{-0.13}^{+0.13}$ & $37.16_{-3.34}^{+3.30}$ & 0.88 & 2601.14 & $-1.24_{-0.14}^{+0.14}$ & $69.83_{-15.51}^{+15.71}$ & 0.68 & 2667.76 & -66.62 \\ \hline
\end{longtable}
\end{center}
\end{scriptsize}
\end{landscape}

\begin{table*}
        \centering
    \caption{{\it ASIM} time-resolved spectral analysis results for \thisgrb . Time intervals have been referred with respect to the ASIM reference time (T$_{\rm 0, ASIM}$).}
    \label{tab:asim}
    \begin{tabular}{ccccc}
    \hline
    Interval      & Time interval & Photon Index &  ${\chi}^2$\,(d.o.f.) &  Flux ($0.5-10$\,MeV)  \\
                  & (s)           &              &                       &  ($10^{-5}\,{\rm erg}\,{\rm cm}^{-2}{\rm s}^{-1}$)                 \\
    \hline
1                 & $0.00-1.00$      & $2.11^{+0.12}_{-0.11}$       &  11.69\,(16)      &  $1.31^{+0.23}_{-0.19}$  \\
2                 & $1.00-1.70$      & $2.25^{+0.03}_{-0.03}$    &  15.92\,(12)       &  $6.89^{+0.27}_{-0.26}$    \\
3                 & $1.70-2.18$      & $2.49^{+0.06}_{-0.06}$    &  6.99\,(12)       &  $4.02^{+0.30}_{-0.28}$   \\
4                 & $2.18-2.79$      & $2.60^{+0.09}_{-0.08}$    &  7.43\,(12)       &  $2.56^{+0.56}_{-0.46}$  \\
5                 & $2.79-3.60$      & $2.56^{+0.11}_{-0.10}$       &  13.31\,(12)      &  $1.81^{+0.51}_{-0.40}$   \\
6                 & $3.60-4.50$      & $2.97^{+0.37}_{-0.30}$       &  9.31\,(12)       &  $0.69^{+0.12}_{-0.11}$    \\
    \hline
    \end{tabular}
\end{table*}

\begin{table*}
\centering
\caption{The results of energy-resolved spectral lag analysis for \thisgrb. We considered 15-25 \keV and 8-30 \keV energy channels as a reference light curve for the \swift BAT and \fermi GBM observations, respectively. The each lag values is calculated for 50000 numbers of simulations to fit the cross-correlation function.}
\begin{tabular}{cc|cc}
\hline
\bf  &  \bf {\swift BAT} &  &  \bf {\fermi GBM} \\
\hline
\bf {Energy range \ (\keV)} &  \bf {Spectral lag (ms)} & \bf {Energy range \ (\keV)} &  \bf {Spectral lag (ms)} \\
\hline 
25-50  &  $-38.28_{-73.51}^{+73.46}$ & 30--50  & $-37.16_{-68.66}^{+68.89}$  \\ 
50-100 &  $-160.25_{-66.42}^{+66.53}$&  50--100  &  $-95.46_{-60.67}^{+61.31}$ \\ 
100-200 & $-239.73_{-46.06}^{+46.26}$ &   100--150 &  $-207.11_{-60.08}^{+60.83}$ \\ 
200-350   &$-386.78_{-36.90}^{+37.23}$ &  150--200 &  $-308.86_{-49.87}^{+50.02}$ \\ 
  &  &  200-250 & $-378.55_{-44.99}^{+44.84}$  \\
 &&  250--300  & $-405.58_{-43.91}^{+44.32}$  \\
 &&  300--400  & $-473.87_{-32.91}^{+32.98}$  \\
 &&  400--500  & $-545.70_{-31.87}^{+32.09}$  \\
 &&  500--600  & $-555.08_{-27.89}^{+27.65}$  \\
 &&  600--700  & $-579.31_{-27.19}^{+27.29}$  \\
 &&  700--800  &  $-511.07_{-26.68}^{+26.40}$ \\
 &&  800--900  & $-786.49_{-21.40}^{+21.37}$  \\
\hline
\end{tabular}
\label{tab:spectral lag}
\end{table*}

\begin{table*}
\centering
    \caption{Our optical observations log of \thisgrb. No extinction corrections are applied in the reported magnitude values.}
    \label{tab:opt}
    \begin{tabular}{ccccccc}
    \hline
    Start time& End time & Exposure & Filter & Mag  & error& Telescope\\
(UT) &(UT)&(s)&&&&\\
    \hline
    $2021-06-20T14:20:15.554$ & $2021-06-20T15:06:12.596$ & 60$\times$15 &Clear& $>18.54$& nan&BOOTES-4\\
    \hline
    $2021-06-20T22:40:39.21$ & $2021-06-20T23:05:57.03$ & 90$\times$5 &B& $>20.76$& nan&OSN\\
    $2021-06-20T22:42:16.57$ & $2021-06-20T23:07:40.52$ & 90$\times$5  &V& 20.54 & 0.20&OSN\\
    $2021-06-20T22:33:32.16$ & $2021-06-20T22:39:39.78$ & 90$\times$4 &R& 19.52 & 0.12&OSN\\
    $2021-06-20T22:43:54.15$ & $2021-06-20T23:09:24.76$ & 90$\times$5  &I& 18.94& 0.12&OSN\\
    $2021-06-22T02:07:27.55$ & $2021-06-22T02:54:19.07$ & 300$\times$5 &R& 20.87 & 0.11&OSN\\ 
    $2021-06-22T02:12:36.78$ & $2021-06-22T02:59:26.09$ & 300$\times$5  &I& 20.59& 0.17&OSN\\
    $2021-06-22T23:16:32.19$ & $2021-06-23T01:03:12.29$ & 300$\times$9 &R& 21.93 & 0.20&OSN\\ 
    $2021-06-22T23:21:42.05$ & $2021-06-23T01:08:24.00$ & 300$\times$9  &I& 21.15& 0.17&OSN\\
    $2021-06-24T23:35:35.23$ & $2021-06-25T00:27:24.91$ & 300$\times$10  &I& 21.77& 0.46&OSN\\
    \hline
    $2021-06-27T02:59:56$ & $2021-06-27T03:25:28$ & 180$\times$6  &r& 23.04& 0.29&CAHA\\
    \hline
 $2021-06-25T04:12:48.732$ & $2021-06-25T04:16:19.954$ & 20$\times$2  &r& 22.21& 0.12&GTC\\
    $2021-07-10T03:17:43.533$ & $2021-07-10T03:40:05.462$ & 90$\times$12 &g& 25.74& 0.26&GTC\\
    $2021-07-10T03:40:36.943$ & $2021-07-10T04:02:59.557$ & 90$\times$12 &r& 24.75& 0.26&GTC\\
    $2021-07-10T04:03:31.040$ & $2021-07-10T04:18:53.471$ & 60$\times$12 &i& 23.89& 0.09&GTC\\
    $2021-07-10T04:20:25.973$ & $2021-07-10T04:32:21.711$ & 50$\times$10 &z& 23.85& 0.21&GTC\\
    $2021-07-30T02:56:04.292$ & $2021-07-30T03:18:26.445$ & 90$\times$12 &g& $>$26.7& nan &GTC\\
    $2021-07-30T03:18:56.263$ & $2021-07-30T03:40:18.501$ & 90$\times$12 &r& $>$25.7& nan &GTC\\
    $2021-07-30T03:41:48.283$ & $2021-07-30T03:58:09.245$ & 60$\times$12 &i& $>$26.4& nan &GTC\\
    $2021-07-30T03:58:40.515$ & $2021-07-30T04:10:35.009$ & 50$\times$10 &z& $>$24.8& nan &GTC\\
    \hline
    \end{tabular}
\end{table*}

\begin{table*}
\caption{The optical observations log of \thisgrb taken from literature. No extinction corrections are applied in the reported magnitude values (in AB system).}
\begin{center}
\begin{tabular}{c c c c c}
\hline
\bf $\rm \bf T_{mid}$ (s) & \bf Magnitude  &\bf Filter & \bf Telescope & \bf References\\
\hline
9307 & 18.58 $\pm$ 0.03 & g& Liverpool & \cite{2021GCN.30271....1P} \\
13755 & 18.93 $\pm$ 0.03 & g& Liverpool & \cite{2021GCN.30271....1P} \\
72180 & 21.14$\pm$ 0.07 &g & GIT & \cite{2021GCN.30286....1K} \\
65700 & 21.13 $\pm$ 0.07  & g& GIT & \cite{2021GCN.30286....1K}\\
2771 & 17.80 $\pm$ 0.05& g& D50 & \cite{2021arXiv210900010O}\\
2902 & 17.93 $\pm$ 0.06 & g& D50 & \cite{2021arXiv210900010O}\\
3061 & 17.87 $\pm$ 0.05 & g& D50 & \cite{2021arXiv210900010O}\\
5384 & 18.14 $\pm$ 0.10 & g& D50 & \cite{2021arXiv210900010O}\\
\hline
13848&  18.11$\pm$ 0.03 &r & Liverpool & \cite{2021GCN.30271....1P} \\
9399& 17.82 $\pm$ 0.03&r & Liverpool & \cite{2021GCN.30271....1P} \\
1318& 16.10 $\pm$ 0.10 &r & Liverpool & \cite{2021GCN.30280....1S} \\
67284& 20.18 $\pm$ 0.05&r & GIT & \cite{2021GCN.30286....1K} \\
22104& 18.52 $\pm$  0.12&r & REM & \cite{2021GCN.30288....1D} \\
84154&  20.50 $\pm$ 0.40 & r& iTelescope & \cite{2021GCN.30292....1R}\\
236914& 21.82 $\pm$ 0.26 & r& AZT-20 & \cite{2021GCN.30791....1B} \\
256068& 21.45 $\pm$ 0.33 & r& RC80 & \cite{2021GCN.30320....1V} \\
529067&  22.56 $\pm$ 0.16 & r& CAHA & \cite{2021GCN.30338....1K}\\
3442&  17.20 $\pm$ 0.05& r&D50 & \cite{2021arXiv210900010O}\\
3505& 17.20 $\pm$ 0.06& r&D50 & \cite{2021arXiv210900010O}\\
3569& 17.19 $\pm$ 0.06 & r&D50 & \cite{2021arXiv210900010O}\\
3633& 17.30 $\pm$ 0.06 & r& D50 & \cite{2021arXiv210900010O}\\
3696& 17.30 $\pm$ 0.07 & r&D50 & \cite{2021arXiv210900010O}\\
3760&  17.29 $\pm$ 0.06 & r& D50 & \cite{2021arXiv210900010O}\\
3823&  17.23 $\pm$ 0.06 & r& D50 & \cite{2021arXiv210900010O}\\
3886&  17.27 $\pm$ 0.06& r&D50 & \cite{2021arXiv210900010O}\\
5799& 17.57 $\pm$ 0.08& r&D50 & \cite{2021arXiv210900010O}\\
5988& 17.57 $\pm$ 0.09& r&D50 & \cite{2021arXiv210900010O}\\
6177& 17.56$\pm$ 0.10  & r&D50 & \cite{2021arXiv210900010O}\\
84656& 20.16 $\pm$ 0.19& r&D50 & \cite{2021arXiv210900010O}\\
163842& 21.12 $\pm$ 0.48  & r&D50 & \cite{2021arXiv210900010O}\\
\hline
9490& 17.44 $\pm$ 0.04  & i& Liverpool & \cite{2021GCN.30280....1S} \\
13939&17.74  $\pm$ 0.04&i & Liverpool & \cite{2021GCN.30280....1S} \\
3952& 16.88 $\pm$ 0.05 & i& D50 & \cite{2021arXiv210900010O}\\
4014&  17.05 $\pm$ 0.07 & i&D50 & \cite{2021arXiv210900010O}\\
4078&  16.93 $\pm$ 0.06& i&D50 & \cite{2021arXiv210900010O}\\
4142& 17.00 $\pm$ 0.07 & i&D50 & \cite{2021arXiv210900010O}\\
4206&  16.91$\pm$ 0.07 & i&D50 & \cite{2021arXiv210900010O}\\
4271& 17.00 $\pm$ 0.07& i& D50 & \cite{2021arXiv210900010O}\\
4335& 17.03 $\pm$ 0.07& i&D50 & \cite{2021arXiv210900010O}\\
4399& 17.01 $\pm$ 0.08 & i& D50 & \cite{2021arXiv210900010O}\\
4462& 17.02 $\pm$ 0.08 & i&D50 & \cite{2021arXiv210900010O}\\
\hline
14149& 17.47 $\pm$ 0.04& z& Liverpool & \cite{2021GCN.30280....1S} \\
9700& 17.17 $\pm$ 0.04 &z & Liverpool & \cite{2021GCN.30280....1S} \\
4589& 16.77 $\pm$ 0.06 &z &D50 & \cite{2021arXiv210900010O}\\
4781&  16.81 $\pm$ 0.08 &z & D50 & \cite{2021arXiv210900010O}\\
4971& 16.68 $\pm$ 0.06 & z&D50 & \cite{2021arXiv210900010O}\\
6865& 16.90 $\pm$ 0.11 & z&D50 & \cite{2021arXiv210900010O}\\
\hline
9498& 17.90 $\pm$ 0.20 & V& iTelescope & \cite{2021GCN.30265....1K} \\
\hline
51948&  19.21 $\pm$ 0.10 &R & TNT & \cite{2021GCN.30277....1X} \\
85679&  20.01 $\pm$ 0.10 & R& SAO RAS& \cite{2021GCN.30291....1M} \\
81215&  20.11 $\pm$ 0.20& R& OSN &  \cite{2021GCN.30293....1H}\\
241580& 21.42 $\pm$ 0.09&R & AZT-22 &  \cite{2021GCN.30791....1B}\\
593153&  22.76 $\pm$ 0.24& R& AZT-22 &  \cite{2021GCN.30791....1B}\\
1111648& 23.71 $\pm$ 0.30 & R& AZT-22 &  \cite{2021GCN.30791....1B}\\
69088& 19.68 $\pm$ 0.06&  R& AZT-22 &  \cite{2021GCN.30791....1B}\\
159856& 20.64 $\pm$ 0.05& R& AZT-22 &  \cite{2021GCN.30791....1B}\\
102244& 20.13 $\pm$ 0.02& R& NOT & \cite{2021GCN.30294....1Z} \\
171876& 20.81 $\pm$ 0.30 & R& AS-32 & \cite{2021GCN.30299....1B} \\
171289& 20.94 $\pm$ 0.08 & R& SAO RAS & \cite{2021GCN.30303....1M} \\
99792& 20.21 $\pm$ 0.20 & R& Schmidt-Cassegrain &  \cite{2021GCN.30305....1R}\\
255989& 21.61  $\pm$ 0.10 & R& SAO RAS  & \cite{2021GCN.30309....1M} \\
\hline
5042& 16.75 $\pm$ 0.05 & I& CAHA & \cite{2021GCN.30275....1K}\\
13373& 17.45 $\pm$ 0.05 & I& CAHA &  \cite{2021GCN.30275....1K}\\ \hline
\end{tabular}
\end{center}
\label{tab:observationslog}
\end{table*}
\bsp	
\label{lastpage}
\end{document}